\documentclass[prb,reprint,a4paper,citeautoscript]{revtex4-2}

\pdfoutput=1

\usepackage[charter]{mathdesign}
\usepackage{amsmath}
\let\mathcal\relax
\usepackage[cal=boondox,scr=boondoxo]{mathalfa}

\usepackage{mathtools}
\usepackage[]{stackengine}
\usepackage[pdftex]{graphicx}
\usepackage{microtype}
\usepackage{bm}\let\vec\bm
\usepackage{gensymb}
\usepackage{mdframed}
\usepackage[bbgreekl]{mathbbol}
\usepackage[svgnames]{xcolor}
\usepackage[
    colorlinks=True,
    linkcolor=DarkRed,
    citecolor=ForestGreen,
    urlcolor=MediumBlue,
    pdfstartview=FitH,
    bookmarks=False,
    pdfpagemode=UseNone
]{hyperref}
\usepackage{feynmp-auto}
\usepackage{natbib}
\usepackage{breqn}
\usepackage{tcolorbox}

\makeatletter
\let\cat@comma@active\@empty
\makeatother

\begin{document}

\title{The flow of local quantum fluids: Conservation laws and vertex corrections from many-body linear-response theory with local self-energy}

\author{D. Valentinis}

\affiliation{Institut f\"{u}r Quantenmaterialien und Technologien, Karlsruher Institut f\"{u}r Technologie, 76131 Karlsruhe, Germany}
\affiliation{Institut f\"{u}r Theorie der Kondensierten Materie, Karlsruher Institut f\"{u}r Technologie, 76131 Karlsruhe, Germany}
\date{\today}

\begin{abstract}
In non-diffusive conduction regimes of strongly correlated quantum electron systems, electromagnetic perturbations simultaneously probe the electronic dynamics in time and space: the exchanged energy $\hbar \omega$ excites retarded, i.e.\@, frequency-dependent, many-body interactions, while the probing spatial modulation renders the response spatially nonlocal, i.e.\@, dependent on the external wave vector $\vec{q}$. 
This work derives the nonlocal electrodynamic response of such dynamical quantum fluids under the assumptions of local but frequency-dependent self-energies, local but frequency-dependent particle-hole irreducible two-particle vertex functions, and charge/mass conservation. The latter is ensured by Bethe-Salpeter equations for renormalized interaction vertices, entering the Kubo formalism for two-particle correlation functions (e.g.\@, for density, currents, momentum, stress). 
Within such a framework, exact symmetry criteria for the presence or absence of vertex corrections are inferred. In particular, it is shown that vertex corrections generally vanish at $q=0$ for single-particle dispersions that are even and bare interaction vertices that are odd with respect to specific point group transformations in momentum space, including inversion for vector vertices, and mirror reflections or two- or higher-fold rotations for tensor vertices. If the dispersion is isotropic, then the transverse part (orthogonal to $\vec{q}$) of vectorial vertices (such as for electric/particle current, momentum, and energy current) is devoid of vertex corrections at any finite $\hbar \omega$ and $\vec{q}$. Moreover, for quadratic and isotropic dispersion the full current-current correlation function can be computed with the bare (unrenormalized) current vertex, at any momentum $\vec{q}$ and frequency $\omega$, and still preserving charge conservation. The robustness of these criteria against further symmetry breaking, multiband effects, and additionally imposing momentum conservation, is discussed, with application to the Hall viscosity of Landau levels. Explicit expressions for generic nonlocal correlation functions are derived for Fermi liquids (with well-defined quasiparticle peaks) and non-Fermi liquids (devoid of quasiparticles), for arbitrary local self-energies. 
\end{abstract}

\maketitle

\section{Introduction: local self-energies and nonlocal electrodynamics in strongly correlated systems}

Correlated quantum electron systems, particularly in the regime of strong interactions, are characterized by intrinsically inhomogeneous and dynamical single-particle properties, which nontrivially evolve in real space $\vec{r}$ and time $t$, or equivalently in momentum $\vec{k}$ and frequencies $\omega$ after Fourier transformation. Such dynamics is reflected by a non-negligible frequency dependence of, e.g.\@, quasiparticle scattering rates/lifetimes and effective masses, which accompany the $\vec{k}$ dependence of the same properties \cite{Mahan-2000,Bruus-2004mb,Berthod-2018}. 

Microscopically, inhomogeneity and dynamics of one-particle features stem from nonlocal, i.e.\@, momentum-dependent, and retarded, i.e.\@, frequency-dependent, many-body interactions. In particular, retardation effects are essential to effectively describe the low-energy physics of ground and excited states in many classes of intensely investigated quantum phases of matter. These phases include condensed-matter systems like strange metals \cite{Chowdhury-2022,Phillips-2022,Checkelsky-2024}, bad metals \cite{Emery-1995,Lee-2006,Vucicevic-2015,Lucas-2017b}, Mott insulators \cite{Lee-2006,Roy-2019,Schaefer-2021}, and unconventional/high-temperature superconductors \cite{Emery-1995,Lee-2006,Chowdhury-2022,Pickett-2023}, but also quantum-simulator platforms such as unitary Fermi gases at Feshbach resonances \cite{Randeria-2014,Frank-2020,Li-2022,Fujii-2024}, as well as high-energy/astrophysics phenomena like the quark-gluon plasma \cite{Kasmaei-2016,Wang-2022}, neutron stars \cite{Blaschke-2004,Pecak-2024,Malik-2024}, and colour superconductivity in quark matter \cite{Son-1999,Brown-2000,Geissel-2024}. In the language of many-body diagrammatic theory, interactions are encoded in the retarded self-energy $\Sigma^R=\Sigma^R(\vec{k},\omega)$\footnote{The superscript ``R'' denotes retarded quantities, corresponding to the analytic continuation $i\omega_m \rightarrow \hbar\omega+i 0^+$ of the Matsubara forms of self-energies and correlation functions, e.g.\@, $\Sigma(i\omega_m)$ for the fermionic self-energy; see also App.\@ \ref{Many_body_def}.}, which quantifies the effect of many-body processes on single-particle propagation. 

Valuable simplifications to many-body calculations arise when one of the functional dependencies of the self-energy can be shown to be subdominant, at least in specific regimes/configurations. 
In particular, local self-energies $\Sigma^R(\vec{k},\omega)\equiv \Sigma^R(\omega)$ formally arise in the limit of infinite dimensionality $d\rightarrow +\infty$ or infinite coordination number $z\rightarrow+\infty$, where Dynamical Mean-Field Theory (DMFT) becomes exact \cite{Metzner-1989,Muller-1989,Khurana-1990,Georges-1996, Janis-2001, Kotliar-2004,Berthod-2013,Schaefer-2021}.
Physically, the frequency dependence of the self-energy can be pictured to dominate over the momentum dependence in multiple regimes: close to phase transitions and critical phenomena, where dynamical correlations and critical slowing down of fluctuations become especially important; in quantum critical systems, where the dynamics of fluctuating, Landau-damped soft bosonic modes generates non-Fermi liquid fermionic self-energies, as illustrated by patch theories of critical Fermi surfaces -- see App.\@ \ref{App:QCP_patch}; in high-energy processes, relevant for optical or inelastic photon scattering experiments. More details are discussed in App.\@ \ref{App:frequency_over_momentum}. 

Concurrently, two-particle correlation functions $\chi_{\hat{\Theta}\hat{\Theta}}^R(\vec{r},\vec{r}',t,t')$ for a space-time fluctuating observable $\Theta(\vec{r},t)$, measured in the same strongly interacting systems, are also nonlocal in space and retarded in time: they depend on coordinates $\left\{\vec{r},t\right\}$ different from the ones of the probing field $F(\vec{r}',t')$ \cite{Mahan-2000,Vignale-2005,Berthod-2018}. Assuming translational invariance in space and time, the differences in relative coordinates and times in $\chi_{\hat{\Theta}\hat{\Theta}}^R(\vec{r}-\vec{r}',t-t')$ translate into dependences of its Fourier modes $\chi_{\hat{\Theta}\hat{\Theta}}^R(\vec{q},\omega)$ on transferred momentum $\vec{q}$ and exchanged frequency $\omega$.

Such nonlocal and dynamical two-particle correlations are experimentally measurable, and have indeed been detected with progressively higher accuracy and in an increasingly broad parameter space by state-of-the-art setups. These breakthroughs have been propelled by striking technological upgrades in both measurement sensitivity and sample preparation. In fact, contemporary experimental probes reach the nanometer scale and ultrafast time domains; concomitantly, striking progress in material synthesis allows for fabrication of ultraclean quantum materials, which show intriguing properties beyond the standard regime of diffusive conduction. 

Hence, the technical advancements in state-of-the-art experimental setups are able to effectively reveal strong fluctuations in time and space of the probed observables. Here, it is crucial to realize that, even with a retarded but local self-energy $\Sigma^R(\omega)$, the two-particle response $\chi_{\hat{\Theta}\hat{\Theta}}(\vec{q},\Omega)$ can be, and is in practice, spatially nonlocal. 
This nonlocality manifests itself in a plethora of condensed-matter domains: electromagnetic phenomena like anomalous skin effect in isotropic \cite{Reuter-1948,Chambers-1952, Silin-1958b, Casimir-1967a,Casimir-1967b,Casimir-1967c,Wooten-1972opt,Sondheimer-2001,Dressel-2001,Valentinis-2021a,Valentinis-2021b,Maslov-2016} and anisotropic \cite{Matus-2022,Vagov-2023,Valentinis-2023,Zhou-2024,Baker-2024} materials, plasmon dispersion and damping \cite{DeCeglia-2018,Yoo-2019,Nag-2020,Haas-2023,Sharma-2024,Fiore-2024}, negative refraction in metamaterials \cite{Veselago-1968,Veselago-2006,Forcella-2014,FZVM2-2014,Forcella-2017,Zanotto-2022,Zhang-2022,Sternbach-2023,Hu-2023}, Raman scattering \cite{Weber-1980a,Weber-1980b,Tanaka-2002,Deveraux-2007,Maiti-2017,Henstridge-2022}, and magnetic penetration depth \cite{Suter-2004,Suter-2005,Kiefl-2010,Suter-2011,Roising-2024,McFadden-2026}; non-diffusive DC and AC transport regimes such as hydrodynamic \cite{Gurzhi-1959,Gurzhi-1968,Gurzhi-1972,Molenkamp-1994,deJong-1995,Conti-1999,Tokatly-2000,Fritz-2008,Muller-2009, Andreev-2011,FZVM2-2014,Schaefer-2014,Briskot-2015, Levitov-2016, Moll-2016,Crossno-2016,Bandurin-2016,Narozhny-2017,Krishna-Kumar-2017,Gorbar-2018, Gooth-2018,Jaoui-2018,Nandi-2018,Cook-2019, Kiselev-2019,Narozhny-2019,Sulpizio-2019,Berdyugin-2019,Mendl-2020,Ku-2020,Levchenko-2020,Gromov-2020,Jaoui-2021,Vool-2021,Baghramyan-2021,Cook-2021,Narozhny-2021,DeLuca-2021,Aharon-2022,Belitz-2022,Jenkins-2022,Palm-2024,Geurs-2025_preprint,DiSalvo-2025_preprint}, tomographic \cite{Guo-2018,Ledwith-2019,Nazaryan-2024, Zeng-2024_preprint, Kryhin-2025,Estrada-2025,Nilsson-2025,Moiseenko-2025,Musser-2026,Ben-Shachar-2025_preprint,Thuillier-2026_preprint}, and directional ballistic \cite{Heiblum-1985,Sondheimer-2001,Du-2008,Mackenzie-2017,Freuenfeld-2020,McGuinness-2021,Bachmann-2022,Valentinis-2023,Ingla-2023} flow of electronic carriers; hydrodynamic relaxation of non-Fermi liquids \cite{Patel-2017,Link-2020,Else-2023} and strongly coupled electron-boson systems \cite{Hosseinabadi-2023,Qiu-2025}; anomalous collective modes, like magnons \cite{Uhrig-2004,Yuan-2021,Wang-2024,Prichard-2025}, paramagnons \cite{Millis-1990,Guarise-2014,Scalapino-2012}, charge density waves/fluctuations \cite{Torchinsky-2013,Arpaia-2019,Kim-2021,Arpaia-2023,Merzoni-2024,Lacmann-2026}, and excitons with spatial structure in Mott insulators \cite{Zhang-1998,Essler-2002,Kim-2002,Hill-2008}. 

Importantly, the phenomena enumerated in the above non-exhaustive list have been experimentally observed across a broad range of material classes, including ultraclean delafossites (e.g.\@, PdCoO$_2$ \cite{Moll-2016,Bachmann-2022,Baker-2024} and PtCoO$_2$ \cite{Mackenzie-2017,Nandi-2018}), graphene \cite{Fritz-2008,Muller-2009,Crossno-2016,Bandurin-2016,Krishna-Kumar-2017,Sulpizio-2019, Berdyugin-2019} and other two-dimensional electron systems, Weyl semimetals \cite{Gooth-2018,Jaoui-2018,Vool-2021}, and low- and high-$T_c$ superconductors (Pb \cite{Suter-2004}, Nb \cite{Suter-2005,McFadden-2026}, Sr$_2$RuO$_4$ \cite{Roising-2024}, cuprates \cite{Kiefl-2010,Suter-2011}). These observations provide direct access to regimes in which the spatial structure of the response cannot be neglected. This makes the treatment of spatially dispersive response particularly timely, both for interpreting existing experiments and for emerging probes of strongly correlated and quantum-critical matter, where nonlocal transport and electrodynamics (at finite momentum $\vec{q}$ and frequency $\omega$) may provide information on spatially extended critical fluctuations and collective dynamics that is inaccessible in the local, $q\rightarrow 0$ limit.

A central challenge in the analysis of two-particle response functions $\chi^R_{\hat{\Theta}\hat{\Theta}}(\vec{q},\Omega)$ is the construction of a \emph{conserving} theory, which respects fundamental conservation laws for quantities like electric charge, current, energy, and momentum flow. In the many-body setting, these continuity equations for densities and currents translate into constraints on vertex corrections for correlation functions, enforced by Ward-Takahashi identities \cite{Ward-1950,Takahashi-1957,Schrieffer-1963th,Behn-1978,Vollhardt-1980,Nieh-1998,Mahan-2000,He-2014,Li-2023,Narozhny-2023} and governed by the local irreducible two-particle vertex $\Lambda^R$; see Sec.\@ \ref{Correl_conservations_gen}. In order to obey conservation laws, the self-energy $\Sigma^R$ and vertex corrections have to be consistently chosen, which poses a considerable hindrance to analytical progress, and in multiple cases becomes a formidable problem. 
A systematic resolution of these difficulties is provided by deriving both self-energy and vertex corrections from a Luttinger-Ward functional, whenever available \cite{Luttinger-1960,Baym-1962}; see Sec.\@ \ref{Correl_conservations_gen}. However, closed-form solutions for conserving exact propagators and vertices are rare, while uncontrolled approximations which neglect seemingly unimportant terms can be misleading and are still debated \cite{Palle-2024,Liu-2024}. For these reasons, it is desirable to devise well defined, general criteria to discern the presence/absence and relative importance of vertex corrections for many-body theories, depending on the set of imposed conservation laws. 
In the case of a local self-energy, common wisdom is that vertex corrections vanish from two-particle correlation functions at $q=0$ \cite{Pakhira-2015}, as it has been explicitly shown for the conductivity in DMFT \cite{Khurana-1990,Pakhira-2015}, but one realizes that there are exceptions to this rule \cite{Georges-1996,Kotliar-2004}; see Sec.\@ \ref{Charge_cons_vertex_vanish}. For instance, vertex corrections are essential to establish continuity equations for charge and spin, as enforced by the related Ward identities, together with self-consistency at the two-particle level between the lattice and impurity formulations of DMFT \cite{Hafermann-2014,Krien-2017}. 
This distinction calls for a systematic analysis of vertex renormalizations for both traditional and recently emerging probes of nonlocal correlations, to ascertain the presence or absence of vertex corrections even in the local limit. 

In this work, the nonlocal electrodynamic response of correlated-electron systems is addressed by assuming a framework of local, frequency-dependent, and charge-conserving interactions. In the following I employ the terms ``DMFT-like'' and ``local approximation'' to denote the locality assumptions employed here, namely a local self-energy and a local particle-hole irreducible two-particle vertex function; this terminology does not imply that every approximation satisfying these assumptions is a self-consistent DMFT solution. This DMFT-like approximation allows for a rigorous derivation of the conditions under which vertex corrections -- often the most computationally intractable part of many-body theory -- can be precisely determined, or shown to vanish.
Clear selection rules emerge based on the parity of the electronic dispersion $\xi_{\vec{k}}$ and on the bare interaction vertices $\Gamma_{\hat{\Theta},\vec{\alpha}}^{(0)}(\vec{k},\vec{q})$: at $q=0$, a symmetry point-group operator $\mathscr{g}$ should exist that leaves the dispersion invariant, $\xi_{\mathscr{g} \vec{k}}=\xi_{\vec{k}}$, and under which the bare vertex is odd, $\Gamma_{\hat{\Theta},\vec{\alpha}}^{(0)}(\mathscr{g}\vec{k},0)=-\Gamma_{\hat{\Theta},\vec{\alpha}}^{(0)}(\vec{k},0)$. 
Based on these criteria, vertex corrections are shown to be essential for density and bulk kinetic stress responses even at $q=0$ and for any frequency $\omega$, while they identically vanish for electric, momentum, and thermal currents, and in general for any vector-like vertex that is odd under $\vec{k}\mapsto -\vec{k}$ inversion. 
Conversely, nonuniversal cases arise for rank-2, tensorial vertices with spatial indices $\vec{\alpha}=\left\{\alpha,\beta\right\}$ for $\alpha \neq \beta$, such as the ones pertaining to shear stresses: these are devoid of vertex corrections under specific symmetry conditions. Namely, the components $\alpha \beta$ of shear-stress vertices are nonrenormalized (bare) if one of the following point-group operations exist: a mirror symmetry plane $\mathscr{M}_\alpha$ orthogonal to $\alpha$, a two-fold rotation $\mathscr{C}_\alpha$ about the axis $\alpha$, or an improper rotation $\mathscr{S}_n$ that involves a rotation plus a mirror plane perpendicular to $\alpha$. Moreover, higher-order rotations $\mathscr{C}_3$, $\mathscr{C}_4$, $\mathscr{C}_6 \ldots$ suppress all off-diagonal components of shear-stress vertex corrections. 

Most notably, for the case of an isotropic dispersion $\epsilon_{\vec{k}}=\epsilon_{\left|\vec{k}\right|}$ with conservation of charge but not momentum, a proof that vertex corrections to the transverse part of correlation functions for vectorial vertices (such as the current vertex) vanish at arbitrary momentum $\vec{q}$ and frequency $\omega$ is presented; this provides a rare exact result valid for both Fermi and non-Fermi liquids. If, in addition to being isotropic, the dispersion is quadratic, then it is shown that the \emph{full} current vertex can be chosen to be bare (unrenormalized), and still fulfill the charge-conservation Ward identity. 

The paper is organized as follows: Sec.\@ \ref{Correl_conservations_gen} introduces the two-point correlation functions, the renormalized interaction vertex, the DMFT-like locality assumptions, the concept of conserving approximations, and many-body continuity equations for charge/mass and momentum in the form of Ward-Takahashi identities. Sec.\@ \ref{Vertex_corrections_q0} derives and discusses symmetry-based criteria in momentum space for vanishing vertex corrections in charge-conserving many-body theories and at $q=0$; these criteria are applied to different classes of scalar, vector, and tensorial vertices, such as momentum, charge, and energy currents, as well as the kinetic stress tensor. Here we focus exclusively on the kinetic part of the stress tensor, and not on interaction contributions; see App.\@ \ref{App:conductivity_viscosity_explicit}. 
The case of nonvanishing vertex corrections is explicitly solved in Sec.\@ \ref{Charge_cons_vertex_vanish_q} for the special case of density-like interactions, whereby the irreducible two-particle vertex is only a function of exchanged frequency $i\Omega_n$. Sec.\@ \ref{Quadratic_vanish} hosts a demonstration of vanishing vertex corrections in the local, DMFT-like limit for current-like vertices, proportional to momentum $\vec{k}$, at any transferred wave vector $\vec{q} \perp \vec{k}$ (in the transverse direction with respect to $\vec{q}$) and frequency $i\Omega_n$, if the single-particle dispersion is isotropic; with the additional assumption of quadratic \emph{and} isotropic dispersion, it is shown that the full current-like vertex can be chosen to be bare while still obey the charge-conservation Ward identity. The above results lead to the analysis of linear-response Kubo formulae in Sec.\@ \ref{Kubo_charge_conserving}, where explicit expressions for Kubo correlators in the absence of vertex corrections are derived for systems endowed with (coherent), or devoid of (incoherent), well-defined quasiparticle excitations. Sec.\@ \ref{Momentum_cons_renorm} addresses the issue of momentum conservation, which can be imposed at $q=0$ in the DMFT-like limit, but cannot hold at generic $\vec{q}$ and $i\Omega_n$ unless the assumptions of local self-energy and/or local two-particle vertex are relaxed; a solution is proposed based on the direct application of the momentum-conservation Ward identity, which leads to the known relation between momentum-momentum and stress-stress correlation functions. Sec.\@ \ref{Discussion} hosts a discussion of multiple applications and generalizations of the present theory: interactions breaking parity and time-reversal symmetry (Sec.\@ \ref{Parity_time_reversal_breaking}), the case of the Hall viscosity in the absence of time reversal symmetry, and specifically for Landau diamagnetism in a single parabolic isotropic band (Sec.\@ \ref{Hall_viscosity}), phase-space enlarging required by pairing interactions (Sec.\@ \ref{Discussion_pairing}), multiband effects (Sec.\@ \ref{Multiband}), the validity of the quadratic-band approximation for low-energy electrodynamics at finite momentum (Sec.\@ \ref{Discussion_quadratic}), and experimental implications for nonlocal electrodynamics in correlated systems and for tunable vertex corrections to shear stresses induced by mechanical deformations (Sec.\@ \ref{Discussion_experiments}). A conclusive summary appears in Sec.\@ \ref{Conclusions}. 

Multiple appendixes report details on technical aspects and explicit derivations of analytical results: App.\@ \ref{Many_body_def} specifies basic definitions of many-body Green's functions, two-particle vertex functions (specifically in particle-hole notation) and their reducibility, three-point bare and renormalized vertices, Bethe-Salpeter equations, and spectral functions. App.\@ \ref{App:frequency_over_momentum} argues about conditions for the self-energy to be dominated by its frequency evolution, with subdominant momentum dependence, in specific classes of strongly correlated and quantum critical systems. Apps.\@ \ref{app:Charge_conservation} and \ref{app:Momentum_conservation} report the derivations of the many-body Ward identities for mass/charge and momentum conservation, respectively. App.\@ \ref{App:corr_Theta_bubble_q0_deriv} includes details of the derivation of two-body correlation functions at $q=0$ in the coherent and incoherent regimes. Finally, App.\@ \ref{app:GaAs_tight_binding} specifies the tight-binding model for the GaAs example, employed in the discussion of parity and time-reversal symmetry breaking.

\section{Correlation functions for nonlocal electrodynamics}\label{Correl_conservations_gen}

Let us first consider the nonlocal and dynamical two-point response function for a generic observable $\hat{\Theta}$: 
\begin{widetext} 
\begin{equation}\label{eq:chi_Theta_gen}
\chi_{\hat{\Theta} \hat{\Theta}}^{\vec{\alpha}\vec{\beta}}(\vec{q},i\Omega_n)=k_B T \sum_{i\omega_m} \frac{1}{\mathscr{V}} \sum_{\vec{k}} \Gamma_{\hat{\Theta},\vec{\alpha}}(\vec{k},\vec{q},i\omega_m,i\Omega_n) G(\vec{k},i\omega_m)G(\vec{k}+\vec{q},i\omega_m+i\Omega_n) \Gamma_{\hat{\Theta},\vec{\beta}}^{(0)}(\vec{k}+\vec{q},-\vec{q}). 
\end{equation}
The indices $\vec{\alpha}=\left\{\alpha_1,\alpha_2,\ldots \alpha_d\right\}$ and $\vec{\beta}=\left\{\beta_1,\beta_2,\ldots \beta_d \right\}$ run over spatial components $1\ldots d$ with $d$ dimensionality. For two-dimensional (2D) systems we have $\vec{\alpha}=\left\{x,y\right\}$, while in three dimensions (3D) $\vec{\alpha}=\left\{x,y,z\right\}$. The intensive form of the correlation function in Eq.\@ (\ref{eq:chi_Theta_gen}) implies the presence of the system volume $\mathscr{V}$ at the denominator. The Matsubara energies $i\Omega_n=2\pi n k_B T$ and $i\omega_m=(2 m+1) \pi k_B T$ are bosonic and fermionic, respectively \cite{Berthod-2018}. For completeness, Eq.\@ (\ref{eq:chi_Theta_gen}) is derived in App.\@ \ref{Many_body_def}; see also Fig.\@ \ref{fig:Diag_Xthetatheta_VC} for a graphical representation of Eq.\@ (\ref{eq:chi_Theta_gen}). 
As applications, this work mainly focuses on correlation functions of the kind (\ref{eq:chi_Theta_gen}) related to the electromagnetic response at zero and finite momentum, in particular the stress-stress correlation function ($\hat{\Theta}\equiv \hat{\textit{T}}$), the momentum-momentum function ($\hat{\Theta}\equiv \hat{\pi}$), and the current-current correlation function ($\hat{\Theta}\equiv \hat{J}$). These correlation functions are linked to their respective observables, given by the rank-4 viscosity tensor $\eta_{\alpha\beta\gamma\delta}(\vec{q},\omega)$ and the rank-2 conductivity tensor $\sigma_{\alpha \beta}(\vec{q},\omega)$, after analytic continuation ($i\Omega_n \rightarrow \hbar \omega+i0^+$ \footnote{In this work, bosonic and fermionic Matsubara frequencies on the imaginary axis have the units of energy (I use the terms ``Matsubara frequencies'' or ``Matsubara energies'' interchangeably, implicitly assuming a conversion factor $\hbar$ between the two terminologies), while on the real axis retarded functions depend on the energy $\hbar \omega$, which results from the analytic continuation $i\Omega_n \rightarrow \hbar \omega+i0^+$.}) in accordance with the general Kubo formula (\ref{eq:Kubo_general_Theta}); see also App.\@ \ref{Many_body_def}.  However, I will also comment on the density-density correlation function ($\hat{\Theta}\equiv \hat{n}$), specifically on its relation with the current-current correlation function entailed by the charge continuity equation. Moreover, I will comment on vertex corrections at $q=0$ for other kinds of correlation functions and related observables, such as the heat current and the thermal conductivity, specifically in Sec.\@ \ref{Vertex_corrections_q0}. 
Our analysis will mainly concern systems that preserve time-reversal symmetry (e.g.\@, in the absence of applied magnetic fields and with trivial topology); however, in Sec.\@ \ref{Hall_viscosity} I will comment on situations involving broken time-reversal symmetry, which produces a nonzero Hall viscosity \cite{Avron-1995,Read-2009,Bradlyn-2012,Hoyos-2014a, Hoyos-2014b,Burmistrov-2019}; in particular, I will consider how such symmetry breaking impacts the renormalization of the Hall component of the stress vertex. 
\begin{figure*}[ht] \centering
\includegraphics[width=1.0\linewidth]{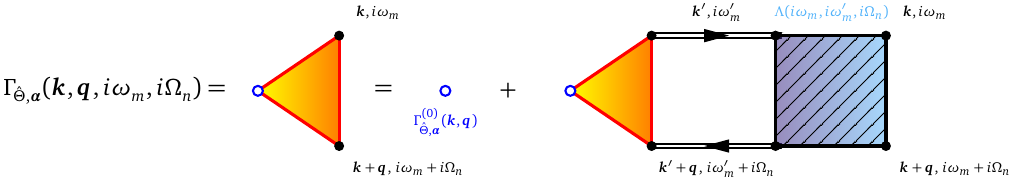}
\caption{\label{fig:Gamma_VC} Graphical representation of the Bethe-Salpeter equation (\ref{eq:Bethe_Salpeter_gen}) for the renormalized vertex $\Gamma_{\hat{\Theta},\vec{\alpha}}(\vec{k},\vec{q},i\omega_m,i\Omega_n)$ associated with the observable $\hat{\Theta}$, assuming a local irreducible two-particle vertex $\Lambda(i\omega_m, i\omega_m',i \Omega_n)$. 
 }
\end{figure*}

In general, the bare (i.e.\@, unrenormalized) vertex $\Gamma_{\hat{\Theta},\vec{\alpha}}^{(0)}(\vec{k},\vec{q})$ is renormalized by interactions as $\Gamma_{\hat{\Theta},\vec{\alpha}}(\vec{k},\vec{q},i\omega_m,i\Omega_n)$, according to the inhomogeneous Bethe-Salpeter equation in the particle-hole channel \cite{Mahan-2000,Bruus-2004mb,Rohringer-2012,Hafermann-2014,Rohringer-2018,Raines-2024}
\begin{align}\label{eq:Bethe_Salpeter_gen_nonlocal}
\Gamma_{\hat{\Theta},\vec{\alpha}}(\vec{k},\vec{q},i\omega_m,i\Omega_n)&=\Gamma_{\hat{\Theta},\vec{\alpha}}^{(0)}(\vec{k},\vec{q})+\frac{1}{\mathscr{V}}\sum_{\vec{k}'} k_B T\sum_{i\omega_m'} \Lambda(\vec{k}',\vec{k}+\vec{q},\vec{q},i\omega_m',i\omega_m+i \Omega_n, i\Omega_n) G(\vec{k}',i\omega_m') \nonumber \\ &\times G(\vec{k}'+\vec{q},i\omega_m'+i\Omega_n) \Gamma_{\hat{\Theta},\vec{\alpha}}(\vec{k}',\vec{q},i\omega_m',i\Omega_n).
\end{align}
The term $\Lambda(\vec{k},\vec{q},\vec{q'},i\omega_m,i \omega_m', i\Omega_n)$ in Eq.\@ (\ref{eq:Bethe_Salpeter_gen_nonlocal}) is the particle-hole irreducible two-particle vertex function, which depends on the underlying electronic interaction mechanism \cite{Mahan-2000,Bruus-2004mb,Berthod-2018}; see App.\@ \ref{Many_body_def} for details and derivations. 

In the DMFT limit, the irreducible two-particle vertex is a local function, i.e.\@, it only depends on the fermionic and exchanged frequencies: 
\begin{equation}\label{eq:Lambda_DMFT}
\Lambda(\vec{k},\vec{k}',\vec{q'},i\omega_m,i \omega_m', i\Omega_n) \equiv \Lambda(i\omega_m,i \omega_m', i\Omega_n) \, \forall \left\{\vec{k},\vec{k}',\vec{q}\right\}.
\end{equation}
Eq.\@ (\ref{eq:Lambda_DMFT}) implies a local (momentum-independent) self-energy $\Sigma(\vec{k},i\omega_n)\equiv \Sigma(i\omega_n)$. In addition, in the spirit of infinite-dimensional DMFT we will also assume ladder (non-crossing) vertex renormalization, as crossing diagrams are generically suppressed in the fully connected limit as $1/z$ where $z$ is the lattice coordination number. 

Within the assumption (\ref{eq:Lambda_DMFT}), Eq.\@ (\ref{eq:Bethe_Salpeter_gen_nonlocal}) becomes 
\begin{align}\label{eq:Bethe_Salpeter_gen}
\Gamma_{\hat{\Theta},\vec{\alpha}}(\vec{k},\vec{q},i\omega_m,i\Omega_n)&=\Gamma_{\hat{\Theta},\vec{\alpha}}^{(0)}(\vec{k},\vec{q})+ k_B T\sum_{i\omega_m'} \Lambda(i\omega_m, i\omega_m',i \Omega_n) \frac{1}{\mathscr{V}}\sum_{\vec{k}'} G(\vec{k}',i\omega_m') G(\vec{k}'+\vec{q},i\omega_m'+i\Omega_n) \nonumber \\ &\times \Gamma_{\hat{\Theta},\vec{\alpha}}(\vec{k}',\vec{q},i\omega_m',i\Omega_n).
\end{align}
\end{widetext}
Fig.\@ \ref{fig:Gamma_VC} shows a graphical representation of Eq.\@ (\ref{eq:Bethe_Salpeter_gen}).
In principle, the ladder equation (\ref{eq:Bethe_Salpeter_gen}) could contain renormalizations of every vertex for nonlocal electrodynamics, e.g.\@, for density, momentum, current, and stress. At this stage, Eq.\@ (\ref{eq:Bethe_Salpeter_gen}) does not encode conservation laws (e.g.\@, charge conservation or momentum conservation) by itself. However, continuity equations are preserved in the many-body theory if the approximation scheme for the two-particle vertex $\Lambda(i\omega_m, i\omega_m',i \Omega_n)$ and the self-energy $\Sigma(i\omega_n)$ is \emph{conserving}, i.e.\@, it can be derived from the same Luttinger-Ward functional $\Phi[G]$ \cite{Luttinger-1960,Baym-1962,Eder-2019}. The latter is the sum of all closed, two-particle-irreducible skeleton diagrams (with no self-energy insertions) built with the full (interacting) Green's function $G(\vec{k},i\omega_n)$ and with bare interaction vertices. 
In particular, at equilibrium ($\mathrm{eq}$) for a local self-energy $\Phi=\Phi[\mathscr{G}]$
\begin{align}\label{eq:Sigma_Phi_loc_noneq}
\Sigma(i\omega_m,i\omega_m+i\Omega_n)&=\left.\frac{\delta \Phi[\mathscr{G}]}{\delta \mathscr{G}(i\omega_m,i\omega_m+i\Omega_n)}\right|_{\mathrm{eq}} \nonumber \\ & \equiv \frac{\delta \Phi[\mathscr{G}]}{\delta \mathscr{G}(i\omega_m)}. 
\end{align}
The general definition (\ref{eq:Sigma_Phi_loc_noneq}) of the functional derivative involves the nonequilibrium local self-energy $\Sigma(i\omega_m,i\omega_m+i\Omega_n)$ and the nonequilibrium local Green's function $\mathscr{G}(i\omega_m,i\omega_m+i\Omega_n)$, which both depend on two fermionic frequencies and are here written in the particle-hole notation with transferred frequency $i\Omega_n$ (see App.\@ \ref{Many_body_def}). The local Green's function is integrated over internal momentum, according to $\mathscr{G}(i\omega_m, i\omega_m+i\Omega_n)=\mathscr{V}^{-1} \sum_{\vec{k}} G(i\omega_m, i\omega_m+i\Omega_n)$ where $\mathscr{V}$ is the system volume.

At equilibrium we have $\Sigma(i\omega_m,i\omega_m+i\Omega_n)=\Sigma(i\omega_m)\delta_{\omega_m,\omega_m+\Omega_n}$ and $\mathscr{G}(i\omega_m,i\omega_m+i\Omega_n)=\mathscr{G}(i\omega_m)\delta_{\omega_m,\omega_m+\Omega_n}$, so that
\begin{equation}\label{eq:Sigma_Phi_loc}
\Sigma(i\omega_m)=\frac{\delta \Phi[\mathscr{G}]}{\delta \mathscr{G}(i\omega_m)};
\end{equation}
that is, the Luttinger-Ward functional is the generating functional for the self-energy, and it depends on the local (momentum-integrated) equilibrium Green's function $\mathscr{G}(i\omega_n)=\mathscr{V}^{-1} \sum_{\vec{k}} G(\vec{k},i\omega_n)$. 

Formally, $\Phi[\mathscr{G}]$ is connected to the thermodynamic grand potential $\mathscr{\Omega}$ in a generic, locally interacting theory:
\begin{align}\label{eq:Omega_grand_potential_Phi}
\Omega&=-k_B T \ln \mathrm{det}\left\{-G^{-1}(\vec{k},i\omega_n)\right\}\nonumber \\ &-k_B T \mathrm{Tr}\left\{\Sigma(i\omega_n) G(\vec{k},i\omega_n)\right\}+k_B T \Phi[\mathscr{G}],
\end{align}
where the $\mathrm{Tr}\ln\left\{\cdot\right\}=\ln \mathrm{det}\left\{\cdot\right\}$ and $\mathrm{Tr}\left\{\cdot\right\}$ are traces over the internal fermionic momenta $\vec{k}$ and frequencies $i\omega_m$, while the on-shell stationarity condition $\delta \Omega/\delta G(\vec{k},i\omega_n)=0$ produces Dyson's equation for the interacting Green's function. \\
Within this framework, the Kadanoff-Baym conserving prescription for the particle-hole irreducible two-particle vertex is \cite{Baym-1961,Baym-1962,Kadanoff-1962qsm,Haensch-1983}
\begin{align}\label{eq:Lambda_Sigma_conserving}
&\Lambda(i\omega_m, i\omega_m',i\Omega_n)=\frac{\delta \Sigma(i\omega_m+i\Omega_n, i\omega_m)}{\delta \mathscr{G}(i\omega_m',i\omega_m'-i\Omega_n)} \nonumber \\ &=\frac{\delta^2 \Phi[\mathscr{G}]}{\delta \mathscr{G}(i\omega_m,i\omega_m+i\Omega_n)\delta\mathscr{G}(i\omega_m',i\omega_m'-i\Omega_n)}.
\end{align}
Here it is essential to keep the nonequilibrium notation, since the variation of the local Green's functions at the denominator of Eq.\@ (\ref{eq:Lambda_Sigma_conserving}) need not be on the equilibrium manifold. This feature is reflected by the local two-particle vertex $\Lambda(i\omega_m, i\omega_m',i\Omega_n)$ depending on three distinct frequency arguments, one of which is the transferred frequency in particle-hole notation in accordance with App.\@ \ref{Many_body_def}.

Combining the Bethe-Salpeter ladder equation (\ref{eq:Bethe_Salpeter_gen}) and the conserving relation (\ref{eq:Lambda_Sigma_conserving}) for the two-particle vertex, one derives the Ward-Takahashi identities for charge conservation and momentum conservation, which constitute the continuity equations for charge/mass density and for momentum density in the diagrammatic, many-body language \cite{Ward-1950,Takahashi-1957,Schrieffer-1963th,Vollhardt-1980,Metzner-1993,Nieh-1998,Mahan-2000,He-2014,Li-2023,Narozhny-2023}. For completeness, these derivations are reviewed in Apps.\@ \ref{app:Charge_conservation} and \ref{app:Momentum_conservation} for a local interacting theory at equilibrium. For charge conservation and a generic dispersion $\epsilon_{\vec{k}}$, we obtain 
\begin{align}\label{eq:Ward_identity_charge}
& i \Omega_n \Gamma_{\hat{n}}(\vec{k},\vec{q},i\omega_n, i\Omega_n)-\hbar \vec{q} \cdot \vec{\Gamma}_{\hat{J}}(\vec{k},\vec{q},i\omega_n, i\Omega_n) \nonumber \\ &=G^{-1}(\vec{k}+\vec{q},i\omega_n+i\Omega_n) -G^{-1}(\vec{k},i\omega_n), 
\end{align}
while for momentum conservation one has 
\begin{align}\label{eq:Ward_identity_momentum}
& i \Omega_n \Gamma_{\hat{\pi},\alpha}(\vec{k},\vec{q},i\omega_n, i\Omega_n)-\sum_\beta q_\beta \, \Gamma_{\hat{\textit{T}},\alpha\beta}(\vec{k},\vec{q},i\omega_n, i\Omega_n) \nonumber \\ &=\left(k_\alpha+\frac{q_\alpha}{2}\right)\left[G^{-1}(\vec{k}+\vec{q},i\omega_n+i\Omega_n) -G^{-1}(\vec{k},i\omega_n)\right].
\end{align}

The Ward identity (\ref{eq:Ward_identity_charge}) entails a relation between the density-density correlation function $\chi_{\hat{n}\hat{n}}(\vec{q},i\Omega_n)$, corresponding to $\hat{\Theta}=\hat{n}$ in Eq.\@ (\ref{eq:chi_Theta_gen}), and the current-current correlation function $\chi_{\hat{J}\hat{J}}^{\alpha \beta}(\vec{q},i\Omega_n)$, obtained through $\hat{\Theta}=\hat{J}$ in Eq.\@ (\ref{eq:chi_Theta_gen}). This relation reads 
\begin{align}\label{eq:Chi_nn_Chi_JJ}
(\Omega_n)^2 \chi_{\hat{n}\hat{n}}(\vec{q},i\Omega_n)&+\hbar^2 \sum_{\alpha,\beta} q_\alpha q_\beta \chi_{\hat{J}\hat{J}}^{\beta \alpha}(\vec{q},i\Omega_n) \nonumber \\ & =-\frac{\hbar}{\mathscr{V}} \sum_{\alpha} q_\alpha \left\langle\left[\hat{n}_{\vec{q}}(0),\hat{J}_\alpha(-\vec{q},0)\right]\right\rangle,
\end{align} 
where the ``contact term'' $\left\langle\left[\hat{n}_{\vec{q}}(0),\hat{J}_\alpha(-\vec{q},0)\right]\right\rangle$ corresponds to the average ($\left\langle \right\rangle$) static commutator between the density operator $\hat{n}_{\vec{q}}(\tau)$ and the current density operator $\hat{J}_{\alpha}(\vec{q},\tau)$ in the spatial direction $\alpha$. In the simple case of isotropic quadratic dispersion (\ref{eq:xi_quadr}), one has $\left\langle\left[\hat{n}_{\vec{q}}(0),\hat{J}_\alpha(-\vec{q},0)\right]\right\rangle=\hbar q_\alpha/m \mathscr{N}$, where $\mathscr{N}=\int d\vec{r} \left\langle \hat{n}(\vec{r},0)\right\rangle$ is the total particle number \cite{Berthod-2018}, corresponding to the density $n=\mathscr{N}/\mathscr{V}$. The volume factor $\mathscr{V}^{-1}$ at the right-hand side of Eq.\@ (\ref{eq:Chi_nn_Chi_JJ}) appears because the correlation functions on the left-hand side are in their intensive form, consistently with Eq.\@ (\ref{eq:chi_Theta_gen}). 

Analogously, the Ward identity (\ref{eq:Ward_identity_momentum}) implies a relation between the momentum-momentum correlation function $\chi_{\hat{\pi}\hat{\pi}}^{\alpha \beta}(\vec{q},i\Omega_n)$, stemming from $\hat{\Theta}=\hat{\pi}$ in Eq.\@ (\ref{eq:chi_Theta_gen}), and the stress-stress correlation function $\chi_{\hat{\textit{T}}\hat{\textit{T}}}^{\alpha \beta \gamma \delta}(\vec{q},i\Omega_n)$, according to $\hat{\Theta}=\hat{\textit{T}}$ in Eq.\@ (\ref{eq:chi_Theta_gen}). Such interrelation is
\begin{align}\label{eq:Chi_JJ_Chi_TT}
(\Omega_n)^2 \chi_{\hat{\pi}\hat{\pi}}^{\alpha \beta}(\vec{q},&i\Omega_n)+\sum_{\gamma,\delta} q_\gamma q_\delta \chi_{\hat{\textit{T}}\hat{\textit{T}}}^{\delta \alpha \gamma \beta}(\vec{q},i\Omega_n)\nonumber \\ & =-i\Omega_n \mathscr{V}^{-1}\left\langle\left[\hat{\pi}_{\alpha}(\vec{q},0),\hat{\pi}_\beta(-\vec{q},0)\right]\right\rangle \nonumber \\ & -\mathscr{V}^{-1} \sum_{\gamma} q_\gamma \left\langle\left[\hat{\pi}_\alpha(\vec{q},0),\hat{\textit{T}}_{\gamma \beta}(-\vec{q},0)\right]\right\rangle,
\end{align} 
where there are two static ``contact terms'' $\left\langle\left[\hat{\pi}_{\alpha}(\vec{q},0),\hat{\pi}_\beta(-\vec{q},0)\right]\right\rangle$ and $\left\langle\left[\hat{\pi}_\alpha(\vec{q},0),\hat{\textit{T}}_{\gamma \beta}(-\vec{q},0)\right]\right\rangle$, involving the momentum operator component $\hat{\pi}_{\alpha}(\vec{q},\tau)$ and the stress operator component $\hat{\textit{T}}_{\alpha \beta}(\vec{q},\tau)$ \cite{Taylor-2010,Bradlyn-2012,Pakhira-2015,Frank-2020}. Notice that the Ward identity (\ref{eq:Ward_identity_momentum}) and the associated Eq.\@ (\ref{eq:Chi_JJ_Chi_TT}) are rigorously valid in the presence of continuous translational invariance. While such invariance holds for interacting quantum fluids in the continuum, for a crystalline solid-state system translational symmetry implies conservation of crystal momentum $\hbar \vec{k}$ modulo reciprocal lattice vectors $\vec{G}$\cite{Ashcroft-1976}. In regimes where Umklapp and other momentum-relaxing processes are negligible, crystal momentum is approximately conserved and may be treated as an effective hydrodynamic momentum density at long wavelengths. I refer to this regime as the effective momentum-conserving limit. The exact continuum momentum Ward identity is recovered when continuous translational invariance is present.
For isotropic quadratic dispersion (\ref{eq:xi_quadr}), the momentum-momentum commutator $\left\langle\left[\hat{\pi}_{\alpha}(\vec{q},0),\hat{\pi}_\beta(-\vec{q},0)\right]\right\rangle$ in Eq.\@ (\ref{eq:Chi_JJ_Chi_TT}) is linked to the total momentum of the system \cite{Bradlyn-2012}, while the momentum-stress commutator $\left\langle\left[\hat{\pi}_\alpha(\vec{q},0),\hat{\textit{T}}_{\gamma \beta}(-\vec{q},0)\right]\right\rangle$ is proportional to the inverse compressibility $\kappa_T^{-1}$ in the $q\rightarrow 0^+$ limit, often associated with electron hydrodynamics. Moreover, for the dispersion (\ref{eq:xi_quadr}) current density $\hat{\vec{J}}(\vec{q},\tau)$ is proportional to momentum $\hat{\vec{\pi}}(\vec{q},\tau)$, so that the current-current and momentum-momentum correlation functions are equal up to a multiplicative constant: $\chi_{\hat{J}\hat{J}}^{\alpha \beta}(\vec{q},i\Omega_n)=(\hbar^2/m^2) \chi_{\hat{\pi}\hat{\pi}}^{\alpha \beta}(\vec{q},i\Omega_n)$. This is the key connection that enables a relation between the $q^2$ term in the $q\rightarrow 0^+$ expansion of the conductivity tensor and the $q=0$ viscosity tensor \cite{Taylor-2010,Bradlyn-2012,He-2014}. We will comment on such relation in Sec.\@ \ref{Momentum_cons_renorm}, with reference to the necessity of relaxing our DMFT-like assumptions in order for the Ward identity (\ref{eq:Ward_identity_momentum}) to hold regardless of electronic dispersion $\xi_{\vec{k}}$, for any momentum $\vec{q}$ and frequency $i\Omega_n$. Derivations of Eqs.\@ (\ref{eq:Chi_nn_Chi_JJ}) and (\ref{eq:Chi_JJ_Chi_TT}) are rooted in the charge and momentum conservation laws respectively, and are developed in the literature at the level of operator equations \cite{Taylor-2010,He-2014}, strain generators \cite{Bradlyn-2012,Link-2018a}, or field theory \cite{Rose-2020,Raines-2024}. Alternative derivations based on the Ward identities (\ref{eq:Ward_identity_charge}) and (\ref{eq:Ward_identity_momentum}) will be reported elsewhere \cite{Valentinis-2026_unpublished_2}. 

Our objective is to construct a many-body theory of nonlocal (finite-$\vec{q}$, finite-$i\Omega_n$) electrodynamics with a local self-energy and two-particle vertex (in the DMFT-like/local limit), which obeys charge conservation and (optionally) momentum conservation. We will see that our local theory can obey charge conservation at each $\vec{q}$ and $i\Omega_n$, consistently with the ladder Bethe-Salpeter Eq.\@ (\ref{eq:Bethe_Salpeter_gen}) and the Ward identity (\ref{eq:Ward_identity_charge}), but it does not allow for momentum conservation at every $\vec{q}$, since this last property would require a momentum-dependent two-particle vertex $\Lambda(\vec{k},\vec{q},\vec{q'},i\omega_m,i \Omega_n', i\Omega_n)$ , not included in the DMFT-like approach.

Let us analyze in which cases the vertex corrections in Eq.\@ (\ref{eq:Bethe_Salpeter_gen}) are finite, and when they vanish by parity cancellation \cite{Khurana-1990}, when the theory is charge-conserving but not momentum-conserving. 

\section{Vertex corrections, and parity-cancellation thereof, from charge conservation at zero momentum}\label{Charge_cons_vertex_vanish}\label{Vertex_corrections_q0}

Let us assume a generic one-band single-particle dispersion 
\begin{equation}\label{eq:dispersion_even_parity}
\epsilon_{\mathscr{g}\vec{k}}=\epsilon_{\vec{k}},
\end{equation}
even with respect to a symmetry point-group operation $\mathscr{g}$ of the underlying crystalline lattice. In addition, consider a generic but local irreducible two-particle vertex (\ref{eq:Lambda_DMFT}). 
The structure of the ladder vertex equation (\ref{eq:Bethe_Salpeter_gen}) at $q=0$ implies that the renormalized vertex is $\Gamma_{\hat{\Theta},\vec{\alpha}}(\vec{k},0,i\omega_m,i\Omega_n)=\Gamma_{\hat{\Theta},\vec{\alpha}}^{(0)}(\vec{k},0)+ C^{\vec{\alpha}}(i\omega_m,i\Omega_n)$, where 
\begin{align}\label{eq:Bethe_Salpeter_C}
C^{\vec{\alpha}}(i\omega_m,i\Omega_n)=k_B T \sum_{i\omega_m'} \Lambda(i\omega_m, i\omega_m',i \Omega_n) X^{\vec{\alpha}}(i\omega_m', i\Omega_n),
\end{align}
and
\begin{align}\label{eq:Bethe_Salpeter_X}
X^{\vec{\alpha}}(i\omega_m',i\Omega_n)&= \frac{1}{\mathscr{V}}\sum_{\vec{k}'} G(\vec{k}',i\omega_m') G(\vec{k}',i\omega_m'+i\Omega_n) \nonumber \\ & \times \Gamma_{\hat{\Theta},\vec{\alpha}}(\vec{k}',0,i\omega_m',i\Omega_n).
\end{align}
The full ladder (\ref{eq:Bethe_Salpeter_gen}) collapses to zero if $X^{\vec{\alpha}}(i\omega_m',i\Omega_n)=0$ for each combination of its frequency arguments, which also implies $C^{\vec{\alpha}}(i\omega_m, i\Omega_n)=0$. Also, notice that the ladder correction $C^{\vec{\alpha}}(i\omega_m, i\Omega_n)$ is only a function of frequencies $\left\{i\omega_m,i\Omega_n\right\}$, but it does not modify the direction of the renormalized vertex in momentum space $\vec{k}$, meaning that the momentum dependence (in magnitude and direction of $\vec{k}$) of the renormalized vertex $\Gamma_{\hat{\Theta},\vec{\alpha}}(\vec{k},0,i\omega_m,i\Omega_n)$ is entirely dictated by the bare vertex $\Gamma_{\hat{\Theta},\vec{\alpha}}^{(0)}(\vec{k},0)$. 
Hence, if the bare vertex is odd with respect to $\vec{k}$ under the point-group symmetry operation $\mathscr{g}$, 
\begin{equation}\label{eq:bare_vertex_odd}
\Gamma_{\hat{\Theta},\vec{\alpha}}^{(0)}(\mathscr{g}\vec{k},0)=-\Gamma_{\hat{\Theta},\vec{\alpha}}^{(0)}(\vec{k},0),
\end{equation}
then by self-consistency, we have $C^{\vec{\alpha}}(i\omega_m, i\Omega_n)=0 \, \forall \left\{i\omega_m,i\Omega_n\right\}$ and, in turn, the renormalized vertex $\Gamma_{\hat{\Theta},\vec{\alpha}}(\mathscr{g}\vec{k},0,i\omega_m,i\Omega_n)=-\Gamma_{\hat{\Theta},\vec{\alpha}}(\vec{k},0,i\omega_m,i\Omega_n)$ is also odd in $\vec{k}$. 

To explicitly prove the last statement, combine Eqs.\@ (\ref{eq:Bethe_Salpeter_C}) and (\ref{eq:Bethe_Salpeter_X}), to obtain
\begin{align}\label{eq:Bethe_Salpeter_comb}
C^{\vec{\alpha}}(i\omega_m, i\Omega_n)&= k_B T \sum_{i\omega_m'} \Lambda(i\omega_m, i\omega_m',i \Omega_n) \nonumber \\ & \times \frac{1}{\mathscr{V}}\sum_{\vec{k}'} G(\vec{k}',i\omega_m') G(\vec{k}',i\omega_m'+i\Omega_n) \nonumber \\ & \times \Gamma_{\hat{\Theta},\vec{\alpha}}^{(0)}(\vec{k}',0)+ k_B T \sum_{i\omega_m'} \Lambda(i\omega_m, i\omega_m',i \Omega_n) \nonumber \\ \times & \frac{1}{\mathscr{V}}\sum_{\vec{k}'} G(\vec{k}',i\omega_m') G(\vec{k}',i\omega_m'+i\Omega_n)  \nonumber \\ \times & C^{\vec{\alpha}}(i\omega_m', i\Omega_n).
\end{align}
The first term in Eq.\@ (\ref{eq:Bethe_Salpeter_comb}) vanishes by the bare-vertex oddness of Eq.\@ (\ref{eq:bare_vertex_odd}), while the second term is a linear matrix equation: 
\begin{subequations}\label{eq:canc_C_matrix}
\begin{align}\label{eq:C_K}
C^{\vec{\alpha}}(i\omega_m, i\Omega_n)&=k_B T \sum_{i\omega_m'} K(i\omega_m,i\omega_m',i\Omega_n) \nonumber \\ & \times C^{\vec{\alpha}}(i\omega_m', i\Omega_n),
\end{align}
\begin{align}\label{eq:K_collective_mode}
K(i\omega_m,i\omega_m', i\Omega_n)&= \Lambda(i\omega_m, i\omega_m',i \Omega_n) \frac{1}{\mathscr{V}}\sum_{\vec{k}'} G(\vec{k}',i\omega_m') \nonumber \\ & \times G(\vec{k}',i\omega_m'+i\Omega_n).
\end{align}
\end{subequations}
A nontrivial solution of the homogeneous Eq.\@ (\ref{eq:C_K}), namely $\mathrm{det}\left\{\delta_{m m'}- k_B T K(i\omega_m,i\omega_m', i\Omega_n)\right\}=0$, where the determinant is taken over the internal fermionic Matsubara frequencies $\left(i\omega_m, i\omega_m'\right)$, would signal a uniform density collective mode (after appropriate analytic continuation of bosonic Matsubara frequencies $i\Omega_n\rightarrow \hbar \omega+i0^+$). Equivalently, the collective mode corresponds to the operator $\mathbb{I}-k_B T \underline{\underline{K}}(i\Omega_n)$, constructed from the fermionic Matsubara components $\underline{\underline{K}}(i\Omega_n)=\left\{K(i\omega_m,i\omega_m', i\Omega_n)\right\}$, containing $1$ in its spectrum. Away from this condition, the operator $\mathbb{I}-k_B T \underline{\underline{K}}(i\Omega_n)$ is invertible: then, the only other solution of Eq.\@ (\ref{eq:canc_C_matrix}) is $C^{\vec{\alpha}}(i\omega_m, i\Omega_n)=0 \, \forall \left\{i\omega_m,i\Omega_n\right\}$.

However, independently from the existence of a collective mode, the correlation function in Eq.\@ (\ref{eq:chi_Theta_gen}) has no contribution from vertex corrections, if Eqs.\@ (\ref{eq:dispersion_even_parity}) and (\ref{eq:bare_vertex_odd}) are satisfied. This is because, even at a collective mode whereby $C^{\vec{\alpha}}(i\omega_m, i\Omega_n)\neq0$, inserting the renormalized vertex $\Gamma_{\hat{\Theta},\vec{\alpha}}(\vec{k},0,i\omega_m,i\Omega_n)=\Gamma_{\hat{\Theta},\vec{\alpha}}^{(0)}(\vec{k},0)+ C^{\vec{\alpha}}(i\omega_m,i\Omega_n)$ in Eq.\@ (\ref{eq:chi_Theta_gen}) one sees that the vertex-correction part depending on $C^{\vec{\alpha}}(i\omega_m,i\Omega_n)$ vanishes by the same parity arguments applied to the Bethe-Salpeter vertex equation. Physically, the collective mode has no coupling to the odd-parity observable $\hat{\Theta}$. 

Therefore, vertex corrections at $q=0$ vanish from the correlation function of the observable $\hat{\Theta}$ if the associated bare vertex $\Gamma_{\hat{\Theta},\vec{\alpha}}^{(0)}(\vec{k},0)$ is odd with respect to a point-group operation $\mathscr{g}$ acting on $\vec{k}$: this property holds for generic anisotropic dispersion $\epsilon_{\vec{k}}$, provided that the dispersion (\ref{eq:dispersion_even_parity}) is even under the same operation $\mathscr{g}$. For current-like vertices, depending on one spatial component $\alpha$, the operation $\mathscr{g}$ corresponds to inversion $\mathscr{I}$: $\vec{k}\mapsto -\vec{k}$, as detailed in Sec.\@ \ref{Vertex_corrections_vanish_currents_alpha}; for tensorial vertices of rank 2, depending on the spatial components $\left\{\alpha,\beta\right\}$ with $\alpha \neq \beta$, suitable $\mathscr{g}$ are mirror symmetries with respect to planes orthogonal to the direction $\alpha$, and twofold or higher-order rotations around an axis $\alpha$, as described in Sec.\@ \ref{Shear_stresses_vertex_corrections}. 

To explicitly see this, consider the example of the current response on a 2D square lattice with dispersion $\epsilon_{\vec{k}}=-2 t \left[\cos(k_x a) +\cos(k_y a)\right]$. The bare current vertex is $\Gamma_{\hat{\vec{J}},\alpha}^{(0)}(\vec{k},0)=\hbar^{-1}\partial \epsilon_{\vec{k}}/\partial k_{\alpha}=2 t a \sin(k_\alpha x)/\hbar$, from Eq.\@ (\ref{eq:current_vertex_gen}). Since the lattice has inversion symmetry, then the current vertex is odd with respect to $k_x \mapsto -k_x$ and $k_y \mapsto -k_y$, while the product of Green's functions $G(\vec{k}',i\omega_m') G(\vec{k}',i\omega_m'+i\Omega_n)=G(-\vec{k}',i\omega_m') G(-\vec{k}',i\omega_m'+i\Omega_n)$ is even with respect to momentum inversion since the dispersion respects parity, while the local self-energy is independent of $\vec{k}$. Hence, the term (\ref{eq:Bethe_Salpeter_X}) vanishes by angular average, because the term under the sum over $\vec{k}$ is globally odd with respect to momentum inversion. Therefore, vertex corrections vanish from the current response at $q=0$ on the square lattice in the DMFT-like limit \cite{Khurana-1990}. 

\subsection{Observables universally devoid of vertex corrections in DMFT limit: currents}\label{Vertex_corrections_vanish_currents_alpha}

Linear-response observables $\hat{\Theta}$ for which vertex corrections universally vanish at $q=0$ are the ones proportional to the velocity operator $\hat{\vec{v}}_{\vec{k}}=\hbar^{-1}\nabla_{\vec{k}} \epsilon_{\vec{k}}$, i.e.\@, to the momentum gradient of the dispersion $\epsilon_{\vec{k}}=\epsilon_{-\vec{k}}$. The latter has even symmetry under $\vec{k}\mapsto -\vec{k}$ if the lattice has inversion symmetry (even parity $\mathscr{P}$), or if time-reversal symmetry ($\mathscr{T}$) is present and the dispersion is spin-isotropic.

These vectorial vertices are current densities (of mass, charge, momentum, or energy): 
\begin{itemize}
	\item{Particle (or electric) current, $\hat{\Theta}\equiv \hat{J}$, for which the spatial components of the bare current vertex $\vec{\Gamma}_{\hat{J}}^{(0)}(\vec{k},0)=\left\{ \Gamma_{\hat{J},\alpha}^{(0)}(\vec{k},0) \right\}$ are
	\begin{align}\label{eq:current_vertex_gen}
	\Gamma_{\hat{J},\alpha}^{(0)}(\vec{k},0)&=v_{\vec{k},\alpha}  =\frac{1}{\hbar}\frac{\partial \epsilon_{\vec{k}}}{\partial k_\alpha},
	\end{align}
	with an additional multiplicative factor of electric charge $e$ for electric current;
	}
		\item{Momentum density (in units of $\hbar$), $\hat{\Theta}\equiv \hat{\vec{\pi}}$, for which the spatial components of the bare momentum vertex $\vec{\Gamma}_{\hat{\pi}}^{(0)}(\vec{k},0)=\left\{ \Gamma_{\hat{\vec{\pi}},\alpha}^{(0)}(\vec{k},0) \right\}$ are
	\begin{align}\label{eq:momentum_vertex_gen}
	\Gamma_{\hat{\vec{\pi}},\alpha}^{(0)}(\vec{k},0)&=k_{\alpha};
	\end{align}
	}
		\item{Thermal (heat) current, $\hat{\Theta}\equiv \hat{\vec{K}}_T$, for which the spatial components of the bare heat current vertex $\vec{\Gamma}_{\hat{K}_T}^{(0)}(\vec{k},i\omega_m,0)=\left\{ \Gamma_{\hat{K}_T,\alpha}^{(0)}(\vec{k},i\omega_m,0) \right\}$ are
	\begin{align}\label{eq:thermal_current_vertex_gen}
	\Gamma_{\hat{K}_T,\alpha}^{(0)}(\vec{k},i\omega_m,0)&=(\epsilon_{\vec{k}}-\mu)\frac{1}{\hbar}\frac{\partial \epsilon_{\vec{k}}}{\partial k_\alpha}. 
	\end{align}
	Notice that for the heat current, the bare vertex (\ref{eq:thermal_current_vertex_gen}) depends on the energy $\epsilon_{\vec{k}}$ of the fermion relative to the chemical potential $\mu$. However, this feature does not affect the overall oddness with respect to $\vec{k}$ inversion, if $\epsilon_{\vec{k}}=\epsilon_{-\vec{k}}$ \footnote{For heat currents, another relevant continuity equation is the one for energy conservation \cite{He-2014}. This aspect and the resulting Ward identity will be addressed in further work \cite{Valentinis-2026_unpublished_2}}. 
	}
\end{itemize}
In general, a vertex is odd under $\vec{k} \mapsto -\vec{k}$, and therefore does not lead to vertex corrections in the DMFT-like limit, if it contains an odd number of momentum vectors $\vec{k}$ or velocity factors $\vec{v}_{\vec{k}}$, or if it transforms as a polar vector.

\subsection{Observables with vanishing vertex corrections in DMFT limit under symmetry constraints: off-diagonal (shear) stresses}\label{Shear_stresses_vertex_corrections}

Although not globally odd with respect to $\vec{k} \mapsto -\vec{k}$ transformations, there are components of the stress tensor, proportional to products of momentum components $k_\alpha$ and first momentum derivatives of the dispersion $\epsilon_{\vec{k}}$, which yield a null vertex correction once summed over all momenta with $\mathscr{V}^{-1}\sum_{\vec{k}}$, upon vanishing angular integration. 
These components include: 
\begin{itemize}
	\item{The shear components of the kinetic stress tensor, $\hat{\Theta}\equiv \hat{\textit{T}}$, for which the spatial components of the bare stress vertex $\underline{\underline{\Gamma}}_{\hat{\textit{T}}}^{(0)}(\vec{k},0)=\left\{ \Gamma_{\hat{\textit{T}},\alpha \beta}^{(0)}(\vec{k},0) \right\}$ are
	\begin{align}\label{eq:stress_vertex_gen_shear}
	\Gamma_{\hat{\textit{T}},\alpha \beta}^{(0)}(\vec{k},0)&=\hbar v_{\vec{k},\alpha} k_\beta=\frac{\partial \epsilon_{\vec{k}}}{\partial k_\alpha} k_\beta;
	\end{align}
	}
		\item{The Hall (or odd, antisymmetric) components of the kinetic stress tensor, $\hat{\Theta}\equiv \hat{\textit{T}}$, which arise in the absence of time-reversal symmetry and involve the spatial components
	\begin{align}\label{eq:stress_vertex_gen_Hall}
	\Gamma_{\hat{\textit{T}},\alpha \alpha}^{(0)}(\vec{k},0)-\Gamma_{\hat{\textit{T}},\beta \beta}^{(0)}(\vec{k},0)&=\hbar \left(v_{\vec{k},\alpha} k_\alpha-v_{\vec{k},\beta} k_\beta\right),
	\end{align}
	in addition to the shear components (\ref{eq:stress_vertex_gen_shear}); see also Sec.\@ \ref{Hall_viscosity}. 
	}
\end{itemize}
Such vanishing of vertex corrections occurs for mixed spatial components $\left\{\alpha, \beta\right\}$ with $\alpha \neq \beta$, if the underlying lattice, and hence the dispersion $\epsilon_{\vec{k}}$, is symmetric ($\epsilon_{\mathscr{g}\vec{k}}=\epsilon_{\vec{k}}$) under specific point-group transformations $\mathscr{g}$ under which the bare vertex is odd: $\Gamma_{\hat{\Theta},\alpha\beta}^{(0)}(\mathscr{g}\vec{k},0)=-\Gamma_{\hat{\Theta}}^{(0)}(\vec{k},0)$. 
Specifically, all off-diagonal components involving the spatial direction $\alpha$ are bare (unrenormalized) if the lattice point group allows for at least one of the following symmetry operations: a mirror symmetry plane $\mathscr{M}_\alpha$ orthogonal to $\alpha$, a two-fold rotation $\mathscr{C}_\alpha$ about the axis $\alpha$, or an improper rotation $\mathscr{S}_n$ that involves a rotation plus a mirror plane perpendicular to $\alpha$. 
However, in 3D the presence of the above symmetries does not annihilate vertex corrections in the direction $\gamma \neq \alpha \neq \beta$, unless another one of such symmetries also exists in the direction $\gamma$. 

Moreover, higher-order rotations $\mathscr{C}_3$, $\mathscr{C}_4$, $\mathscr{C}_6 \ldots$ suppress all off-diagonal vertex corrections in $\Gamma_{\hat{\Theta},\alpha\beta}^{(0)}(\vec{k},0)$, thus guaranteeing that all shear vertex corrections identically vanish in the local DMFT-like limit. 

In 2D, examples of point-group symmetries yielding shear-stress vertex-corrections cancellations include $2mm$ (rectangular), $4mm$ (square), $6mm$ (triangular or hexagonal); by contrast, groups like $2$ (oblique) devoid of mirror symmetry acquire vertex corrections for shear stresses.  

In 3D, the criterion holds whenever the crystal point group contains any symmetry that forbids a rank-2 shear tensor invariant. Concretely, this ensemble includes 3-fold, 4-fold, or 6-fold rotation symmetry, systems with at least a mirror plane, and in general any inversion-symmetric and isotropically connected Fermi surface. Examples then include cubic point groups (e.g.\@, $O_h$, $O$, $T_d$, $T$), hexagonal/trigonal (e.g.\@, $D_{6h}$, $C_{6v}$, $D_{3d}$, $C_{3v}$), tetragonal (e.g.\@, $D_{4h}$, $C_{4v}$), and orthorhombic (e.g.\@, $D_{2h}$, $C_{2v}$); conversely, triclinic systems (e.g.\@, $C_1$, $C_i$) have no rotational or mirror symmetries, and thus have finite vertex corrections for shear stresses even in the DMFT-like limit. 

Thus, inversion symmetry $\mathscr{I}$ alone is not enough to suppress shear-stress vertex corrections, while additional rotational and/or mirror transformations provide enough symmetry for null angular averages of momentum sums for the shear stress components. 

\begin{figure}[ht] \centering
\includegraphics[width=1.0\linewidth]{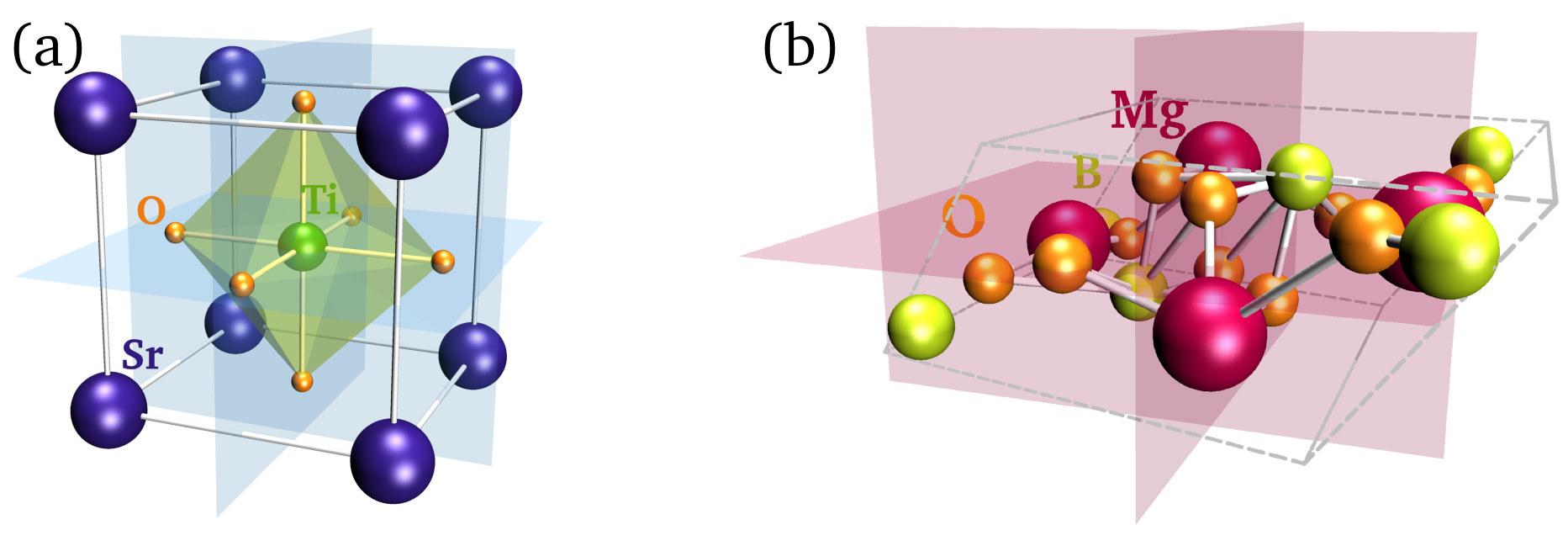}
\caption{\label{fig:mirrors_symmetry} Schematic illustration of the relation between symmetry of a 3D crystal structure and vanishing criteria for shear-stress vertex corrections in the DMFT-like limit, according to the Bethe-Salpeter equation (\ref{eq:Bethe_Salpeter_gen}) and Eq.\@ (\ref{eq:stress_vertex_gen_shear}). (a) Unit cell of cubic SrTiO$_3$, showing mirror symmetry planes (light-blue shaded planes), and therefore yielding vanishing vertex corrections for kinetic shear stresses. (b) Unit cell of triclinic Mg$_2$B$_2$O$_5$, possessing no mirror symmetry planes perpendicular to any principal axis (as highlighted by the light-red shaded planes), as relevant for Eq.\@ (\ref{eq:stress_vertex_gen_shear}); a hypothetical one-band metal of the same symmetry, or a symmetry-degraded metallic crystal subject to mechanical deformation -- see also Sec.\@ \ref{Discussion_experiments} -- acquires vertex corrections for shear components of the stress tensor. 
 }
\end{figure}
Fig.\@ \ref{fig:mirrors_symmetry} illustrates the crystal structure of two exemplary 3D lattices for which the $\vec{k}$-parity based criteria for shear-stress null vertex corrections hold or fail, respectively: Fig.\@ \ref{fig:mirrors_symmetry}(a) shows the perovskite SrTiO$_3$ in its cubic phase (with space group $Pm \overline{3}m$) at temperatures $T\gtrapprox 105$ K \footnote{At temperatures $T\lessapprox 105$ K, SrTiO$_3$ undergoes an antiferrodistortive structural transition to a tetragonal phase (with space group $I4/mcm$), caused by the rotation of the TiO$_6$ octahedra \cite{Unoki-1967}; see yellow-shaded surface in Fig.\@ \ref{fig:mirrors_symmetry}(a).}, which possesses mirror symmetries with respect to all principal axes (light-blue shaded planes) and therefore obeys the $\vec{k}$-parity criteria for vanishing vertex corrections for shear stresses in the DMFT-like limit; by contrast, Fig.\@ \ref{fig:mirrors_symmetry}(b) displays the triclinic lattice structure of the inorganic ceramic insulator Mg$_2$B$_2$O$_5$, which does not have mirror symmetry with respect to any principal axis. 
In this regard, we notice that triclinic crystal structures in pure metals are extremely rare, and generally non-existent under standard conditions: metallic systems typically crystallize in high-symmetry structures, such as body-centered cubic (BCC), face-centered cubic (FCC), or hexagonal close-packed (HCP), to maximize atomic density and metallic bonding stability \cite{Ashcroft-1976}. 

Therefore, the $\vec{k}$-based symmetry criteria outlined above, for null vertex corrections to shear stresses, are expected to generally apply to extended classes of metallic systems with DMFT-like retarded interactions. However, suppose one applies external shear mechanical deformations to a mirror-symmetric metallic crystal, thus degrading its symmetry class down to triclinic: assuming that the applied strain preserves the DMFT-like character of interactions, then the deformation enables vertex corrections to shear stresses according to Eq.\@ (\ref{eq:stress_vertex_gen_shear}); implications of this mechanical tunability for transport/spectroscopic experiments are discussed in Sec.\@ \ref{Discussion_experiments}. 

\subsection{Observables with universal vertex corrections in DMFT limit: density and bulk stresses}

Nevertheless, there are notable counterexamples for which the $\vec{k}$-inversion/$\mathscr{g}$-symmetry argument cannot hold, and thus vertex corrections arise even at $q=0$: 
\begin{itemize}
	\item{Particle density, $\hat{\Theta}\equiv \hat{n}$, for which the spatial components of the bare density vertex are simply
	\begin{equation}\label{eq:density_vertex}
	\Gamma_{\hat{n}}^{(0)}(\vec{k},0)=1;
	\end{equation}
	}
	\item{The bulk (diagonal) components of the stress tensor, $\hat{\Theta}\equiv \hat{\textit{T}}$, for which the spatial components of the bare stress vertex $\underline{\underline{\Gamma}}_{\hat{\textit{T}}}^{(0)}(\vec{k},0)=\left\{ \Gamma_{\hat{\textit{T}},\alpha \alpha}^{(0)}(\vec{k},0) \right\}$ are
	\begin{align}\label{eq:stress_vertex_gen}
	\Gamma_{\hat{\textit{T}},\alpha \alpha}^{(0)}(\vec{k},0)&=\hbar v_{\vec{k},\alpha} k_\alpha =\frac{\partial \epsilon_{\vec{k}}}{\partial k_\alpha} k_\alpha 
	\end{align}
	}
\end{itemize}
Therefore, when calculating linear response functions in the DMFT-like setting, one has to take into account vertex corrections even at $q=0$ for any correlation function for which the bare vertex is not odd with respect to momentum inversion (for polar vector-like vertices) and is not symmetric with respect to mirror planes or $\mathscr{C}_n$ rotations with $n\geq 2$, i.e.\@, when it does not obey Eq.\@ (\ref{eq:bare_vertex_odd}). This observation refines the common wisdom according to which for $q=0$ response function vertex corrections vanish in the DMFT limit: in fact, this property holds only insofar as the bare vertex satisfies Eq.\@ (\ref{eq:bare_vertex_odd}) and the dispersion has the symmetry of Eq.\@ (\ref{eq:dispersion_even_parity}). 

The underlying structural principle for presence/absence of vertex corrections is that the DMFT-like vertex renormalization (\ref{eq:Bethe_Salpeter_gen}) at $q=0$ is momentum-independent, hence it can only renormalize the uniform component of a vertex. This can only occur if the given bare vertex is not odd with respect to the symmetry transformation $\vec{k} \mapsto \mathscr{g} \vec{k}$ that leaves the dispersion $\epsilon_{\vec{k}}$ invariant. Notice that, in general, $\mathscr{g}$-odd vertices in the $q=0$ limit acquire a nonzero non-odd component for $q \neq 0$: this is why vertex-corrections cancellations in general do not extend to finite $\vec{q}$, unless the dispersion is isotropic and one focuses on vectorial vertices like the current vertex (\ref{eq:current_vertex_gen}), as detailed in Sec.\@ \ref{Quadratic_vanish}. 

The symmetry-based mechanism for cancellation of vertex corrections is especially transparent in the special case of a separable form of the renormalized vertex, $\Gamma_{\hat{\Theta},\vec{\alpha}}(\vec{k},\vec{q},i\omega_m,i\Omega_n)=\Gamma_{\hat{\Theta},\vec{\alpha}}^{(0)}(\vec{k},\vec{q}) \mathscr{F}(i\omega_m,i\Omega_n)$, which is the starting assumption for many closed-form solutions of the Bethe-Salpeter equation (\ref{eq:Bethe_Salpeter_gen}), for instance in the case of impurity scattering \cite{Bruus-2004mb}. This emphasizes that the $\vec{k}$-oddness/null angular average of the bare vertex $\Gamma_{\hat{\Theta},\vec{\alpha}}^{(0)}(\vec{k},0)$ and the $\vec{k}$-even parity of the dispersion $\epsilon_{\vec{k}}$ are the key properties for vertex-correction cancellations. 

\section{Vertex corrections from charge conservation at zero and finite momentum for RPA-like two-particle vertex}\label{Charge_cons_vertex_vanish_q}

In the special case when the irreducible two-particle vertex depends only on the exchanged energy $i\Omega_n$, but not on the incoming and outgoing fermionic frequencies $\omega_m$, $\omega_m'$, Eq.\@ (\ref{eq:Lambda_DMFT}) specializes to the ``Random Phase Approximation (RPA)-like'' form
\begin{equation}\label{eq:Lambda_DMFT_special}
\Lambda(i\omega_m,i \Omega_n', i\Omega_n) \equiv \Lambda(i\Omega_n) \, \forall \left\{i\omega_m,i\omega_m'\right\}.
\end{equation}
The case (\ref{eq:Lambda_DMFT_special}) arises in approximation schemes where the Luttinger–Ward functional $\Phi[\mathscr{G}]$ is quadratic in the density operator $\hat{n}$, or equivalently, if the interaction is Gaussian in auxiliary bosonic fields. This condition includes: density-density interactions with a dispersionless bosonic mode $\phi_i$ (e.g.\@, an Einstein phonon or a local collective mode), where the interaction Hamiltonian is $\hat{H}=g \sum_{i}\hat{n}_i \phi_i$, the corresponding fermionic action is $\mathscr{S}= 1/2 \sum_{i}k_B T \sum_{ i\Omega_n}\hat{n}_i(-i\Omega_n) g^2 D_0(i\Omega_n) \hat{n}_i(i\Omega_n)$, the self-energy is $\Sigma(i\omega_n)=g^2 k_B T \sum_{i\omega_m} \mathscr{V}^{-1} \sum_{\vec{k}} G(\vec{k},i\omega_m) D_0(i\omega_n-i\omega_m)$, and the two-particle vertex is the bare interaction $\Lambda(i\Omega_n) \equiv g^2 D_0(i\Omega_n)$; dynamically screened Hubbard-interaction models $U(i\Omega_n)$, where the two-particle vertex $\Lambda(i\Omega_n)=U(i\Omega_n)$; infinite-range density-density interactions, where the Hamiltonian is $\hat{H}=1/(2\mathscr{N}) \sum_{i,j} V(\tau-\tau')\hat{n}_i(\tau) \hat{n}_j(\tau')$, and in the large-$\mathscr{N}$ limit the self-averaging interaction produces $\Lambda(i\Omega_n)\equiv V(i\Omega_n)$; dynamically screened RPA interactions. 

The special case (\ref{eq:Lambda_DMFT_special}) allows us not only to retrieve the general conditions for cancellation of vertex corrections of Sec.\@ \ref{Charge_cons_vertex_vanish}, but also to obtain explicit forms of vertex corrections, whenever the latter do not vanish. In the following, we formally proceed at finite $\vec{q}$ and comment on the $q=0$ limit at the end of the derivations. 

Under the assumption (\ref{eq:Lambda_DMFT_special}), the Bethe-Salpeter equation (\ref{eq:Bethe_Salpeter_gen}) admits a closed-form solution for the renormalized vertex, which is obtained after multiplying the ladder equation (\ref{eq:Bethe_Salpeter_gen}) by $G(\vec{k},i\omega_m)G(\vec{k}+\vec{q},i\omega_m+i\Omega_n)$ and summing over momenta $\vec{k}$ and frequencies $i\omega_m$. We have 
\begin{align}\label{eq:Gamma_Theta_renorm}
\Gamma_{\hat{\Theta},\vec{\alpha}}(\vec{k},\vec{q},i\omega_m,i\Omega_n)&=\Gamma_{\hat{\Theta},\vec{\alpha}}^{(0)}(\vec{k},\vec{q})+\Lambda(i\Omega_n) \nonumber \\ &\times \frac{B_0^{\vec{\alpha}}}{1-\Lambda(i\Omega_n) A_G},
\end{align}
where 
\begin{subequations}\label{eq:local_ladder_gen}
\begin{align}\label{eq:A_G_const_gen}
A_G&=A_G(\vec{q},i\Omega_n)=\frac{1}{\mathscr{V}}\sum_{\vec{k}'} k_B T \sum_{i\omega_m} G(\vec{k}',i\omega_m) \nonumber \\ & \times G(\vec{k}'+\vec{q},i\omega_m+i\Omega_n),
\end{align}
\begin{align}\label{eq:B_0_gen}
B_0^{\vec{\alpha}}&=B_0^{\vec{\alpha}}(\vec{q},i\Omega_n)=\frac{1}{\mathscr{V}}\sum_{\vec{k}'} k_B T \sum_{i\omega_m} G(\vec{k}',i\omega_m)\nonumber \\ &\times G(\vec{k}'+\vec{q},i\omega_m+i\Omega_n)  \Gamma_{\hat{\Theta},\vec{\alpha}}^{(0)}(\vec{k}',\vec{q}).
\end{align}
\end{subequations}
The correlation function (\ref{eq:chi_Theta_gen}) then comprises a ``renormalized bubble'' term stemming from self-energy effects, and a vertex-correction term with scalar but frequency-dependent two-particle vertex function:
\begin{subequations}\label{eq:chi_Theta_decomp}
\begin{equation}
\chi_{\hat{\Theta} \hat{\Theta}}^{\vec{\alpha}\vec{\beta}}(\vec{q},i\Omega_n)=\chi_{\hat{\Theta} \hat{\Theta},(0)}^{\vec{\alpha}\vec{\beta}}(\vec{q},i\Omega_n)+\delta \chi_{\hat{\Theta} \hat{\Theta}}^{\vec{\alpha}\vec{\beta}}(\vec{q},i\Omega_n),
\end{equation}
\begin{align}\label{eq:chi_Theta_0}
\chi_{\hat{\Theta} \hat{\Theta},(0)}^{\vec{\alpha}\vec{\beta}}(\vec{q},&i\Omega_n)=k_B T \sum_{i\omega_m} \frac{1}{\mathscr{V}} \sum_{\vec{k},\sigma} \Gamma_{\hat{\Theta},\vec{\alpha}}^{(0)}(\vec{k},\vec{q}) G(\vec{k},i\omega_m) \nonumber \\ & \times G(\vec{k}+\vec{q},i\omega_m+i\Omega_n) \Gamma_{\hat{\Theta},\vec{\beta}}^{(0)}(\vec{k}+\vec{q},-\vec{q}),
\end{align}
\begin{align}\label{eq:chi_Theta_VC}
\delta\chi_{\hat{\Theta} \hat{\Theta}}^{\vec{\alpha}\vec{\beta}}(\vec{q},i\Omega_n)&=\Lambda(i\Omega_n)\frac{B_0^{\vec{\alpha}}B_0^{\vec{\beta}}}{1-\Lambda(i\Omega_n) A_G}. 
\end{align}
\end{subequations}
From Eqs.\@ (\ref{eq:Gamma_Theta_renorm}) and (\ref{eq:B_0_gen}), we see that the renormalized vertex for the observable $\hat{\Theta}$ is proportional to $B_0^{\vec{\alpha}}$, which involves a sum over momenta $\vec{k}'$ of the particle-hole bubble multiplied by the \emph{bare vertex} $\Gamma_{\hat{\Theta},\vec{\alpha}}^{(0)}(\vec{k}',\vec{q})$. Therefore, if the bare vertex is odd with respect to momentum inversion $\left\{\vec{k}'\right\} \mapsto \left\{-\vec{k}'\right\}$ and the dispersion satisfies $\epsilon_{\vec{k}}=\epsilon_{-\vec{k}}$ (for vectorial vertices), or if the vertex yields a null sum upon angular average due to the respectively odd and even symmetry of the bare vertex and the dispersion under the same symmetry operation $\mathscr{g}$, then $B_0^{\vec{\alpha}}$ is null at $q=0$ due to the sum over $\vec{k}'$, and vertex corrections vanish identically at zero external momentum, $q=0$, consistently with Sec.\@ \ref{Charge_cons_vertex_vanish}. For the same reason, $B_0^{\vec{\alpha}}=0$ implies that vertex corrections (\ref{eq:chi_Theta_VC}) to the two-particle correlation function are also null, $\delta\chi_{\hat{\Theta} \hat{\Theta}}^{\vec{\alpha}\vec{\beta}}(0,i\Omega_n)=0 \, \forall i\Omega_n$, and the entirety of the linear response is given by the interacting particle-hole bubble $\chi_{\hat{\Theta} \hat{\Theta},(0)}^{\vec{\alpha}\vec{\beta}}(0,i\Omega_n)$ in accordance with Eq.\@ (\ref{eq:chi_Theta_0}). As derived more generally in Sec.\@ \ref{Charge_cons_vertex_vanish}, correlation functions that enjoy absence of vertex corrections include the current-current, momentum-momentum, and heat-heat ones, while the ones pertaining to shear stresses have no vertex corrections under additional rotational/mirror symmetry assumptions, and density-density and bulk stress-stress correlation functions always include nonzero corrections (\ref{eq:chi_Theta_VC}). 

However, there are situations in which vertex corrections vanish from the finite-$\vec{q}$ and finite-$i\Omega_n$ response in the DMFT-like limit, whenever the bare vertex is $\vec{k}$-odd: this is the case of the transverse ($\vec{k}\perp \vec{q}$) response for vectorial vertices, previously treated in Sec.\@ \ref{Vertex_corrections_vanish_currents_alpha}, whenever the dispersion $\epsilon_{\vec{k}}$ is isotropic. Another notable case arises for quadratic, but possibly anisotropic, dispersion (\ref{eq:quadr_gen_anis}), whereby the finite-$\vec{q}$, static current/momentum response is devoid of vertex corrections. Finally, for quadratic and isotropic dispersion, one can show that the full current/momentum vertex can be chosen to be bare, while still obeying the charge-conservation Ward identity (\ref{eq:Ward_identity_charge}). All these instances are discussed in the following sections. 

\section{Absence of vertex corrections for vectorial vertices with isotropic dispersion and generic two-particle vertex}\label{Isotropic_vanish}

If we now additionally assume isotropic dispersion, 
\begin{equation}\label{eq:epsilonk_isotropic}
\epsilon_{\vec{k}}=\epsilon_{\left|\vec{k}\right|},
\end{equation}
then vertex corrections vanish from the vectorial response vertices, treated in Sec.\@ \ref{Vertex_corrections_vanish_currents_alpha}, \emph{even at finite momentum $\vec{q}$ and finite frequency}, in directions orthogonal to $\vec{q}$ and for a local self-energy. These vertices include the ones for current, momentum, and heat current, respectively given by Eqs.\@ (\ref{eq:current_vertex_gen}), (\ref{eq:momentum_vertex_gen}), and (\ref{eq:thermal_current_vertex_gen}). 

To see this property, working at finite $\vec{q}$ and finite $i\Omega_n$ with the local irreducible two-particle vertex (\ref{eq:Lambda_DMFT}), the vertex equation (\ref{eq:Bethe_Salpeter_gen}) reads $\Gamma_{\hat{\Theta},\vec{\alpha}}(\vec{k},\vec{q},i\omega_m,i\Omega_n)=\Gamma_{\hat{\Theta},\vec{\alpha}}^{(0)}(\vec{k},\vec{q})+ C^{\vec{\alpha}}(\vec{q},i\omega_m,i\Omega_n)$, where we have finite-$\vec{q}$ generalizations of Eqs.\@ (\ref{eq:Bethe_Salpeter_C}) and (\ref{eq:Bethe_Salpeter_X}):
\begin{align}\label{eq:Bethe_Salpeter_C_q}
C^{\vec{\alpha}}(\vec{q},i\omega_m,i\Omega_n)&=k_B T \sum_{i\omega_m'} \Lambda(i\omega_m, i\omega_m',i \Omega_n) \nonumber \\ & \times X^{\vec{\alpha}}(\vec{q},i\omega_m', i\Omega_n),
\end{align}
and
\begin{align}\label{eq:Bethe_Salpeter_X_q}
X^{\vec{\alpha}}(\vec{q},i\omega_m',i\Omega_n)&= \frac{1}{\mathscr{V}}\sum_{\vec{k}'} G(\vec{k}',i\omega_m') \nonumber \\ & \times G(\vec{k}'+\vec{q},i\omega_m'+i\Omega_n) \nonumber \\ & \times \Gamma_{\hat{\Theta},\vec{\alpha}}(\vec{k}',\vec{q},i\omega_m',i\Omega_n).
\end{align}
Again, if $X^{\vec{\alpha}}(\vec{q},i\omega_m',i\Omega_n)$ is identically null, then self-consistently $C^{\vec{\alpha}}(\vec{q},i\omega_m,i\Omega_n)$ is also null, and vertex corrections vanish at finite $\vec{q}$ and finite $i\Omega_n$ as well. Combining Eqs.\@ (\ref{eq:Bethe_Salpeter_C_q}) and (\ref{eq:Bethe_Salpeter_X_q}) for spatial coordinates $\vec{\alpha} \equiv \alpha$, we have
\begin{align}\label{eq:Bethe_Salpeter_comb_q}
C^\alpha(\vec{q},i\omega_m,i\Omega_n)&= B^\alpha(\vec{q},i\omega_m,i\Omega_n)\nonumber \\  &+ k_B T \sum_{i\omega_m'} \Lambda(i\omega_m, i\omega_m',i \Omega_n) \nonumber \\ \times & \frac{1}{\mathscr{V}}\sum_{\vec{k}'} G(\vec{k}',i\omega_m') G(\vec{k}'+\vec{q},i\omega_m'+i\Omega_n) \nonumber \\ &\times C^\alpha(\vec{q},i\omega_m',i\Omega_n),
\end{align}
where 
\begin{align}\label{eq:Bethe_Salpeter_B_alpha}
B^\alpha(\vec{q},i\omega_m,i\Omega_n)&=k_B T \sum_{i\omega_m'} \Lambda(i\omega_m, i\omega_m',i \Omega_n) \nonumber \\ & \times \frac{1}{\mathscr{V}}\sum_{\vec{k}'} G(\vec{k}',i\omega_m') G(\vec{k}'+\vec{q},i\omega_m'+i\Omega_n) \nonumber \\ & \times \Gamma_{\hat{J},\vec{\alpha}}^{(0)}(\vec{k}',\vec{q}).
\end{align}

For generic $\epsilon_{\vec{k}}$, Eq.\@ (\ref{eq:Bethe_Salpeter_B_alpha}) need not vanish, and is generally nonzero. 
However, if we show that, for the dispersion (\ref{eq:epsilonk_isotropic}), Eq.\@ (\ref{eq:Bethe_Salpeter_B_alpha}) only admits the null solution for $B^\alpha(\vec{q},i\omega_m,i\Omega_n)$, then Eq.\@ (\ref{eq:Bethe_Salpeter_comb_q}) becomes a homogeneous equation for $C^\alpha(\vec{q},i\omega_m,i\Omega_n)$, which has the solution $C^\alpha(\vec{q},i\omega_m,i\Omega_n)=0$ away from collective modes, similarly to the $q=0$ case of Eq.\@ (\ref{eq:canc_C_matrix}).
To show $B^\alpha(\vec{q},i\omega_m,i\Omega_n)=0$, we invoke the isotropy of the dispersion (\ref{eq:epsilonk_isotropic}), by which the only vectorial argument $\vec{q}$ of $\vec{B}(\vec{q},i\omega_m,i\Omega_n)$ dictates the direction of the latter, in other words it must be
\begin{equation}\label{eq:B_alpha_isotropic}
B^\alpha(\vec{q},i\omega_m,i\Omega_n)=b(q,i\omega_m,i\Omega_n) q_\alpha,
\end{equation}
with $b(q,i\omega_m,i\Omega_n)$ a scalar quantity. 
On the other hand, for the isotropic dispersion (\ref{eq:epsilonk_isotropic}), one can always decompose the response into two parts, respectively longitudinal and transverse with respect to $\vec{q}$ \cite{Dressel-2001}: 
\begin{align}\label{eq:Gamma_theta_isotropic}
\vec{\Gamma}_{\hat{\Theta}}(\vec{k},\vec{q},i\omega_m,i\Omega_n)&=\vec{\Gamma}_{\hat{\Theta},\mathrm{T}}(\vec{k},\vec{q},i\omega_m,i\Omega_n) \nonumber \\ & +\vec{\Gamma}_{\hat{\Theta},\mathrm{L}}(\vec{k},\vec{q},i\omega_m,i\Omega_n),
\end{align} 
where
\begin{equation}\label{eq:Gamma_T}
\vec{q} \cdot \vec{\Gamma}_{\hat{\Theta},\mathrm{T}}(\vec{k},\vec{q},i\omega_m,i\Omega_n)=0,
\end{equation}
and 
\begin{equation}\label{eq:Gamma_L}
\vec{q} \times \vec{\Gamma}_{\hat{\Theta},\mathrm{L}}(\vec{k},\vec{q},i\omega_m,i\Omega_n)=0. 
\end{equation}
For instance, in a three-dimensional (3D) system, one can choose the $z$ axis of a Cartesian reference system to align with $\vec{q}=q \hat{u}_z$, so that the longitudinal response pertains to $\alpha=z\equiv\mathrm{L}$ in Eq.\@ (\ref{eq:Bethe_Salpeter_comb_q}), while the transverse response involves $\alpha=\left\{x,y\right\}\equiv \mathrm{T}$.

In the transverse sector $\alpha \equiv \mathrm{T}$, from Eq.\@ (\ref{eq:B_alpha_isotropic}) we see that indeed $B^{\mathrm{T}}(\vec{q},i\omega_m,i\Omega_n)=0$, so the Bethe-Salpeter equation is homogeneous: the second term in Eq.\@ (\ref{eq:Bethe_Salpeter_comb_q}) yields, similarly to Eqs.\@ (\ref{eq:canc_C_matrix}), a matrix equation that reads
\begin{subequations}\label{eq:canc_C_matrix_q}
\begin{align}
C^{\vec{\alpha}}(\vec{q},i\omega_m, i\Omega_n)&=k_B T \sum_{i\omega_m'} K(\vec{q},i\omega_m,i\omega_m',i\Omega_n) \nonumber \\ & \times C^{\vec{\alpha}}(\vec{q},i\omega_m', i\Omega_n),
\end{align}
\begin{align}\label{eq:K_collective_mode_q}
K(\vec{q},i\omega_m,i\omega_m', i\Omega_n)&=\Lambda(i\omega_m, i\omega_m',i \Omega_n) \frac{1}{\mathscr{V}}\sum_{\vec{k}'} G(\vec{k}',i\omega_m') \nonumber \\ & \times G(\vec{k}'+\vec{q},i\omega_m'+i\Omega_n).
\end{align}
\end{subequations}
The condition $\mathrm{det}\left\{\delta_{m m'}-k_B T  K(\vec{q},i\omega_m,i\omega_m', i\Omega_n)\right\}=0$, where the determinant is taken over the internal fermionic Matsubara frequencies $\left(i\omega_m, i\omega_m'\right)$, signals a genuine nonuniform density collective mode at external momentum $\vec{q}$ and frequency $i\Omega_n$, stemming from Eq.\@ (\ref{eq:canc_C_matrix_q}) (technically, the collective mode results after analytic continuation to the retarded function with $i\Omega_n \rightarrow \hbar\omega+i0^+$). In other words, the collective mode appears when the operator $\mathbb{I}-k_B T\underline{\underline{K}}(\vec{q},i\Omega_n)$, constructed from the fermionic Matsubara components $\underline{\underline{K}}(\vec{q},i\Omega_n)=\left\{K(\vec{q},i\omega_m,i\omega_m', i\Omega_n)\right\}$, includes $1$ in its spectrum.
Assuming that the system lies away from such a resonance, the operator $\mathbb{I}-k_B T\underline{\underline{K}}(\vec{q},i\Omega_n)$ is invertible, and the only other solution is $C^\alpha(\vec{q},i\omega_m,i\Omega_n)=0 \, \forall \left\{\vec{q},\omega_m,\Omega_n\right\}$. 

Notice that, similarly to the $q=0$ case of Eq.\@ (\ref{eq:canc_C_matrix}), regardless of the existence of a density collective mode, the transverse part of the correlation function in Eq.\@ (\ref{eq:chi_Theta_gen}) contains no vertex corrections, if Eqs.\@ (\ref{eq:dispersion_even_parity}), (\ref{eq:bare_vertex_odd}), and (\ref{eq:epsilonk_isotropic}) hold. This cancellation is realized by inserting the renormalized vertex $\Gamma_{\hat{\Theta},\vec{\alpha}}(\vec{k},\vec{q},i\omega_m,i\Omega_n)=\Gamma_{\hat{\Theta},\vec{\alpha}}^{(0)}(\vec{k},\vec{q})+ C^{\vec{\alpha}}(\vec{q},i\omega_m,i\Omega_n)$, which might include a contribution $C^{\vec{\alpha}}(\vec{q},i\omega_m,i\Omega_n)$ at a collective mode, into Eq.\@ (\ref{eq:chi_Theta_gen}): the putative transverse vertex-correction contribution reads $\chi_{\hat{\Theta} \hat{\Theta}}^{T}(\vec{q},i\Omega_n)=k_B T\sum_{i\omega_m} C^{\vec{\alpha}}(\vec{q},i\omega_m,i\Omega_n) \mathscr{V}^{-1} \sum_{\vec{k}} \Gamma_{\hat{\Theta},T}^{(0)}(\vec{k},0) G(\vec{k},i\omega_m) G(\vec{k}+\vec{q},i\omega_m+i\Omega_n)$, which is still globally odd with respect to transverse components of the wave vector $\vec{k}$. 

Then, for the transverse response of vectorial (current, momentum, heat-current) vertices and the isotropic dispersion (\ref{eq:epsilonk_isotropic}), the corresponding two-point correlation function (\ref{eq:chi_Theta_gen}) is devoid of vertex corrections in the local DMFT-like limit, even at finite $\vec{q}$ and finite $i\Omega_n$. This property is very useful to calculate, for instance, the nonlocal electromagnetic response of isotropically dispersing systems with local self-energy. 

\section{Absence of static vertex corrections for vectorial vertices with quadratic dispersion and generic two-particle vertex}\label{Quadratic_vanish}

Let us now assume quadratic but anisotropic dispersion of the kind
\begin{equation}\label{eq:quadr_gen_anis}
\epsilon_{\vec{k}}=k_0+\frac{1}{2}\sum_{i,j} k_i A_{ij} k_j,
\end{equation}
where $\left\{k_0, A_{ij}\right\} \in \mathbb{R}$.
Then, vertex corrections vanish from any component of vectorial response vertices that are linear in momentum $\vec{k}$, at finite momentum $\vec{q}$ but null frequency $i\Omega_n=0$, for a local self-energy. Such vanishing occurs because the dispersion (\ref{eq:quadr_gen_anis}) makes the static source term $B^\alpha(\vec{q},i\omega_m,0)$ in Eq.\@ (\ref{eq:Bethe_Salpeter_B_alpha}) identically null. 

In the following, as concrete illustration I treat the case of the current response, for which the bare vertex reads
\begin{align}\label{eq:Gamma_J_quadr}
\Gamma_{\hat{J},\alpha}^{(0)}(\vec{k},\vec{q})&=\frac{1}{\hbar} \frac{\partial \epsilon_{\vec{k}+\vec{q}/2}}{\partial k_\alpha} \nonumber \\ &=\frac{1}{\hbar} \sum_{\beta} A_{\alpha \beta} \left(k_\beta+\frac{q_\beta}{2}\right). 
\end{align}
Hence, the bare vertex (\ref{eq:Gamma_J_quadr}) is a linear combination of terms which are proportional to momenta components $k_{\beta}$, i.e.\@, current density components $\hat{J}_\alpha$ are a linear combination of momentum components $\hat{\pi}_\alpha$. The source term in the static version of Eq.\@ (\ref{eq:Bethe_Salpeter_comb_q}) then reads
\begin{align}\label{eq:Bethe_Salpeter_B_alpha_static}
B^\alpha(\vec{q},i\omega_m,0)&=k_B T \sum_{i\omega_m'} \Lambda(i\omega_m, i\omega_m',0) \nonumber \\ & \times \frac{1}{\mathscr{V}}\sum_{\vec{k}'} G(\vec{k}',i\omega_m') G(\vec{k}'+\vec{q},i\omega_m') \nonumber \\ & \times \Gamma_{\hat{J},\vec{\alpha}}^{(0)}(\vec{k}',\vec{q}).
\end{align}
A momentum shift $\vec{k}' \mapsto \vec{k}'-\vec{q}/2$ in Eq.\@ (\ref{eq:Bethe_Salpeter_B_alpha_static}) produces
\begin{align}\label{eq:B_0_alpha_current_static}
B^\alpha(\vec{q},i\omega_m,0)&=k_B T \sum_{i\omega_m'} \Lambda(i\omega_m, i\omega_m',0) \nonumber \\ & \times\frac{1}{\mathscr{V}} \sum_{\vec{k}'} F_G(\vec{k}',\vec{q},i\omega_m') \frac{1}{\hbar}\sum_\beta A_{\alpha \beta} k_\beta',
\end{align}
where 
\begin{align}\label{eq:F_G_q}
&F_G(\vec{k}',\vec{q},i\omega_m') \nonumber \\ &=G(\vec{k}'-\vec{q}/2,i\omega_m')G(\vec{k}'+\vec{q}/2,i\omega_m') \nonumber \\ &=\left[ i\omega_m'-\epsilon_{\vec{k}-\vec{q}/2}+\mu -\Sigma(i\omega_m')\right]^{-1} \nonumber \\ &\times \left[i\omega_m'-\epsilon_{\vec{k}+\vec{q}/2}+\mu -\Sigma(i\omega_m')\right]^{-1}.
\end{align}
Eq.\@ (\ref{eq:F_G_q}) is even under inversion $\vec{k}\mapsto \vec -\vec{k}$, while $k_\alpha'$ in Eq.\@ (\ref{eq:B_0_alpha_current_static}) is odd, hence $B^\alpha(\vec{q},i\omega_m,0)$ vanishes identically. Away from a static collective mode, the $\Omega_n=0$ version of Eq.\@ (\ref{eq:canc_C_matrix_q}) then gives zero vertex corrections for any spatial component $\alpha$. 
This result is relevant for nonlocal DC transport (e.g., size-dependent conductivity in narrow channels) of systems in the DMFT-like limit. 

\subsection{Charge conservation without vertex corrections for current response with isotropic quadratic dispersion and generic two-particle vertex}\label{Quadratic_iso_vanish}

Finally, if we combine the assumptions of isotropy and quadratic dispersion, we have
\begin{equation}\label{eq:xi_quadr}
\xi_{\vec{k}}= \epsilon_{\vec{k}}-\mu=\frac{\hbar^2 k^2}{2 m}-\mu,
\end{equation}
where $m$ is the band mass and $\mu$ is the chemical potential. The same arguments of Sec.\@ \ref{Isotropic_vanish} then imply that the transverse current response is devoid of vertex corrections, at finite $\vec{q}$ and $i\Omega_n$. 

Nevertheless, Eq.\@ (\ref{eq:Bethe_Salpeter_comb_q}) does not rule out longitudinal vertex corrections: these are in general nonzero, and contribute to establishing the charge continuity equation, which in turn generates both the charge-conservation Ward identity (\ref{eq:Ward_identity_charge}), and the relation (\ref{eq:chinn_sigmaL}) between the retarded density-density correlation function $\chi_{\hat{n}\hat{n}}^R(\vec{q},\omega)$ and the longitudinal conductivity $\sigma_L(\vec{q},\omega)$ \cite{Conti-1999,Dressel-2001,Vignale-2005}:
\begin{equation}\label{eq:chinn_sigmaL}
\chi_{\hat{n}\hat{n}}^R(\vec{q},\omega)=\frac{q^2}{i e^2 (\omega+i0^+)} \sigma_L(\vec{q},\omega).
\end{equation}
Eq.\@ (\ref{eq:chinn_sigmaL}) can be straightforwardly verified in the case of an imaginary constant self-energy, $\Sigma^R(\omega)\equiv -i \Gamma$ with $\Gamma>0$, by independently computing $\chi_{\hat{n}\hat{n}}^R(\vec{q},\omega)$ and $\chi_{\hat{J}\hat{J}}^{\alpha \alpha,R}(\vec{q},\omega)$ in the longitudinal channel with $\vec{q}=q \hat{u}_\alpha$ \cite{Chen-1989,Kim-2019}. 
Moreover, the charge-conservation Ward identity (\ref{eq:Ward_identity_charge}) by itself just specifies the relation between the density vertex $\Gamma_{\hat{n}}(\vec{k},\vec{q},i\omega_m,i\Omega_n)$ and the longitudinal part of the current vertex $\vec{\Gamma}_{\hat{J}}(\vec{k},\vec{q},i\omega_m,i\Omega_n)$, without individually determining both quantities (except for the current vertex at $q=0$). Hence, both vertices could in principle be renormalized, as long as they obey the Ward identity (\ref{eq:Ward_identity_charge}). 

However, for the dispersion (\ref{eq:xi_quadr}) one can employ the bare (unrenormalized) current vertex, and still preserve the charge-conservation Ward identity (\ref{eq:Ward_identity_charge}) at any momentum $\vec{q}$ and frequency $i\Omega_n$, if we correspondingly renormalize the density vertex. Explicitly, the charge-conserving choices for the density and current vertices read 
\begin{subequations}\label{eq:density_current_charge_conserving}
\begin{equation}\label{eq:density_vertex_quadr_0}
\Gamma_{\hat{n}}(\vec{k},\vec{q},i\omega_m,i\Omega_n)=1-\frac{\Sigma(i\omega_m+i\Omega_n)-\Sigma(i\omega_m)}{i\Omega_n},
\end{equation}
\begin{align}\label{eq:current_vertex_quadr_0}
\Gamma_{\hat{J},\alpha}(\vec{k},\vec{q},i\omega_m,i\Omega_n)&\equiv\Gamma_{\hat{J},\alpha}^{(0)}(\vec{k},\vec{q}) \nonumber \\ &=\frac{\hbar}{m} \left(k_\alpha +\frac{q_\alpha}{2}\right). 
\end{align}
\end{subequations}
Using Eqs.\@ (\ref{eq:density_current_charge_conserving}), vertex corrections to the current-current correlation function $\chi_{\hat{J} \hat{J}}^{\alpha \beta}(\vec{q},i\Omega_n)$, which enters into the conductivity tensor $\sigma_{\alpha \beta}(\vec{q},\omega)$ through the Kubo formula (\ref{eq:Kubo_conductivity_gen}), vanish identically at any external momentum and frequency. In this case, since the density response is affected by vertex corrections, while the current response is not, it is practically advantageous to derive the longitudinal conductivity $\sigma_L(\vec{q},\omega)$ from the current response, rather than from the density response. Then, the density susceptibility $\chi_{\hat{n}\hat{n}}(\vec{q},\omega)$ can then be deduced through Eq.\@ (\ref{eq:chinn_sigmaL}). 

\section{Charge-conserving correlation functions and Kubo formulae at zero and finite momentum}\label{Kubo_charge_conserving}

The locality of the self-energy and of the irreducible two-particle vertex not only impact the correlation function (\ref{eq:chi_Theta_gen}) for the observable $\hat{\Theta}$, but also the associated linear-response function $\Theta_{\vec{\alpha}\vec{\beta}}(\vec{q},\omega)$ if the latter follows from a Kubo formula \cite{Kubo-1957a,Kubo-1957b,Mahan-2000,Bruus-2004mb}:
\begin{equation}\label{eq:Kubo_general_Theta}
\Theta_{\vec{\alpha}\vec{\beta}}(\vec{q},\omega)=\frac{i \mathscr{C}}{\omega+i0^+} \left[\chi_{\hat{\Theta} \hat{\Theta}}^{R,\vec{\alpha}\vec{\beta}}(\vec{q},\omega)- \chi_{\hat{\Theta} \hat{\Theta}}^{R,\vec{\alpha}\vec{\beta}}(0,0)\right],
\end{equation}
where $\mathscr{C}$ is a momentum- and frequency-independent constant. Examples of Eq.\@ (\ref{eq:Kubo_general_Theta}) include the optical conductivity tensor $\sigma_{\alpha \beta}(\vec{q},\omega)$ for the current response $\hat{\Theta}\equiv \hat{J}$, and the viscosity (or viscoelasticity) tensor $\eta_{\alpha \beta \gamma \delta}(\omega)$ for the stress response $\hat{\Theta}\equiv 	\hat{\textit{T}}$ at $q=0$. 

Eq.\@ (\ref{eq:Kubo_general_Theta}) assumes a well-defined static and uniform (long-wavelength) limit $\chi_{\hat{\Theta} \hat{\Theta}}^{R,\vec{\alpha}\vec{\beta}}(0,0)$ for the retarded form of the correlation function (\ref{eq:chi_Theta_gen}). However, notice that $\chi_{\hat{\Theta} \hat{\Theta}}^{R,\vec{\alpha}\vec{\beta}}(0,0)$ may in general depend on the order of limits $\omega\rightarrow 0$, $q\rightarrow 0$ or $q\rightarrow 0$, $\omega\rightarrow 0$, or may contain a zero-frequency singularity, as in the case of the conductivity of systems with conserved quantities (e.g.\@, ballistic transport, superfluids). In general, Hermiticity of the operator $\hat{\Theta}$ prescribes $\left[\chi_{\hat{\Theta} \hat{\Theta}}^{R,\vec{\alpha}\vec{\beta}}(\vec{q},\omega)\right]^\ast=\chi_{\hat{\Theta} \hat{\Theta}}^{R,\vec{\beta}\vec{\alpha}}(-\vec{q},\omega)$, from which we deduce that the static uniform limit (assuming analyticity) of the diagonal part is real-valued: $\chi_{\hat{\Theta} \hat{\Theta}}^{R,\vec{\alpha}\vec{\alpha}}(0,0) \in \mathbb{R}$ for autocorrelation functions of a Hermitian observable $\hat{\Theta}$ \cite{Mahan-2000,Bruus-2004mb}. In the absence of other symmetries, the off-diagonal part $\chi_{\hat{\Theta} \hat{\Theta}}^{R,\vec{\alpha}\vec{\beta}}(0,0)$, with $\alpha \neq \beta$, does not have to be real-valued. 
One can also define the symmetric and antisymmetric parts of the response function, 
\begin{subequations}\label{eq:chi_Theta_Theta_symmetry}
\begin{equation}\label{eq:chi_decomp_symmetry}
\chi_{\hat{\Theta} \hat{\Theta}}^{R,\vec{\alpha}\vec{\beta}}(\vec{q},\omega)=\chi_{\hat{\Theta} \hat{\Theta}}^{R,\vec{\alpha}\vec{\beta},S}(\vec{q},\omega)+\chi_{\hat{\Theta} \hat{\Theta}}^{R,\vec{\alpha}\vec{\beta},A}(\vec{q},\omega),
\end{equation}
\begin{equation}\label{eq:chi_symmetric}
\chi_{\hat{\Theta} \hat{\Theta}}^{R,\vec{\alpha}\vec{\beta},S}(\vec{q},\omega)=\frac{\chi_{\hat{\Theta} \hat{\Theta}}^{R,\vec{\alpha}\vec{\beta}}(\vec{q},\omega)+\chi_{\hat{\Theta} \hat{\Theta}}^{R,\vec{\beta}\vec{\alpha}}(\vec{q},\omega)}{2},
\end{equation}
\begin{equation}\label{eq:chi_antisymmetric}
\chi_{\hat{\Theta} \hat{\Theta}}^{R,\vec{\alpha}\vec{\beta},A}(\vec{q},\omega)=\frac{\chi_{\hat{\Theta} \hat{\Theta}}^{R,\vec{\alpha}\vec{\beta}}(\vec{q},\omega)-\chi_{\hat{\Theta} \hat{\Theta}}^{R,\vec{\beta}\vec{\alpha}}(\vec{q},\omega)}{2},
\end{equation}
\end{subequations}
so that $\chi_{\hat{\Theta} \hat{\Theta}}^{R,\vec{\alpha}\vec{\beta},S}(0,0)\in \mathbb{R}$ (real-valued) and $\chi_{\hat{\Theta} \hat{\Theta}}^{R,\vec{\alpha}\vec{\beta},A}(0,0)\in \mathbb{I}$ (imaginary-valued).
If we additionally assume a well-defined time-reversal parity for the operator $\hat{\Theta}$, then Onsager reciprocity also implies $\chi_{\hat{\Theta} \hat{\Theta}}^{R,\vec{\alpha}\vec{\beta}}(0,0)=\chi_{\hat{\Theta} \hat{\Theta}}^{R,\vec{\beta}\vec{\alpha}}(0,0)$; combining the latter with Hermiticity, one concludes that all static uniform components are real-valued: $\chi_{\hat{\Theta} \hat{\Theta}}^{R,\vec{\alpha}\vec{\beta}}(0,0)\in \mathbb{R}$ \footnote{Real-valuedness of the static, uniform part does not have to hold for correlation functions $\chi_{\hat{\Theta} \hat{\Theta}'}^{R,\vec{\alpha}\vec{\beta}}(0,\omega)$ of different Hermitian operators $\hat{\theta}$ and $\hat{\Theta}'$, for which the general relation $\left[\chi_{\hat{\Theta} \hat{\Theta}'}^{R,\vec{\alpha}\vec{\beta}}(\vec{q},\omega)\right]^\ast=\chi_{\hat{\Theta}' \hat{\Theta}}^{R,\vec{\alpha}\vec{\beta}}(-\vec{q},\omega)$.}.

Within the DMFT assumptions of local self-energy and two-particle vertex, the correlation function for the observable $\hat{\Theta}$ is decomposed into two parts, following Eqs.\@ (\ref{eq:chi_Theta_gen}) and (\ref{eq:Bethe_Salpeter_gen}): we have
\begin{equation}\label{eq:Chi_Theta_decomp}
\chi_{\hat{\Theta} \hat{\Theta}}^{\vec{\alpha}\vec{\beta}}(\vec{q},i\Omega_n)=\chi_{\hat{\Theta} \hat{\Theta}}^{\vec{\alpha}\vec{\beta},(0)}(\vec{q},i\Omega_n)+\delta \chi_{\hat{\Theta} \hat{\Theta}}^{\vec{\alpha}\vec{\beta}}(\vec{q},i\Omega_n),
\end{equation}
where the two parts pertain to the ``renormalized bubble'' with interacting Green's functions, and to ladder vertex corrections, respectively:
\begin{widetext}
\begin{equation}\label{eq:chi_Theta_gen_bubble}
\chi_{\hat{\Theta} \hat{\Theta}}^{\vec{\alpha}\vec{\beta},(0)}(\vec{q},i\Omega_n)=k_B T \sum_{i\omega_m} \frac{1}{\mathscr{V}} \sum_{\vec{k}} \Gamma_{\hat{\Theta},\vec{\alpha}}^{(0)}(\vec{k},\vec{q}) G(\vec{k},i\omega_m)G(\vec{k}+\vec{q},i\omega_m+i\Omega_n) \Gamma_{\hat{\Theta},\vec{\beta}}^{(0)}(\vec{k}+\vec{q},-\vec{q}),
\end{equation}
\begin{equation}\label{eq:chi_Theta_gen_VC}
\delta \chi_{\hat{\Theta} \hat{\Theta}}^{\vec{\alpha}\vec{\beta}}(\vec{q},i\Omega_n)=k_B T \sum_{i\omega_m} C^{\vec{\alpha}}(\vec{q},i\omega_m,i\Omega_n) \frac{1}{\mathscr{V}} \sum_{\vec{k}} \Gamma_{\hat{\Theta},\vec{\beta}}^{(0)}(\vec{k}+\vec{q},-\vec{q}) G(\vec{k},i\omega_m)G(\vec{k}+\vec{q},i\omega_m+i\Omega_n).
\end{equation} 
\end{widetext}
\begin{figure*}[ht] \centering
\includegraphics[width=0.9\linewidth]{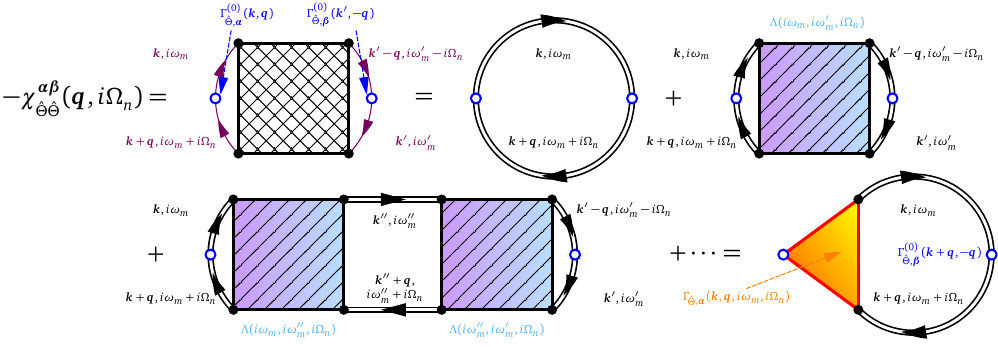}
\caption{\label{fig:Diag_Xthetatheta_VC} Graphical representation of the decomposition (\ref{eq:Chi_Theta_decomp}) for the correlation function $\chi_{\hat{\Theta} \hat{\Theta}}^{\vec{\alpha}\vec{\beta}}(\vec{q},i\Omega_n)$ associated with the observable $\hat{\Theta}$, in terms of the renormalized bubble $\chi_{\hat{\Theta} \hat{\Theta}}^{\vec{\alpha}\vec{\beta},(0)}(\vec{q},i\Omega_n)$ given by Eq.\@ (\ref{eq:chi_Theta_gen_bubble}) (second diagram on the first line) and vertex-correction contributions $\delta \chi_{\hat{\Theta} \hat{\Theta}}^{\vec{\alpha}\vec{\beta}}(\vec{q},i\Omega_n)$ stemming from Eq.\@ (\ref{eq:chi_Theta_gen_VC}) (schematized by the third diagram on the first line, and by the first diagram on the second line). The final diagram on the second line shows the resummation yielding Eq.\@ (\ref{eq:chi_Theta_gen}) with the renormalized vertex $\Gamma_{\hat{\Theta},\vec{\alpha}}(\vec{k},\vec{q},i\omega_m,i\Omega_n)$ given in Eq.\@ (\ref{eq:Bethe_Salpeter_gen}). 
 }
\end{figure*}
In Eq.\@ (\ref{eq:chi_Theta_gen_VC}) we have made use of the definition (\ref{eq:Bethe_Salpeter_C_q}) for the quantity $C^{\vec{\alpha}}(\vec{q},i\omega_m,i\Omega_n)$. Fig.\@ \ref{fig:Diag_Xthetatheta_VC} shows a graphical representation of the decomposition (\ref{eq:Chi_Theta_decomp}), together with the resummation resulting in Eq.\@ (\ref{eq:chi_Theta_gen}) with the renormalized vertex $\Gamma_{\hat{\Theta},\vec{\alpha}}(\vec{k},\vec{q},i\omega_m,i\Omega_n)$ satisfying Eq.\@ (\ref{eq:Bethe_Salpeter_gen}). In the following, our perspective will be to first write explicit expressions for Eqs.\@ (\ref{eq:chi_Theta_gen_bubble}) and (\ref{eq:chi_Theta_gen_VC}) in terms of the self-energy $\Sigma(i\omega_m)$, and proceed analytically as far as possible in two opposite regimes: the one with well-defined, sharp quasiparticle peaks (coherent, or ``Fermi-liquid'') and the one with broad, featureless spectral functions (incoherent, or ``non-Fermi liquid''). 

We will utilize the spectral (Lehmann) representation of the renormalized (interacting) Green's functions, given by Eq.\@ (\ref{eq:spectr_rep}) with the spectral function (\ref{eq:A_G_gen}). 

\subsection{Vertex correction to the correlation function at finite momentum}\label{Corr_Theta_VC}

Let us first treat the vertex-correction part (\ref{eq:chi_Theta_gen_VC}), where comparatively less analytical progress can be made without knowing the explicit expression for the two-particle vertex $\Lambda(i\omega_m,i\omega_m',i\Omega_n)$. Employing the spectral representation (\ref{eq:spectr_rep}) in Eq.\@ (\ref{eq:chi_Theta_gen_VC}), and defining the \emph{$\hat{\Theta}$-vertex transport function} as
\begin{align}\label{eq:vertex_transport_function}
\Upsilon_{\hat{\Theta}}^{\vec{\alpha}}(\xi,\xi',\vec{q})&=\frac{1}{\mathscr{V}}\sum_{\vec{k}} \Gamma_{\hat{\Theta},\vec{\alpha}}^{(0)}(\vec{k}+\vec{q},-\vec{q}) \delta(\xi-\xi_{\vec{k}}) \nonumber \\ & \times \delta(\xi'-\xi_{\vec{k}+\vec{q}}),
\end{align}
Eq.\@ (\ref{eq:chi_Theta_gen_VC}) is recast into
\begin{align}\label{eq:chi_Theta_gen_VC_expl}
\delta \chi_{\hat{\Theta} \hat{\Theta}}^{\vec{\alpha}\vec{\beta}}(\vec{q},i\Omega_n)&=\int_{-\infty}^{+\infty} d \epsilon_1 \int_{-\infty}^{+\infty} d \epsilon_2 \nonumber \\ &\times k_B T  \sum_{i\omega_m} \frac{C^{\vec{\alpha}}(\vec{q},i\omega_m,i\Omega_n)}{(i\omega_m-\epsilon_1)(i\omega_m+i \Omega_n-\epsilon_2)} \nonumber \\ &\times \int_{-\infty}^{+\infty} d\xi \int_{-\infty}^{+\infty} d\xi' \Upsilon_{\hat{\Theta}}^{\vec{\beta}}(\xi,\xi',\vec{q}) \nonumber \\ &\times A_G(\xi,\epsilon_1) A_G(\xi',\epsilon_2). 
\end{align}
\vspace{0.5cm}

\subsection{Renormalized-bubble correlation function at finite momentum}\label{Corr_Theta_bubble}

The ``renormalized-bubble'' part (\ref{eq:chi_Theta_gen_bubble}) can be brought into more compact and explicit form. For vectorial vertices, Eq.\@ (\ref{eq:chi_Theta_gen_bubble}) is the only part of the correlation function that survives at finite $\vec{q}$ and $i\Omega_n$ for isotropic dispersion, as previously shown in Sec.\@ \ref{Isotropic_vanish}, and at $q=0$ for $\vec{k}$-even dispersion and a $\vec{k}$-odd bare vertex, as demonstrated in Sec.\@ \ref{Charge_cons_vertex_vanish}. 

To arrive at the result, we use the spectral representation (\ref{eq:spectr_rep}) in Eq.\@ (\ref{eq:chi_Theta_gen_bubble}), we define the \emph{$\hat{\Theta}$-transport function} as \begin{widetext}
\begin{equation}\label{eq:gen_transport_function}
\Phi_{\hat{\Theta}}^{\vec{\alpha}\vec{\beta}}(\xi,\xi',\vec{q})=\frac{1}{\mathscr{V}}\sum_{\vec{k}} \Gamma_{\hat{\Theta},\vec{\alpha}}^{(0)}(\vec{k},\vec{q})  \Gamma_{\hat{\Theta},\vec{\beta}}^{(0)}(\vec{k}+\vec{q},-\vec{q}) \delta(\xi-\xi_{\vec{k}}) \delta(\xi'-\xi_{\vec{k}+\vec{q}}),
\end{equation}
and we perform the Matsubara sum
\begin{align}\label{eq:Matsubara_2_poles}
k_B T \sum_{i\omega_m} \frac{1}{(i\omega_m-\epsilon_1)(i\omega_m-\epsilon_2)} = \frac{f_{FD}(\epsilon_1)-f_{FD}(\epsilon_2)}{\epsilon_1-\epsilon_2},
\end{align}
employing the invariance property $f_{FD}(i\omega_m)=f_{FD}(i\omega_m+i \Omega_n)$ of the Fermi-Dirac distribution function on the imaginary axis with respect to shifts by a bosonic Matsubara frequency $\Omega_n=2\pi n k_B T$. The final outcome transforms Eq.\@ (\ref{eq:chi_Theta_gen_bubble}) into 
\begin{equation}\label{eq:Chi:theta_bubble_expl}
\chi_{\hat{\Theta} \hat{\Theta}}^{\vec{\alpha}\vec{\beta},(0)}(\vec{q},i\Omega_n)=\int_{-\infty}^{+\infty} d \xi \int_{-\infty}^{+\infty} d \xi' \Phi_{\hat{\Theta}}^{\vec{\alpha}\vec{\beta}}(\xi,\xi',\vec{q}) \int_{-\infty}^{+\infty} d \epsilon_1 \int_{-\infty}^{+\infty} d\epsilon_2 A_G(\xi,\epsilon_1) A_G(\xi',\epsilon_2) \frac{f_{FD}(\epsilon_1)-f_{FD}(\epsilon_2)}{i\Omega_n+\epsilon_1-\epsilon_2}. 
\end{equation}
\end{widetext}
One can further simplify Eq.\@ (\ref{eq:Chi:theta_bubble_expl}) for the interesting case of the isotropic quadratic dispersion (\ref{eq:xi_quadr}), especially in the small-momentum regime where we can linearize $\xi_{\vec{k}+\vec{q}}\approx \xi_{\vec{k}}+\hbar \vec{k} \cdot \vec{q}$. This analysis is performed in the companion paper \cite{Valentinis-2026_unpublished}, and applied to different types of local self-energies. 
Here we stay at the general level of unspecified $\Sigma(i\omega_n)$, and in the following specialize Eqs.\@ (\ref{eq:gen_transport_function}) and (\ref{eq:Chi:theta_bubble_expl}) to $q=0$. 

\subsection{Renormalized-bubble correlation function at zero momentum}\label{Corr_Theta_bubble_q0}

At $q=0$, where according to Sec.\@ \ref{Charge_cons_vertex_vanish} vertex corrections cancel for any $\vec{k}$-odd bare vectorial vertex if the dispersion is $\vec{k}$-even, Eqs.\@ (\ref{eq:gen_transport_function}) and (\ref{eq:Chi:theta_bubble_expl}) specialize to the zero-momentum transport function
\begin{equation}\label{eq:gen_transport_function_q0}
\Phi_{\hat{\Theta}}^{\vec{\alpha}\vec{\beta}}(\xi)=\frac{1}{\mathscr{V}}\sum_{\vec{k},\sigma} \Gamma_{\hat{\Theta},\vec{\alpha}}^{(0)}(\vec{k},0)  \Gamma_{\hat{\Theta},\vec{\beta}}^{(0)}(\vec{k},0) \delta(\xi-\xi_{\vec{k}}),
\end{equation}
and the zero-momentum correlation function
\begin{align}\label{eq:Chi:theta_bubble_expl_q0}
\chi_{\hat{\Theta} \hat{\Theta}}^{\vec{\alpha}\vec{\beta},(0)}(0,i\Omega_n)&=\int_{-\infty}^{+\infty} d \xi \Phi_{\hat{\Theta}}^{\vec{\alpha}\vec{\beta}}(\xi) \nonumber \\ & \times \int_{-\infty}^{+\infty} d \epsilon_1 \int_{-\infty}^{+\infty} d\epsilon_2 A_G(\xi,\epsilon_1) A_G(\xi,\epsilon_2) \nonumber \\ & \times \frac{f_{FD}(\epsilon_1)-f_{FD}(\epsilon_2)}{i\Omega_n+\epsilon_1-\epsilon_2}. 
\end{align}
We now have two opposite regimes in which analytical progress is possible. The first limit is the one in which the transport function (\ref{eq:gen_transport_function_q0}) is sharply peaked around $\xi=0$ in comparison to the broader spectral function $A_G(\xi,\epsilon)$: this is the regime of no sharply defined quasiparticle, ``non-Fermi liquid'', or incoherent; the second limit is the one in which we neglect all energy dependences of the transport function (\ref{eq:gen_transport_function_q0}) in comparison to the sharper spectral function $A_G(\xi,\epsilon)$: this is the regime of well-defined quasiparticle, ``Fermi liquid'', or coherent. In the following we separately analyze these two limits, while in the crossovers the more general expression (\ref{eq:Chi:theta_bubble_expl_q0}) applies. 

\subsubsection{Coherent regime at zero momentum: Allen formula for the linear-response function}

Let us approximate 
\begin{equation}\label{eq:constant_transport_function_q0}
\Phi_{\hat{\Theta}}^{\vec{\alpha}\vec{\beta}}(\xi)\approx \Phi_{\hat{\Theta}}^{\vec{\alpha}\vec{\beta}}(0) \, \forall \xi,
\end{equation}
which places our system in the coherent, or ``Fermi liquid'', regime. Starting from Eq.\@ (\ref{eq:Chi:theta_bubble_expl_q0}) for the correlation function, and assuming Hermiticity and time-reversal symmetry for the operator $\hat{\Theta}$, the derivation reported in App.\@ \ref{App_Kubo_coherent} leads us to the following result for the linear-response Kubo formula given by Eq.\@ (\ref{eq:Kubo_general_Theta}): 
\begin{align}\label{eq:Kubo_Theta_Allen_q0}
&\Theta_{\vec{\alpha}\vec{\beta}}(\omega)=\frac{i \mathscr{C} \Phi_{\hat{\Theta}}^{\vec{\alpha}\vec{\beta}}(0)}{\omega+i0^+} \nonumber \\ &\times \int_{-\infty}^{+\infty} d \epsilon \frac{f_{FD}(\epsilon)-f_{FD}(\epsilon+\hbar\omega)}{\hbar\omega+i0^+ +\left[\Sigma^R(\epsilon)\right]^\ast-\Sigma^R(\epsilon+\hbar\omega)},
\end{align}
where $^\ast$ means complex conjugation. Eq.\@ (\ref{eq:Kubo_Theta_Allen_q0}) is a kind of generalized ``Allen formula'' \cite{Berthod-2013,Allen-2015}: it is the most general expression for the Kubo formula at $q=0$ in coherent regime, when vertex corrections (\ref{eq:chi_Theta_gen_VC_expl}) vanish. 

\subsubsection{Incoherent regime at zero momentum}

When the retarded self-energy $\Sigma^R(\epsilon)$ in the spectral functions (\ref{eq:A_G_gen}) dominates over the dispersion $\xi_{\vec{k}}$, we can approximate $A_G(\epsilon)$ to be momentum-independent:
\begin{equation}\label{eq:broad_AG_q0}
A_G(\epsilon)\approx -\frac{1}{\pi}\frac{\mathrm{Im}\left\{\Sigma^R(\epsilon)\right\}}{\left[\epsilon-\mathrm{Re}\left\{\Sigma^R(\epsilon)\right\}\right]^2+\left[\mathrm{Im}\left\{\Sigma^R(\epsilon)\right\}\right]^2},
\end{equation}
which implies that the system lies in the incoherent, or ``non-Fermi liquid'', regime. In particular, an applicability criterion for Eq.\@ (\ref{eq:broad_AG_q0}) arises when the static self-energy $\left|\mathrm{Im}\left\{\Sigma^R(0)\right\}\right|>W$, where $W$ is the typical bandwidth of the single-particle dispersion $\xi_{\vec{k}}$ \cite{Cha-2020b,Patel-2023,Valentinis-2026}. In this case, the derivation sketched in App.\@ \ref{App_Kubo_incoherent} brings us to another explicit expression for the correlation function (\ref{eq:Chi:theta_bubble_expl_q0}) entering into the linear-response Kubo formula (\ref{eq:Kubo_general_Theta}), namely 
\begin{align}\label{eq:Kubo_bubble_expl_q0_NFL}
\chi_{\hat{\Theta} \hat{\Theta}}^{R,\vec{\alpha}\vec{\beta},(0)}(0,\omega)&=\frac{1}{\pi}\left[\int_{-\infty}^{+\infty} d \xi \Phi_{\hat{\Theta}}^{\vec{\alpha}\vec{\beta}}(\xi)\right] \int_{-\infty}^{+\infty} d \epsilon f_{FD}(\epsilon) \nonumber \\ & \times \frac{-\mathrm{Im}\left\{\Sigma^R(\epsilon)\right\}}{\left[\epsilon-\mathrm{Re}\left\{\Sigma^R(\epsilon)\right\}\right]^2 +\left[\mathrm{Im}\left\{\Sigma^R(\epsilon)\right\}\right]^2} \nonumber \\ & \left\{\frac{1}{\epsilon+\hbar\omega-\Sigma^R(\epsilon+\hbar\omega)} \right. \nonumber \\ & +\left. \frac{1}{\epsilon-\hbar\omega-\left[\Sigma^R(\epsilon)\right]^\ast} \right\}.
\end{align}
Eq.\@ (\ref{eq:Kubo_bubble_expl_q0_NFL}) applies to incoherent, locally interacting non-Fermi liquids, such as strange metals analyzed through DMFT \cite{Metzner-1989,Muller-1989,Khurana-1990,Georges-1996, Janis-2001, Kotliar-2004,Berthod-2013,Schaefer-2021} or through Sachdev-Ye-Kitaev large-$\mathscr{N}$ saddle-point equations \cite{Esterlis-2019,Cha-2020,Esterlis-2021,Patel-2023,Valentinis-2026}. 

\section{Momentum conservation and renormalization of momentum and current vertices}\label{Momentum_cons_renorm}

\subsection{Global momentum conservation at $q=0$}\label{Momentum_cons_q0}

The above derivations ensure charge/mass conservation at any $\vec{q}$ and $i\Omega_n$, but not momentum conservation. In fact, for a local self-energy, momentum conservation in accordance with the Ward identity (\ref{eq:Ward_identity_momentum}) cannot hold locally at generic $\vec{k}$ and $\vec{q}$, since the local irreducible vertex function (\ref{eq:Lambda_DMFT}) cannot possibly carry the required momentum structure to preserve momentum conservation at any $q$. However, at $q=0$ one can still impose the \emph{global} conservation of \emph{total momentum} in the DMFT-like limit, by constraining the renormalized momentum vertex to 
\begin{align}\label{eq:momentum_vertex_momentum_cons}
&\Gamma_{\hat{\pi},\alpha}(\vec{k},0,i\omega_m,i\Omega_n)=\Gamma_{\hat{\pi},\alpha}^{(0)}(\vec{k},0) \nonumber \\ &\times \left[1-\frac{\Sigma(i\Omega_n+i\omega_m)-\Sigma(i\omega_m)}{i\Omega_n}\right]. 
\end{align}
Eq.\@ (\ref{eq:momentum_vertex_momentum_cons}) directly stems from the Ward identity (\ref{eq:Ward_identity_momentum}) at $q=0$ and ensures conservation of total momentum in the system, which is equal to the $q=0$ component. For a general dispersion $\epsilon_{\vec{k}}$, this constraint on the momentum vertex does not propagate to a constraint on the current vertex, since the latter is in general not proportional to momentum \cite{Raines-2024}. 

However, for the isotropic quadratic dispersion (\ref{eq:xi_quadr}) current is indeed proportional to momentum, so Eq.\@ (\ref{eq:momentum_vertex_momentum_cons}) fixes the $q=0$ current vertex through the momentum Ward identity:
\begin{align}\label{eq:current_vertex_quadr_momentum_cons}
&\Gamma_{\hat{J},\alpha}(\vec{k},0,i\omega_m,i\Omega_n)=\Gamma_{\hat{J},\alpha}^{(0)}(\vec{k},0) \nonumber \\ &\times \left[1-\frac{\Sigma(i\Omega_n+i\omega_m)-\Sigma(i\omega_m)}{i\Omega_n}\right] \nonumber \\ & = \frac{\hbar}{m} k_\alpha \left[1-\frac{\Sigma(i\Omega_n+i\omega_m)-\Sigma(i\omega_m)}{i\Omega_n}\right]. 
\end{align}
Inserting Eq.\@ (\ref{eq:current_vertex_quadr_momentum_cons}) into the Kubo formula (\ref{eq:Kubo_general_Theta}) for the current-current correlation function -- $\hat{\Theta}\equiv\hat{J}$ -- at $q=0$ yields a conductivity consistent with the perfect Drude model, $\sigma(\omega)=i e^2 n/\left[m (\omega+ i0^+)\right]$, without momentum dissipation apart from a delta function at $\omega=0$ \cite{Maiti-2017}. In order to ensure the conservation of both total momentum (at $q=0$) and charge/mass (at any $q$), one can use the finite-$\vec{q}$ generalization (with $\Gamma_{\hat{J},\alpha}^{(0)}(\vec{k},0) \mapsto \Gamma_{\hat{J},\alpha}^{(0)}(\vec{k},\vec{q})=\hbar/m(k_\alpha+q_\alpha/2)$) of Eq.\@ (\ref{eq:current_vertex_quadr_momentum_cons}) for the current vertex, and formally modify the density vertex (\ref{eq:density_vertex_quadr_0}) such that the Ward identities (\ref{eq:Ward_identity_charge}) for charge conservation at any $q$ and (\ref{eq:Ward_identity_momentum}) for momentum conservation at $q=0$ are both simultaneously obeyed :
\begin{align}\label{eq:density_vertex_quadr_momentum_cons}
\Gamma_{\hat{n}}(\vec{k},\vec{q},i\omega_m,i\Omega_n)&=1 -\frac{\Sigma(i\Omega_n+i\omega_m)-\Sigma(i\omega_m)}{(i\Omega_n)^2} \nonumber \\ & \times \left[i\Omega_n+\hbar \vec{q} \cdot \frac{\hbar}{m} \left(\vec{k}+\frac{\vec{q}}{2}\right)\right].
\end{align}
A many-body theory constructed with the current vertex (\ref{eq:current_vertex_quadr_momentum_cons}) and the density vertex (\ref{eq:density_vertex_quadr_momentum_cons}) for quadratic isotropic dispersion has only finite-$q$ (inhomogeneous) processes as sources of momentum dissipation. In this sense, 
one could refer to such theory as \emph{emergent viscoelasticity}, or \emph{quasi-hydrodynamics}, meaning that the leading long-wavelength term of momentum dissipation appears at order $q^2$, as it occurs for the viscosity coefficient (viscosity-conductivity relation) in Galilean-invariant systems \cite{Taylor-2010,Bradlyn-2012}. 

Nevertheless, the analogy with emergent viscoelasticity is not complete for the theory (\ref{eq:density_vertex_quadr_momentum_cons}), since the coefficient of the $q^2$ term in the long-wavelength expansion of the conductivity is not the viscosity given by the renormalized bubble (\ref{eq:Chi:theta_bubble_expl_q0}) without vertex corrections. This is because a Galilean-invariant system obeys momentum conservation at any $\vec{q}$ and $i\Omega_n$, as prescribed by Eq.\@ (\ref{eq:Ward_identity_momentum}) at $q\neq 0$, while Eq.\@ (\ref{eq:density_vertex_quadr_momentum_cons}) imposes momentum conservation only at $q=0$. 

Another comment is that, by imposing Eqs.\@ (\ref{eq:density_vertex_quadr_momentum_cons}) and (\ref{eq:current_vertex_quadr_momentum_cons}) as the renormalized density and current vertices, these vertices are no longer consistent with the ladder Bethe-Salpeter equations (\ref{eq:Bethe_Salpeter_gen}). This is because the latter force a frequency-dependent but momentum-independent renormalization of the bare vertex, in accordance with Eq.\@ (\ref{eq:Bethe_Salpeter_C}) for the correction $C^\alpha(i\omega_m,i\Omega_n)$, while Eq.\@ (\ref{eq:current_vertex_quadr_momentum_cons}) for global momentum conservation prescribes a $\vec{k}$-dependent renormalization of the vertex, even at $q=0$. Such discrepancy is linked to the impossibility of imposing momentum conservation with a purely local two-particle vertex (\ref{eq:Lambda_DMFT}), which would instead require a more general nonlocal vertex $\Lambda(\vec{k},\vec{q},\vec{q'},i\omega_m,i \Omega_n', i\Omega_n)$ obeying the Bethe-Salpeter equation (\ref{eq:Bethe_Salpeter_gen_nonlocal}). Implicitly, by imposing Eqs.\@ (\ref{eq:density_vertex_quadr_momentum_cons}) and (\ref{eq:current_vertex_quadr_momentum_cons}), we are relaxing the assumption of a purely local two-particle vertex, including nonlocality in an, as yet unspecified, nonlocal two-particle vertex. 

\subsection{Momentum conservation at any $\vec{q}$, and relation between nonlocal momentum-linear correlation functions and the viscosity tensor}\label{Momentum_cons_q2}

\begin{figure*}[ht] \centering
\includegraphics[width=0.75\linewidth]{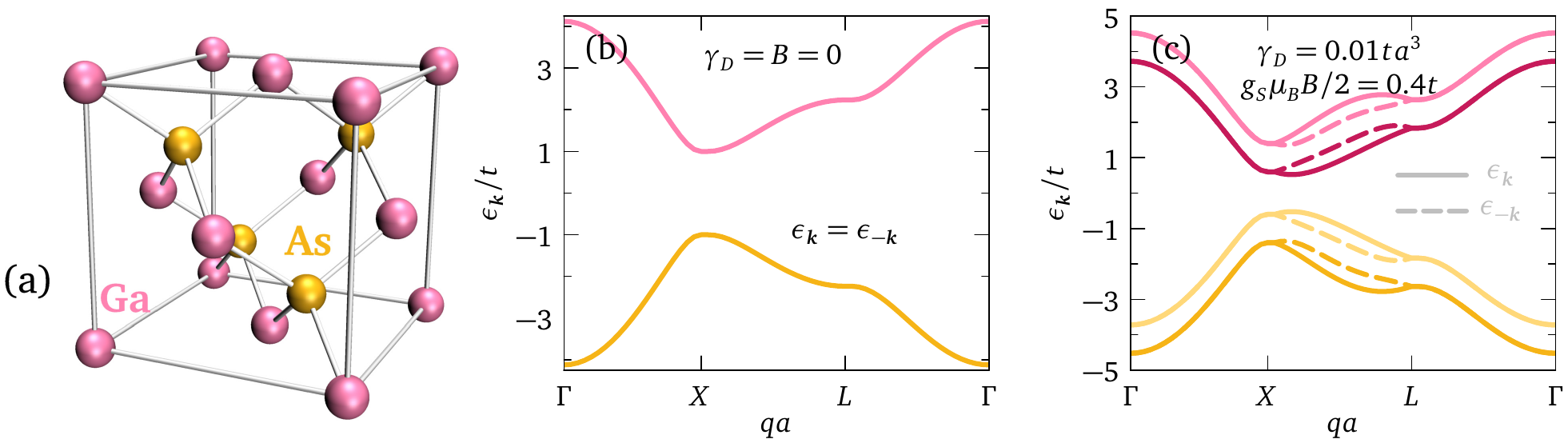}
\caption{\label{fig:GaAs} An example of noncentrosymmetric crystal where breaking of parity and time-reversal symmetry invalidates the $\vec{k}$-inversion criteria (\ref{eq:dispersion_even_parity}) and (\ref{eq:bare_vertex_odd}) for vanishing vertex corrections at $q=0$. (a) Unit cell of GaAs showing the absence of a lattice inversion symmetry. (b) Bandstructure of the 2-band toy-model Hamiltonian (\ref{eq:H_GaAs}) for GaAs in the absence of spin-orbit coupling ($\gamma_D=0$) and of Zeeman splitting ($B=0$), along the reciprocal-space path $\Gamma \rightarrow X \rightarrow L \rightarrow \Gamma$, with $\Gamma=(0,0,0)$, $X=2\pi/a (1,0,0)$, and $L=2\pi/a (1/2,1/2,1/2)$, where $a$ is the lattice parameter: in this situation, $\epsilon_{\vec{k},\sigma} =\epsilon_{-\vec{k},\sigma}$ for both the Ga and As bands. (c) Bandstructure for the GaAs Hamiltonian (\ref{eq:H_GaAs}) with Dresselhaus spin-orbit coupling and Zeeman splitting, which breaks both parity and time-reversal symmetry and shows different dispersions $\epsilon_{\vec{k},\sigma} \neq \epsilon_{-\vec{k},\sigma'}$ along the directions $X \rightarrow L$ (solid curves) and $L \rightarrow X$ (dashed curves) in reciprocal space; here the pink and magenta upper bands refer to the spin-mixed Ga bands split by Dresselhaus spin-orbit coupling, while the yellow and orange lower bands are the corresponding split As bands. 
 }
\end{figure*}

Along the same lines as in the previous Sec.\@ \ref{Momentum_cons_q0}, one can make the theory obey both momentum conservation and charge conservation at any $\vec{q}$, at the unavoidable cost of distancing the theory from the pure DMFT limit and relaxing the locality assumption in the self-energy and/or the two-particle vertex. In the following, we choose to keep a momentum-independent self-energy, but the ensuing conclusions also hold for a nonlocal $\Sigma(\vec{k},i\omega_n)$ upon substitution. 

Technically, when momentum conservation is imposed in general, the system should obey the associated Ward identity (\ref{eq:Ward_identity_momentum}) at any $\vec{q}$ and $i\Omega_n$. This identity is a linear relation between the momentum vertex $\vec{\Gamma}_{\hat{\pi}}(\vec{k},\vec{q},i\omega_n, i\Omega_n)$ and the stress vertex $\underline{\underline{\Gamma}}_{\hat{\textit{T}}}(\vec{k},\vec{q},i\omega_n, i\Omega_n)$ \cite{He-2014}. As such, it does not determine the momentum (except at $q=0$) and stress vertices individually, but only their relation, in a ``gauge freedom'' reminiscent of gauge invariance for electromagnetic vector and scalar potentials. Moreover, the Ward identity (\ref{eq:Ward_identity_momentum}) affects only the \emph{longitudinal} part of the stress vertex (parallel to $\vec{q}$), so that we can add to it a transverse part (orthogonal to $\vec{q}$). Similar considerations were expressed for the charge-conservation Ward identity in Sec.\@ \ref{Quadratic_vanish}. 

Instead of reconstructing full forms of the momentum and stress vertices, we can directly impose the Ward identity (\ref{eq:Ward_identity_momentum}), from which the relation (\ref{eq:Chi_JJ_Chi_TT}) between the momentum-momentum and stress-stress correlation functions straightforwardly stems \cite{Taylor-2010,Bradlyn-2012,Valentinis-2026_unpublished_2}. In doing so, we are assuming effective translational invariance, which holds exactly in the continuum and approximately in crystalline lattices in the ``hydrodynamic'' regime where momentum-relaxing processes and Umklapp scattering are perturbatively weak with respect to the kinetic energy scale of the interacting electrons.

In turn, Eq.\@ (\ref{eq:Chi_JJ_Chi_TT}) implies that any nonlocal correlation function of observables proportional to momentum $\hat{\vec{\pi}}(\vec{q},\tau)$ also enjoys a relation with the double divergence of the stress-stress correlation function $\chi_{\hat{\textit{T}}\hat{\textit{T}}}^{\alpha\beta\gamma\delta}(\vec{q},i\Omega_n)$. In particular, the isotropic quadratic dispersion (\ref{eq:xi_quadr}) implies proportionality between momentum and current densities as previously argued, and thus the Ward identity (\ref{eq:Ward_identity_momentum}) restores the relation between the leading nonlocal $q^2$ part of the nonlocal conductivity tensor $\sigma_{\alpha \beta}(\vec{q},\omega)$ and the $q=0$ viscosity tensor $\eta_{\alpha \beta \gamma \delta}(\omega)$, valid for Galilean-invariant systems \cite{Taylor-2010,Bradlyn-2012,He-2014}. Explicitly, this relation is
\begin{equation}\label{eq:sigma_T_quadr_Galilean}
\sigma_{\alpha 	\beta}(\vec{q},\omega)=\frac{i n e^2}{m (\omega+i0^+)}+\sigma_{\alpha \beta}^{(2)}(\vec{q},\omega)+\mathscr{O}(q^4),
\end{equation}
where 
\begin{align}\label{eq:sigma_2_visc_cond_Galilean}
\sigma_{\alpha \beta}^{(2)}(\vec{q},\omega)&=e^2 \sum_{\mu,\nu} \frac{q_\mu q_\nu}{m^2(\omega+i0^+)^2} \eta_{\mu \alpha \nu \beta}(\omega) \nonumber \\ &+\frac{i e^2}{m^2 (\omega+i0^+)^3} q_\alpha q_\beta \kappa_T^{-1}, 
\end{align}
where $k_T^{-1}$ is the inverse compressibility \cite{Bradlyn-2012}, while $\sigma_{\alpha \beta}(\vec{q},\omega)$ and $\eta_{\mu \alpha \nu \beta}(\omega)$ formally stem from Eq.\@ (\ref{eq:Kubo_general_Theta}), with $\hat{\Theta}=\hat{J}$ at $q \neq0$ and $\hat{\Theta}=\hat{\textit{T}}$ at $q=0$, respectively. The Drude term at $q=0$ in Eq.\@ (\ref{eq:sigma_T_quadr_Galilean}) directly follows from the zero-momentum part of the current vertex, in accordance with Eq.\@ (\ref{eq:current_vertex_quadr_momentum_cons}): this absence of dissipative conductivity at $\omega>0$ makes sense, since when momentum is conserved there are no dissipation channels for the current response and the $q=0$ conductivity reduces to the pure Drude model \cite{Maiti-2017}. The compressibility term at the right-hand side of Eq.\@ (\ref{eq:sigma_2_visc_cond_Galilean}) emerges from the ``contact'' terms, specifically the current-stress commutator, in Eq.\@ (\ref{eq:Ward_identity_momentum}) once the latter is expanded to order $q^2$.

Conversely, Eq.\@ (\ref{eq:Chi_JJ_Chi_TT}) can be connected to correlation functions other than current for non-quadratic dispersion $\epsilon_{\vec{k}}$. For graphene-like system with linear dispersion $\xi_{\vec{k}}\propto \left|\vec{k}\right|$ one expects a relation between the nonlocal (momentum-dependent) part of the thermal conductivity tensor $K_{T,\alpha \beta}(\vec{q},\omega)$, linked to the heat (energy current $\hat{J}_{\mathscr{E}}(\vec{q},\tau)$) correlation function $\chi_{\hat{J}_{\mathscr{E}} \hat{J}_{\mathscr{E}}}^{R,\alpha \beta}(\vec{q},\omega)$, and the $q=0$ viscosity tensor \cite{Link-2018a}. For generic dispersion $\xi_{\vec{k}}$, the homogeneous viscosity tensor will be connected with any linear response function encoding correlations of an observable $\hat{\Theta} \propto \hat{\vec{\pi}}$ proportional to momentum $\hat{\vec{\pi}}(\vec{q},\tau)$ \cite{Valentinis-2026_unpublished_2}. However, this connection cannot hold within the local DMFT-like limit, and requires nonlocality of at least one object between self-energy and irreducible two-particle vertex. 
For translationally invariant non-Fermi liquids, a recent work analyzes the finite-momentum electrodynamics of critical Fermi surfaces, investigating the spectrum of bosonic fluctuations and their impact on the hydrodynamic and tomographic regimes of nonlocal conduction \cite{Guo-2024}.  

A related question is what happens to the viscosity-conductivity relation when just charge conservation, but not momentum conservation, is imposed in our electron liquid with local interactions and quadratic isotropic dispersion. These assumptions are compatible with the DMFT-like limit, as previously shown in Sec.\@ \ref{Quadratic_vanish}. This is a subject of the companion paper \cite{Valentinis-2026_unpublished}, where it is shown that, without momentum conservation, the $q^2$ coefficient in Eq.\@ (\ref{eq:sigma_2_visc_cond_Galilean}) acquires a new multiplicative factor $\Theta_\eta(\omega)$, which depends both on temperature and on the local self-energy $\Sigma^R(\epsilon)$. 

\section{Discussion}\label{Discussion}

\subsection{Broken symmetries}\label{Discussion_broken_symmetries}

\subsubsection{Breaking of parity and time-reversal symmetry}\label{Parity_time_reversal_breaking}

The parity-related conditions for vanishing vertex correction at $q=0$, namely the dispersion (\ref{eq:dispersion_even_parity}) and the related vertex (\ref{eq:bare_vertex_odd}) respectively being even and odd under a point-group symmetry operation $\mathscr{g}$, effectively generalize Khurana's argument for vanishing vertex corrections to the conductivity at $q=0$ in the DMFT limit of the Hubbard model \cite{Khurana-1990}. Within the DMFT-like assumptions of local self-energy and irreducible two-particle vertex (\ref{eq:Lambda_DMFT}), these criteria can be applied to any (possibly anisotropic) dispersion $\epsilon_{\vec{k}}$ and vertex $\Gamma_{\hat{\Theta},\vec{\alpha}}(\vec{k},0,i\omega_m',i\Omega_n)$ for observable $\hat{\Theta}$ for which momentum $\vec{k}$ is a good quantum number. In particular, the even $\vec{k}$-inversion condition $\epsilon_{\vec{k}}=\epsilon_{-\vec{k}}$ is protected by a combination of parity $\mathscr{P}$ and time-reversal $\mathscr{T}$ symmetries: $\mathscr{P}$ specifies the symmetry of the electronic state with respect to space inversion $\vec{r} \mapsto -\vec{r}$ (or equivalently $\vec{k} \mapsto -\vec{k}$ in reciprocal space of momenta), while $\mathscr{T}$ relates the states at $\vec{k}$ and $-\vec{k}$, implying $\epsilon_{n,\vec{k}}=\epsilon_{\bar n,-\vec{k}}$; at time-reversal-invariant momenta this relation becomes the Kramers degeneracy condition. Thus, the simultaneous presence of $\mathscr{P}$ and $\mathscr{T}$ symmetries enforces double degeneracy at generic $\vec{k}$, i.e.\@, the existence of distinct but doubly degenerate fermionic states related by time-reversal symmetry \cite{Kramers-1930, Wigner-1993}. 

If both $\mathscr{P}$ and $\mathscr{T}$ are broken, the criteria on vertex-correction vanishing do not apply and need to be generalized. 
Parity-broken states are realized in non-centrosymmetric crystals, in which electronic bands do not satisfy parity with respect to $\vec{k}$-inversion due to the absence of an inversion symmetry point. As an example, Fig.\@ \ref{fig:GaAs}(a) illustrates the noncentrosymmetric crystal structure of GaAs. The lack of inversion symmetry allows for antisymmetric spin-orbit coupling, which implies a nonzero momentum- and spin-dependent effective internal magnetic field $\vec{B}_i(\vec{k})=-\vec{B}_i(-\vec{k})$, in accordance with the Dresselhaus effect \cite{Dresselhaus-1955,Bychkov-1984}. Hence, for a given spin species $\sigma=\left\{\uparrow,\downarrow\right\}$, in general we have $\epsilon_{\vec{k},\sigma}\neq \epsilon_{-\vec{k},\sigma}$. This property invalidates the $\vec{k}$-inversion arguments leading to vertex-corrections cancellations at $q=0$ as described in Sec.\@ \ref{Charge_cons_vertex_vanish}, and the Bethe-Salpeter equation becomes a matrix in spin space $\left\{\sigma,\sigma'\right\}=\left\{\uparrow,\downarrow \right\}$, in a way similar to the multiorbital case of Sec.\@ \ref{Multiband}. Hence, one expects vertex corrections to not generically vanish anymore for spin-resolved $\vec{k}$-odd vertices in the presence of antisymmetric spin-orbit coupling, leading to phenomena like the spin Hall response \cite{Spinova-2015}. 

For reference, Fig.\@ \ref{fig:GaAs}(b) shows an effective two-band tight-binding bandstructure of GaAs in the absence of spin-orbit coupling, so that $\epsilon_{\vec{k},\sigma}=\epsilon_{-\vec{k},\sigma}$; the exemplary path $\Gamma \rightarrow X \rightarrow L \rightarrow \Gamma$ in reciprocal space is chosen, with $\Gamma=(0,0,0)$, $X=2\pi/a (1,0,0)$, and $L=2\pi/a (1/2,1/2,1/2)$, where $a$ is the lattice parameter. 
However, even in the presence of spin-orbit coupling, Kramers' degeneracy is still preserved through $\epsilon_{\vec{k},\uparrow}=\epsilon_{-\vec{k},\downarrow}$. To break the latter as well, one needs time-reversal symmetry breaking, as realized for instance by the application of an external magnetic field $\vec{B}(\vec{r},t)$. 
Fig.\@ \ref{fig:GaAs}(c) shows the two-band tight-binding model for GaAs along the reciprocal-space path $\Gamma \rightarrow X \rightarrow L \rightarrow \Gamma$ (solid curves) and its reverse $\Gamma \rightarrow L \rightarrow X \rightarrow \Gamma$ (dashed curves), in the presence of both Dresselhaus spin-orbit coupling and a Zeeman spin-splitting magnetic field. The Hamiltonian for this model is specified in App.\@ \ref{app:GaAs_tight_binding}. Here we see that bands along the two inverted paths are not superimposed in the region $X \leftrightarrow L$ where the Dresselhaus spin-orbit coupling term (\ref{eq:Dresselhaus_SOC}) is nonnull, highlighting that $\epsilon_{\vec{k},\sigma}\neq\epsilon_{-\vec{k},\sigma'}$, i.e.\@, both $\mathscr{P}$ and $\mathscr{T}$ are simultaneously broken, thus invalidating any symmetry between $\epsilon_{\vec{k}}$ and $\epsilon_{-\vec{k}}$. 

In relation to the above discussion about time-reversal symmetry breaking, Sec.\@ \ref{Hall_viscosity} comments on the exemplary special case of Hall viscosity due to Landau diamagnetism in an isotropic 2D electron liquid with a local self-energy. 

\subsubsection{Time-reversal symmetry breaking and Hall viscosity}\label{Hall_viscosity}

In the absence of time-reversal symmetry, the Hall viscosity (or odd viscosity) component of the frequency-dependent viscosity tensor $\eta_{\alpha\beta \gamma \delta}(\omega)$ emerges \cite{Avron-1995,Read-2009,Bradlyn-2012,Hoyos-2014a, Hoyos-2014b,Burmistrov-2019}. Focusing on two spatial dimensions for definiteness, the Hall viscosity stems from the Kubo formula 
\begin{align}\label{eq:eta_H_def_gen}
\eta_H &= -\lim_{\omega \rightarrow 0} \frac{1}{2\omega} \mathrm{Im} \left\{ \chi^{R,(xx-yy)xy}_{\hat{\textit{T}}\hat{\textit{T}}}(0,\omega) \right\} \nonumber \\ &= -\lim_{\omega \rightarrow 0} \frac{1}{2\omega} \mathrm{Im} \left\{ \chi^{R,xxxy}_{\hat{\textit{T}}\hat{\textit{T}}}(0,\omega) - \chi^{R,yyxy}_{\hat{\textit{T}}\hat{\textit{T}}}(0,\omega) \right\}. 
\end{align}
The expression (\ref{eq:eta_H_def_gen}) corresponds to the real part of the $\omega \rightarrow 0$ limit of the following complex-valued generalized Hall viscosity: 
\begin{equation}\label{eq:AC_Hall_visc}
\tilde{\eta}_H(\omega) = \frac{i}{2 \omega} \left[ \chi^{R,(xx-yy)xy}_{\hat{\textit{T}}\hat{\textit{T}}}(0,\omega)-\chi^{R,(xx-yy)xy}_{\hat{\textit{T}}\hat{\textit{T}}}(0,0) \right].
\end{equation}
Similar to the Hall conductivity, the Hall viscosity (\ref{eq:eta_H_def_gen}) produces viscous forces that are perpendicular to the velocity gradient, rather than opposite to the flow direction. Because these forces are orthogonal to the flow, they do not produce work and therefore they do not dissipate energy.

The off-diagonal stress-stress correlation function $\chi^{R,\alpha \alpha xy}_{\hat{\textit{T}}\hat{\textit{T}}}(0,\omega)$ in Eqs.\@ (\ref{eq:eta_H_def_gen}) and (\ref{eq:AC_Hall_visc}) may be affected by or immune to vertex corrections, depending on the physical mechanism for time-reversal symmetry breaking. If the latter occurs without external magnetic fields, as in magnetically ordered materials like ferromagnets \cite{Blundell-2001} or altermagnets \cite{Song-2025,Jang-2026_preprint}, then momentum $\vec{k}$ is still a good quantum number; hence, the $\vec{k}$-parity criteria (\ref{eq:dispersion_even_parity}) and (\ref{eq:bare_vertex_odd}), laid out in Sec.\@ \ref{Charge_cons_vertex_vanish} for the absence of vertex corrections at $q=0$ with generic single-band dispersion $\xi_{\vec{k}}$, still hold. 

Specifically, in the DMFT-like limit the Bethe-Salpeter equation (\ref{eq:Bethe_Salpeter_gen}) shows that the Hall viscosity amounts to its renormalized-bubble form (\ref{eq:chi_Theta_gen_bubble}), if the system is symmetric under exchange $k_x \leftrightarrow k_y$. To see this property, recall that at $q=0$ the stress vertex satisfies $\Gamma_{\hat{\textit{T}},\alpha\beta}(\vec{k},0,i\omega_m,i\Omega_n)=\Gamma_{\hat{\textit{T}},\alpha\beta}^{(0)}(\vec{k},0)+ C^{\alpha \beta}(i\omega_m,i\Omega_n)$, where the vertex renormalization $C^{\alpha \beta}(i\omega_m,i\Omega_n)$ (when nonnull) is only a function of frequency and satisfies Eq.\@ (\ref{eq:Bethe_Salpeter_C}). Now, for the stress component $\textit{T}_{xy}$ the vertex renormalization is null upon angular integration, so vertex corrections vanish for this component. If, in addition, the vertex renormalizations for $\textit{T}_{xx}$ and $\textit{T}_{yy}$ are equal, i.e.\@, $C^{xx}(i\omega_m,i\Omega_n)=C^{yy}(i\omega_m,i\Omega_n)$, then such renormalizations cancel in the subtraction $\textit{T}_{xx}-\textit{T}_{yy}$. In turn, these conditions require that the sum over momenta $\vec{k}$ in $C^{\alpha \alpha}(i\omega_m,i\Omega_n)$ is invariant under exchange $k_x \leftrightarrow k_y$, which requires point-group symmetry containing axis exchange, such as $C_4$, $C_{4v}$, and $D_4$. In other words, in group-theoretical language vertex renormalizations for $\textit{T}_{xx}$ and $\textit{T}_{yy}$ are equal if these stress-tensor components belong to the same reducible representation $A_{1g} \oplus G$ where $G\neq A_{1g}$ is another point group. Hence, the Hall viscosity (\ref{eq:eta_H_def_gen}) is devoid of vertex corrections only under the specific requirement of system symmetry under $k_x \leftrightarrow k_y$ exchange. 

On the other hand, if time-reversal symmetry breaking occurs due to the application of an external magnetic field $\vec{B}$, as in the case of quantum Hall states generated by Landau orbital diamagnetism, then $\vec{k}$ is not a good quantum number and the criteria (\ref{eq:dispersion_even_parity}) and (\ref{eq:bare_vertex_odd}) need to be revised. 
For an out-of-plane magnetic field $\vec{B} = B \hat{u}_z$ applied to a 2D electronic system with isotropic quadratic dispersion, in Landau-level basis the Green's function is
\begin{equation}\label{eq:Landau_G}
G_{l l'}(i\omega_n) = \frac{\delta_{l l'}}{i\omega_n + \mu - \epsilon_l- \Sigma_l(i\omega_n)}=G_{l}(i\omega_n)\delta_{l l'},
\end{equation}
where the Landau levels $l \in \mathbb{N}$ are
\begin{equation}\label{eq:E_n_Landau}
\epsilon_l = \hbar \omega_c \left( l + \frac{1}{2} \right),
\end{equation}
and $\omega_c=e \left|B\right|/m$ is the electron cyclotron frequency. 
The stress tensor operator connects different Landau levels; using the Green's functions (\ref{eq:Landau_G}), the uniform part of the stress-stress correlation function per spin is then
\begin{align}\label{eq:stress_stress_corr_Landau}
\chi^{\alpha \beta \gamma \delta}_{\hat{\textit{T}}\hat{\textit{T}}}(0,i\Omega_n)& =n_{LL} k_B T \sum_{i\omega_m} \sum_{l,p} \Gamma_{\hat{\textit{T}},\alpha\beta}(l,p, i\omega_m, i\Omega_n) \nonumber \\ & \times  G_l(i\omega_m + i\Omega_n) G_p(i\omega_m) \nonumber \\ &  \times \Gamma^{(0)}_{\hat{\textit{T}},\gamma\delta}(p,l),
\end{align}
where the Landau-level degeneracy density (per unit area) is
\begin{equation}\label{eq:n_LL}
n_{LL}=\frac{e \left|B\right|}{2 \pi \hbar} =\frac{1}{2 \pi (l_B)^2},
\end{equation}
and $l_B = \sqrt{\hbar/(e\left|B\right|)}=\sqrt{\hbar/(m \omega_c)}$ is the magnetic length. \cite{Blundell-2001}. Therefore, the total Landau-level degeneracy corresponding to Eq.\@ (\ref{eq:n_LL}) is $\mathscr{N}_{\phi}=\mathscr{A} n_{LL}$, where $\mathscr{A}$ is the area of the 2D system. 
The multiplicative factor $n_{LL}$ in Eq.\@ (\ref{eq:stress_stress_corr_Landau}) stems from the trace over the guiding-center degeneracy, which arises due to the assumed magnetic translational invariance \cite{Lozovik-2012,Wang-2019}; see also Eqs.\@ (\ref{eq:Gamma_T_Landau})-(\ref{eq:T_Landau_overlap_diagonal}). 

For convenience, we define the following combinations of kinetic stress-tensor elements, involved in the Hall-viscosity calculation according to Eq.\@ (\ref{eq:AC_Hall_visc}): 
\begin{subequations}\label{eq:stress_Hall_components}
\begin{equation}\label{eq:Hall_xx_yy_stress}
\hat{D}=\hat{\textit{T}}_{xx}-\hat{\textit{T}}_{yy},
\end{equation}
\begin{equation}
\hat{B}=\hat{\textit{T}}_{xy}.
\end{equation}
\end{subequations}
With the definitions (\ref{eq:stress_Hall_components}), the stress-stress correlation function for the Hall viscosity corresponds to
\begin{align}\label{eq:Chi_TT_Hall}
\chi^{(xx-yy) xy}_{\hat{\textit{T}} \hat{\textit{T}}}&(0,i\Omega_n)=n_{LL} k_B T \sum_{i\omega_m} \sum_{l,p} G_l(i\omega_m + i\Omega_n) \nonumber \\ & \times \Gamma_{\hat{\textit{D}}}(l,p, i\omega_m, i\Omega_n) G_p(i\omega_m) \Gamma^{(0)}_{\hat{\textit{B}}}(l,p) \nonumber \\ & = n_{LL} k_B T \sum_{i\omega_m} \sum_{l,p} \left[i\omega_m+i\Omega_n+\mu-\hbar \omega_c \right. \nonumber \\ & \left. \times (l+1/2)-\Sigma_l(i\omega_m+i\Omega_n)\right]^{-1} \left\langle p\right| \hat{\textit{T}}_{xx}-\hat{\textit{T}}_{yy}\left|l\right\rangle \nonumber \\ & \times  \left[i\omega_m+\mu-\hbar \omega_c (p+1/2)-\Sigma_p(i\omega_m)\right]^{-1} \nonumber \\ & \times \left\langle l\right|\hat{\textit{T}}_{xy} \left|p\right\rangle,
\end{align} 
where at the second step we have defined $\Gamma_{\hat{\textit{D}}}(l,p, i\omega_m, i\Omega_n)=\left\langle p\right| \hat{\textit{T}}_{xx}-\hat{\textit{T}}_{yy}\left|l\right\rangle$ and $ \Gamma^{(0)}_{\hat{\textit{B}}}(l,p)=\left\langle l\right|\hat{\textit{T}}_{xy} \left|p\right\rangle_0$, with the subscript $0$ indicating that the corresponding vertex is bare. 

One can show that, for a local self-energy $\Sigma_l(i\omega_m)\equiv\Sigma(i\omega_m)$ (which we assume to not depend on the Landau-level index $l$) and a local irreducible two-particle vertex $\Lambda(i\omega_m,i\omega_m',i\Omega_n)$, vertex corrections still vanish for the stress-stress correlation function (\ref{eq:Chi_TT_Hall}) in Landau-level basis: this is due to the fact that, while the Bethe-Salpeter equation (\ref{eq:Bethe_Salpeter_gen}) generates renormalizations to the Hall-viscosity vertex that are purely diagonal in Landau-level index ($l= l'$) , the transition matrix elements involving the combined stress-tensor components (\ref{eq:stress_Hall_components}) are purely off-diagonal ($l\neq l'$), hence no renormalization to the Hall-viscosity vertex is allowed in the local DMFT-like limit. 

For the purpose of the present discussion, a sketch of the underlying physical argument is given below, while detailed derivations including dispersions other than the quadratic isotropic one will be reported elsewhere \cite{Valentinis-2026_unpublished_4}. 
To show from Eq.\@ (\ref{eq:Bethe_Salpeter_gen}) that no renormalization to the Hall-viscosity kinetic-stress vertex occurs in the DMFT-like limit, we need to assess which Landau-level transitions are allowed to contribute to such renormalization. 
To this purpose, we have first to convert the Bethe-Salpeter equation (\ref{eq:Bethe_Salpeter_gen}) to the Landau-level basis, since  momentum $\vec{k}$ is not a good quantum number when time-reversal symmetry is broken by the magnetic field $B$. This projection can be performed by first transforming the renormalized vertex and the Green's function from the space of momenta $\vec{k}$ to real space of coordinates $\vec{r}$. The basis change is simplified by the locality of the irreducible two-particle vertex (\ref{eq:Lambda_DMFT}), which translates in real space as 
\begin{equation}\label{eq:Lambda_DMFT_real}
\Lambda(\vec{r},i\omega_m,i\omega_m',i\Omega_n)=\delta (\vec{r}) \Lambda (i\omega_m,i\omega_m',i\Omega_n).
\end{equation}
One can then project all vertices (bare and renormalized) and Green's functions onto the Landau-level basis, using the Landau-level index $l \in \mathbb{N}$ and the guiding-center index $X \in \mathbb{R}$ \cite{Lozovik-2012,Wang-2019}. This basis stems from the eigenfunctions for Landau levels, which in the Landau gauge $\vec{A}=\left(0,B_x,0\right)$ are \cite{Guo-2023}
\begin{subequations}\label{eq:Landau_xy_eigenfunctions}
\begin{equation}\label{eq:Landau_eigenfunctions_total}
\psi_{l,X}(\vec{r})=e^{i k_y y} \frac{1}{l_B} \phi_l\left(\frac{x}{l_B}-X\right),
\end{equation}
\begin{equation}\label{eq:Landau_eigenfunctions}
\phi_{l}(x)=\frac{1}{\sqrt{2^l l! \sqrt{\pi}}} H_l(x) e^{-\frac{x^2}{2}},
\end{equation}
\end{subequations}
where $H_n(x)=(-1)^n e^{x^2} d^n (e^{-x^2})/d x^n$ are the ``physicist's'' Hermite polynomials, and $X=k_y l_B$. Projecting the renormalized stress vertex onto the basis (\ref{eq:Landau_xy_eigenfunctions}), we have
\begin{align}\label{eq:Gamma_T_Landau}
\Gamma_{\hat{\textit{T}},\alpha \beta}(l,l',X,X',i\omega_m,i\Omega_n)&=\left\langle l,X\right| \Gamma_{\hat{\textit{T}},\alpha \beta}(\vec{r}_1-\vec{r}_2,i\omega_m,i\Omega_n) \nonumber \\ & \times \left| l',X'\right\rangle, 
\end{align}
where the projection is equivalent to $\left\langle \cdot \right\rangle=\int d^2 r_1 \int d^2 r_2 \psi_{l,X}^\ast(\vec{r}_1) \cdot \psi_{l',X'}(\vec{r}_2)$. Then, assuming magnetic translational invariance, the renormalized stress vertex is simplified to
\begin{equation}\label{eq:Gamma_T_magnetic_translational}
\Gamma_{\textit{T},\alpha \beta}(l,l',X,X',i\omega_m,i\Omega_n)= \Gamma_{\textit{T},\alpha \beta}(l,l',i\omega_m,i\Omega_n) \delta_{X X'}. 
\end{equation}
Inserting Eq.\@ (\ref{eq:Gamma_T_magnetic_translational}) into the Landau level-projected Bethe-Salpeter equation, and exploiting the orthonormality of the eigenfunctions (\ref{eq:Landau_eigenfunctions}) that simplify the Landau-level overlaps in real space, one realizes that
\begin{align}\label{eq:T_Landau_overlap_diagonal}
\Gamma_{\hat{\textit{T}},\alpha \beta}(l,l',i\omega_m,i\Omega_n)&=\Gamma_{\hat{\textit{T}},\alpha \beta}^{(0)}(l,l')+ \delta_{ll'} k_B T \nonumber \\ & \times \sum_{i\omega_m'} \Lambda(i\omega_m,i\omega_m',i\Omega_n) \nonumber \\ & \times  \sum_{l} G_l(i\omega_m')G_l(i\omega_m'+i\Omega_n) \nonumber \\ & \times  \Gamma_{\hat{\textit{T}},\alpha \beta}(l,l,i\omega_m',i\Omega_n). 
\end{align}
Therefore, the vertex renormalization in Eq.\@ (\ref{eq:T_Landau_overlap_diagonal}) is diagonal in the indices $l$, $l'$. 

On the other hand, the matrix elements for the kinetic stress-tensor components (\ref{eq:stress_Hall_components}), calculated through the Landau-level creation and annihilation operators, only involve transitions $l \rightarrow l\pm 2$, with a difference in Landau-level index $\Delta l=\pm 2$. Specifically, 
\begin{subequations}\label{eq:Hall_viscosity_matrix_elements}
\begin{equation}\label{eq:D_lp2}
\left\langle l+2\right| \hat{D}\left| l\right\rangle =\hbar \omega_c \sqrt{(l+1)(l+2)},
\end{equation}
\begin{equation}\label{eq:D_l}
\left\langle l-2\right| \hat{D}\left| l\right\rangle =\hbar \omega_c \sqrt{l (l-1)},
\end{equation}
\begin{equation}\label{eq:B_lp2}
\left\langle l+2\right| \hat{B}\left| l\right\rangle =\frac{i\hbar \omega_c}{2} \sqrt{(l+1)(l+2)},
\end{equation}
\begin{equation}\label{eq:B_l}
\left\langle l-2\right| \hat{B}\left| l\right\rangle =-\frac{i\hbar \omega_c}{2} \sqrt{l (l-1)}.
\end{equation}
\end{subequations}
Therefore, since Eq.\@ (\ref{eq:T_Landau_overlap_diagonal}) implies vertex renormalizations that are diagonal in Landau-level index, but Eq.\@ (\ref{eq:Hall_viscosity_matrix_elements}) involves off-diagonal matrix elements, vertex corrections vanish from the Landau-level Hall viscosity generated by kinetic stresses; hence, $\tilde{\eta}_H(\omega)$ can be calculated with just the renormalized bubble (\ref{eq:chi_Theta_gen_bubble}). This conclusion makes sense in light of the ``d-wave'', or ``$L=2$'' character of the Hall viscosity, which cannot induce a density-like, ``s-wave'', or ``$L=0$'' deformation if the self-energy and the irreducible two-particle vertex are both local. 

Notice that the above discussion does not exclude vertex corrections from the components $\hat{\textit{T}}_{xx}$ and $\hat{\textit{T}}_{yy}$ individually. These stress-tensor components are subject to vertex renormalizations, in general. However, from the discussion following Eq.\@ (\ref{eq:AC_Hall_visc}), if the system is isotropic, the individual renormalizations of these vertices are equal, $C^{xx}(i\omega_m,i\Omega_n) =C^{yy}(i\omega_m,i\Omega_n)$, and the combination (\ref{eq:Hall_xx_yy_stress}) is devoid of vertex corrections. 

In the noninteracting limit (i.e.\@, $\Sigma(i\omega_m)=0 \, \forall i\omega_m$) and in the static, zero-temperature limit ($\omega=0$, $T=0$), Eqs.\@ (\ref{eq:eta_H_def_gen}) and (\ref{eq:Chi_TT_Hall}) reproduce the ``topological'' value of the DC Hall viscosity for integer quantum Hall states, which is \cite{Avron-1995,Burmistrov-2019}
\begin{equation}\label{eq:eta_H_quant}
\eta_H=n_{LL}\frac{\hbar}{4} \sum_{n=0}^{N_L-1} (2n+1)=\frac{\hbar}{4} \overline{n}_S N_L=\frac{\hbar}{2} \overline{n}_S\overline{s},
\end{equation}
where $N_L$ is the number of filled Landau levels, while the particle density of filled Landau levels is
\begin{equation}\label{eq:n_S}
\overline{n}_S=n_{LL} N_L=\frac{N_L}{2 \pi (l_B)^2},
\end{equation}
and $\overline{s}=N_L/2$ is the average orbital spin per particle \cite{Read-2011,Read-2011_erratum,Burmistrov-2019}. 
Physically, the topological value (\ref{eq:eta_H_quant}) stems from transitions $n \rightarrow n\pm 2$ between sharp Landau levels (Dirac delta functions) at energies $\epsilon_l$ with exchanged energy $\pm 2 \hbar \omega_c$. The same result can be reached from a genuinely topological perspective, by twisting the boundary conditions in a two-dimensional quantum Hall fluid placed on the surface of a torus-shaped configuration-space domain \cite{Avron-1995}: in fact, in general the viscosity of quantum fluids with an energy gap at zero temperature is related to the adiabatic curvature on the space parametrizing flat background metrics \cite{Avron-1995,Read-2009,Bradlyn-2012}. 

At finite temperature and frequency, the self-energy $\Sigma(i\omega_m)$ and the Fermi-Dirac distribution function $f_{FD}(\epsilon)$ modify the Hall viscosity even in the local DMFT-like regime, by broadening the single-particle spectral functions and by modifying thermal occupations, respectively. In this case, the associated retarded stress-stress correlation function becomes
\begin{align}\label{eq:Chi_TT_AG_R}
\chi^{R,(xx-yy)xy}_{\hat{\textit{T}} \hat{\textit{T}}}(0,\omega)& = \frac{i n_{LL} (\hbar \omega_c)^2 }{2}   \sum_{l=0}^{+\infty}  (l+1)(l+2) \nonumber \\ & \times   \int_{-\infty}^{+\infty} d\epsilon_1 \int_{-\infty}^{+\infty} d\epsilon_2 A_G(l,\epsilon_1)\nonumber \\ &  \times A_G(l+2,\epsilon_2) \frac{f_{FD}(\epsilon_1)-f_{FD}(\epsilon_2)}{\hbar\omega+i0^+ +\epsilon_1-\epsilon_2}   \nonumber \\ &    \frac{i n_{LL} (\hbar \omega_c)^2 }{2} \sum_{l=0}^{+\infty} l(l-1) \int_{-\infty}^{+\infty} d\epsilon_1   \nonumber \\ &   \times \int_{-\infty}^{+\infty} d\epsilon_2 A_G(l,\epsilon_1) A_G(l-2,\epsilon_2)  \nonumber \\ & \times \frac{f_{FD}(\epsilon_1)-f_{FD}(\epsilon_2)}{\hbar\omega + i0^+ +\epsilon_1-\epsilon_2} .
\end{align}
Eq.\@ (\ref{eq:Chi_TT_AG_R}) tells us that, due to the retarded self-energy $\Sigma^R(\epsilon)$, the Landau-level spectral functions $A_G(l,\epsilon)=-\pi^{-1}\mathrm{Im}\left\{G_l(\epsilon)\right\}$ become Lorentzian-like functions of energy $\epsilon$, broadened with respect to their noninteracting forms $\delta\left[\epsilon-(\epsilon_l-\mu)\right]$. Such broadening implies that, even at $T=0$ and $\omega=0$, the ground-state Hall viscosity differs from its noninteracting, ``topological'' expression (\ref{eq:eta_H_quant}). Nevertheless, in the weakly interacting limit $\left|\Sigma^R(0)\right| \ll \hbar \omega_c$, where the full-width at half maximum (FWHM) of the spectral functions is much smaller than the energy distance $\pm 2 \hbar \omega_c$ between allowed noninteracting-states transitions, the Hall viscosity (\ref{eq:eta_H_def_gen}) should negligibly differ from its noninteracting value (\ref{eq:eta_H_quant}); this robustness is confirmed by similar explicit computations of the Hall viscosity with static impurities treated in self-consistent Born approximation (scBA) \cite{Burmistrov-2019}, where vertex corrections (contrarily to the DMFT limit) have to be considered due to the nonlocal irreducible two-particle vertex for impurity scattering. Continuing this analogy, Eq.\@ (\ref{eq:Chi_TT_AG_R}) qualitatively predicts that the Hall viscosity is strongly modified by interactions in the opposite limit $\left|\Sigma^R(0)\right| \gtrapprox \hbar \omega_c$, where self-energy broadening significantly increases the overlap between Landau-level spectral functions; this qualitative conclusion is shared by the scBA impurity-scattering analysis \cite{Burmistrov-2019}. It would be interesting to estimate the robustness of the ``topological'' value (\ref{eq:eta_H_quant}) at $\omega \neq 0$ with respect to other kinds of frequency-dependent self-energy models, for noninteracting dispersions at $B=0$ other than the quadratic isotropic case of Eq.\@ (\ref{eq:E_n_Landau}), and for fractional quantum Hall states for partially filled Landau levels. 

\subsubsection{Pairing channel and Nambu structure}\label{Discussion_pairing}

Other potentially insightful applications concern spontaneously broken-symmetry states of metals, e.g.\@, electronic charge order (charge/spin density waves) and superconductors. For the latter, the generalization to $2 \times 2$ Nambu space of the propagators (\ref{eq:Green_Sigma_imag}) coincides with Eliashberg theory, where anomalous averages of creation and annihilation operators and the associated anomalous Green's function $F(\vec{k},i\omega_n)$ must be considered. Proceeding along the same DMFT-like perspective as in the normal state, we could assume local irreducible two-particle vertex in the particle-hole and pairing channels, as well as normal and anomalous self-energies, given by $\Sigma^R(\omega)$ and $\Phi^R(\omega)$ respectively; this procedure could allow one to investigate Ward identities for both sectors simultaneously, and derive generalized criteria for presence/absence of vertex corrections in the condensed state, along the lines of Ref.\@ \onlinecite{Raines-2024} which insightfully considered the density and current susceptibilities of an Eliashberg superconductor.

\subsection{Multiband effects}\label{Multiband}

Another practically relevant case that goes beyond the one-band criteria (\ref{eq:dispersion_even_parity}) and (\ref{eq:bare_vertex_odd}) arises in coupled electronic bands, arising for instance in multiband systems or by quantum confinement of bulk metals in clean ultrathin films \cite{Valentinis-2016a,Valentinis-2016b,Valentinis-2020,Zaccone-2025,Wijesinghe-2026} or at conductive interfaces \cite{Gariglio-2015,Choi-2015,Valentinis-2017,Li-2018,Campi-2025,Campi-2026}. There, the definitions in Sec.\@ \ref{Correl_conservations_gen} (e.g.\@, the momentum-conservation Ward identity, the viscosity tensor) have to be generalized to take into account the band/orbital index $\left\{ o_1,o_2\right\}=\left\{1, \ldots N_{o}\right\}$, where $N_{o}$ is the dimension of the Hilbert space in the band/orbital representation. The Green's function becomes a matrix $\underline{\underline{G}}(\vec{k},i\omega_n)=\left\{G_{o_1 o_2}(\vec{k},i\omega_n) \right\}$,
\begin{equation}\label{eq:G_matrix}
\underline{\underline{G}}(\vec{k},i\omega_n)=\left[(i\omega_n+\mu) \underline{\underline{\mathbb{I}}}-\underline{\underline{\hat{H}}}(\vec{k})-\underline{\underline{\Sigma}}(i\omega_n)\right]^{-1},
\end{equation}
where we keep the self-energy matrix $\underline{\underline{\Sigma}}(i\omega_n)=\left\{\Sigma_{o_1 o_2}(i\omega_n)\right\}$ as a local quantity, while the Hamiltonian matrix $\underline{\underline{\hat{H}}}(\vec{k})=\left\{\hat{H}_{o_1 o_2}(\vec{k}) \right\}$ contains both diagonal and off-diagonal (band-coupling) terms in general. 
The bare vertices for correlation functions of an observable $\hat{\Theta}$, for instance the current vertex $\Gamma_{\hat{J},\alpha}^{o_1 o_2}(\vec{k},\vec{q})$ or the stress vertex $\Gamma_{\hat{\textit{T}},\alpha \beta}^{o_1 o_2}(\vec{k},\vec{q})$, now become a matrix in orbital/band space as well. 
Naturally, in this multiband setting quantum geometry effects will appear \cite{Yu-2025}, in addition to interaction effects as governed by the self-energy. 

The Bethe-Salpeter equation for renormalized vertices and local irreducible two-particle vertex becomes
\begin{align}\label{eq:Bethe_Salpeter_gen_multiband}
&\Gamma_{\hat{\Theta},\vec{\alpha}}^{o_1 o_2}(\vec{k},\vec{q},i\omega_m,i\Omega_n)=\Gamma_{\hat{\Theta},\vec{\alpha}}^{(0),o_1 o_2}(\vec{k},\vec{q}) \nonumber \\ &+ \sum_{o_3 o_4 o_5 o_6} k_B T\sum_{i\omega_m'} \Lambda_{o_1 o_2 o_3 o_4}(i\omega_m,i \Omega_n, i\omega_m') \nonumber \\ &\times \frac{1}{\mathscr{V}}\sum_{\vec{k}'} G_{o_3 o_5}(\vec{k}',i\omega_m')  G_{o_6 o_4}(\vec{k}'+\vec{q},i\omega_m'+i\Omega_n) \nonumber \\ & \times  \Gamma_{\hat{\Theta},\vec{\alpha}}^{o_5 o_6}(\vec{k}',\vec{q},i\omega_m',i\Omega_n).
\end{align}
From Eq.\@ (\ref{eq:Bethe_Salpeter_gen_multiband}), we see that in a multiband system the simple $\vec{k}$-symmetry rule (even dispersion $\epsilon_{\vec{k}}=\epsilon_{\mathscr{g}\vec{k}}$, odd bare vertex $\Gamma_{\hat{\Theta},\vec{\alpha}}^{(0)}(\mathscr{g}\vec{k},0)=-\Gamma_{\hat{\Theta},\vec{\alpha}}^{(0)}(\vec{k},0)$) for vanishing vertex corrections at $q=0$ does not hold anymore: it is the whole sum over the orbital indices $\left\{o_3 o_4 o_5 o_6\right\}$ that has to vanish under $\vec{k} \mapsto \mathscr{g}\vec{k}$. 

The two-point correlation function is then the matrix 
\begin{align}\label{eq:Chi_Theta_decomp_multiband}
\chi_{\hat{\Theta} \hat{\Theta}}^{\vec{\alpha}\vec{\beta}}(\vec{q},i\Omega_n)&=\sum_{o_3,o_4}k_B T \sum_{i\omega_m} \frac{1}{\mathscr{V}} \sum_{\vec{k}} \Gamma_{\hat{\Theta},\vec{\beta}}^{(0),o_1 o_2}(\vec{k}+\vec{q},-\vec{q})  \nonumber \\ & \times G_{o_2 o_3}(\vec{k},i\omega_m) \Gamma_{\hat{\Theta},\vec{\alpha}}^{o_3 o_4}(\vec{k},\vec{q},i\omega_m,i\Omega_n)  \nonumber \\ & \times G_{o_4 o_1}(\vec{k}+\vec{q},i\omega_m+i\Omega_n) \nonumber \\ &= k_B T \sum_{i\omega_m} \frac{1}{\mathscr{V}} \sum_{\vec{k}} \mathrm{Tr}\left\{ \underline{\underline{\Gamma}}_{\hat{\Theta},\vec{\beta}}^{(0)}(\vec{k}+\vec{q},-\vec{q}) \right. \ \nonumber \\ & \underline{\underline{G}}(\vec{k},i\omega_m) \underline{\underline{\Gamma}}_{\hat{\Theta},\vec{\alpha}}(\vec{k},i\omega_m,\vec{q},i\Omega_n)  \nonumber \\ & \left. \underline{\underline{G}}(\vec{k}+\vec{q},i\omega_m+i\Omega_n)\right\},
\end{align}
which depends on the trace $\mathrm{Tr}\left\{ \right\}$ \cite{Herasymchuk-2024,Herasymchuk-2025}. 

In the noninteracting case ($\Sigma(i\omega_n)=0 \, \forall i\omega_n$), the correlation function formula (\ref{eq:Chi_Theta_decomp_multiband}) has been employed, among others, to calculate the components of the viscosity tensor for 2-band Weyl semimetals \cite{Herasymchuk-2024} and altermagnets \cite{Herasymchuk-2025}. An open question in this context is how much these multiband results, including effects related to wavefunction overlaps and quantum geometry, are affected by both self-energy and vertex-corrections effects in the interacting case. 
Concerning vertex corrections, the multiband/multiorbital generalizations of the single-band criteria (\ref{eq:dispersion_even_parity}) and (\ref{eq:bare_vertex_odd}) on the dispersion and the bare vertex are
\begin{equation}\label{eq:G_matrix_even}
\underline{\underline{G}}(\mathscr{g}\vec{k},i\omega_m)=\underline{\underline{\mathscr{P}_{\mathscr{g}}}}(\vec{k}) \underline{\underline{G}}(\vec{k},i\omega_m) \underline{\underline{\mathscr{P}_{\mathscr{g}}}}^\dagger(\vec{k}),
\end{equation}
\begin{equation}\label{eq:vertex_matrix_even}
\underline{\underline{\Gamma}}_{\hat{\Theta},\vec{\alpha}}^{(0)}(\mathscr{g}\vec{k},0)=-\underline{\underline{\mathscr{P}_{\mathscr{g}}}}(\vec{k}) \underline{\underline{\Gamma}}_{\hat{\Theta},\vec{\alpha}}^{(0)}(\vec{k},0)\underline{\underline{\mathscr{P}_{\mathscr{g}}}}^\dagger(\vec{k}),
\end{equation} 
where $\underline{\underline{\mathscr{P}_{\mathscr{g}}}}(\vec{k})$ is a unitary transformation acting on $\vec{k}$ in the multiband/multiorbital space. 
In other words, vertex corrections to Eq.\@ (\ref{eq:Bethe_Salpeter_gen_multiband}) vanish when the vertex is odd under the combined momentum-space transformation $\vec{k}\mapsto \mathscr{g}\vec{k}$ and internal-space unitary transformation $\underline{\underline{\mathscr{P}_{\vec{k}}}}$, in the same representation as the one in which the Green's function is even. 
Physical examples where the conditions (\ref{eq:G_matrix_even}) and (\ref{eq:vertex_matrix_even}) are realized are graphene-like Dirac systems with Hamiltonian $\underline{\underline{\hat{H}}}_D=v_D \vec{k}\cdot \vec{\sigma}$, where inversion acts as $\underline{\underline{\mathscr{P}}}\equiv \underline{\underline{\sigma}}_{z}$ and the bare current vertex $\underline{\underline{\Gamma}}_{\hat{J},\vec{\alpha}}^{(0)}(\vec{k},0) \propto \left\{\underline{\underline{\sigma}}_{x}, \underline{\underline{\sigma}}_{y}\right\}$. Other instances where vertex corrections vanish include vertices that mix different orbitals which belong to orthogonal symmetry representations (e.g.\@, one s orbital and one p orbital), and the quantum Hall/Landau level basis of Sec.\@ \ref{Hall_viscosity}.  

\subsection{Generality and applications of vanishing vertex corrections at finite $\vec{q}$ for isotropic quadratic dispersion}\label{Discussion_quadratic}

For the isotropic quadratic dispersion (\ref{eq:xi_quadr}), we have shown that current vertex corrections vanish from the current-current correlation function in the DMFT-like limit not only at $q=0$, but at any momentum $\vec{q}$ and frequency $\omega$ in a charge-conserving (but not momentum-conserving) theory. This result greatly simplifies the calculation to the renormalized-bubble term (\ref{eq:chi_Theta_gen_bubble}), for any local self-energy. The companion paper \cite{Valentinis-2026_unpublished} departs from this result to derive explicit ``Allen-like'' expressions for the shear viscosity and the small-$\vec{q}$, finite-$\omega$ optical conductivity of correlated electrons. 

The quadratic-dispersion case may be appropriate to cold-atom traps and degenerate quantum fluids, but may seem of limited applicability to solid-state lattice systems with anisotropic crystal structures. However, notice that for simple Bravais lattices (e.g.\@, simple cubic, square, and triangular lattices) the long-wavelength (low-energy) limit of the electronic dispersion is well captured by the quadratic form (\ref{eq:xi_quadr}). The same quadratic approximation holds for band edges of lattice structures: most lattice systems, including square lattices, display a quadratic, roughly isotropic dispersion near their band edges before warping occurs at higher energies. Hence, one expects that, whenever the quadratic-dispersion approximation is applicable, vertex corrections in the local DMFT-like limit are qualitatively negligible and a renormalized-bubble form (\ref{eq:chi_Theta_gen_bubble}) of the two-particle correlation function might correctly describe the low-energy physics even at finite $\vec{q}$. 
These considerations are appealing in view of their potential applications to the finite-momentum, finite-frequency electrodynamics of both Fermi liquids and non-Fermi liquids which are not translationally invariant, i.e.\@, where Galilean invariance does not hold. The latter case is especially fascinating due to the finite-momentum charge response of strange metals, which is observed to be essentially momentum-independent in disagreement with standard predictions based on Boltzmann quasiparticle transport \cite{Guo-2024_preprint}. I leave this exciting application to future work. 

\subsection{Experimental implications: tuning of shear-stress vertex corrections by mechanical strain}\label{Discussion_experiments}

The $\vec{k}$-parity conditions (\ref{eq:dispersion_even_parity}) and (\ref{eq:bare_vertex_odd}) in the local DMFT-like limit show that shear stresses are not affected by vertex corrections at $q=0$ for shear components $\left\{\alpha, \beta\right\}$ involving a coordinate $\alpha$ with respect to which the system possesses an orthogonal mirror symmetry plane. For instance, the SrTiO$_3$ cubic unit cell of Fig.\@ \ref{fig:mirrors_symmetry}(a) possesses three of such planes, which means that vertex corrections vanish from all shear components of the kinetic stress tensor. However, suppose a mechanical deformation (strain) is applied to the system, so that the unit-cell symmetry is reduced to a fewer number of symmetry planes, with the triclinic configuration being the most extreme, asymmetric one. Then, assuming that reversible strain adiabatically and perturbatively modifies the dispersion $\epsilon_{\vec{k}}$, by the arguments in Sec.\@ \ref{Shear_stresses_vertex_corrections} the system acquires shear-stress vertex corrections for components $\left\{\alpha, \beta\right\}$ including the coordinate $\alpha$ orthogonal to a strain-induced broken symmetry plane. This way, a tantalizing perspective would arise, namely to controllably tune shear stresses and the ensuing generalized shear viscosity, by acting on the underlying symmetry of the crystalline lattice by mechanical deformations. 
However, the measurement of such shear stresses requires the identification of the finite-$\vec{q}$ correlation function, for the specific dispersion $\epsilon_{\vec{k}}$, which couples to momentum $\vec{k}$ and therefore allows the connection with the stress tensor through Eq.\@ (\ref{eq:Chi_JJ_Chi_TT}) or generalizations thereof \cite{Bradlyn-2012,Link-2018a,Valentinis-2026_unpublished_2}. In general, such correlation function will be affected by vertex corrections at $q\neq 0$; nevertheless, the shear components of the inferred $q=0$ stress tensor should be not affected by vertex corrections in the DMFT-like limit.

\section{Conclusions}\label{Conclusions}

This work presented parity-based criteria for the presence or absence of vertex corrections to two-point correlation functions for observables $\hat{\Theta}$ at finite transferred momentum $\vec{q}$ and frequency $\omega$ for correlated-electron fluids, endowed with a one-band single-particle dispersion $\epsilon_{\vec{k}}$, and assuming local self-energy $\Sigma(i\omega_n)$ and local irreducible two-particle vertex $\Lambda(\omega_m,\omega_m',\Omega_n)$ (a configuration here dubbed ``DMFT-like limit''). The results are derived from the ladder Bethe-Salpeter equation (\ref{eq:Bethe_Salpeter_gen}) for the renormalized vertices, and are consistent with the ensuing Ward identity (\ref{eq:Ward_identity_charge}) for charge conservation (but in the absence of momentum conservation). 

\begin{figure}[ht] \centering
\includegraphics[width=1.0\linewidth]{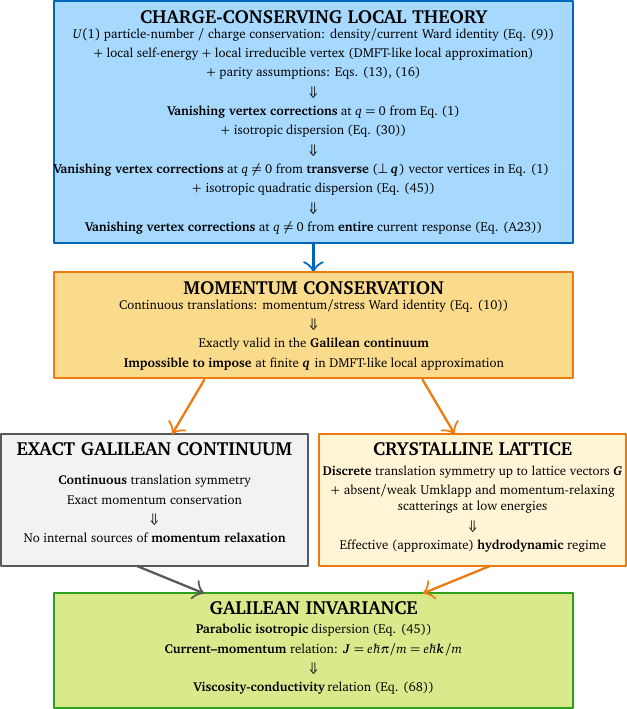}
\caption{\label{fig:scheme_conservation_laws} Schematic flow-chart summary of conservation laws, symmetry-related vertex-correction cancellations, and approximation schemes leading from charge-conserving conditions to momentum conservation (either exact or effective/approximate). 
 }
\end{figure}

For generic dispersion and at $q=0$ with arbitrary frequency $i\Omega_n$, the criteria for vanishing vertex corrections are: even parity of the dispersion with respect to a point-group symmetry operation $\mathscr{g}$ acting on momenta $\vec{k}\mapsto \mathscr{g}\vec{k}$, according to Eq.\@ (\ref{eq:dispersion_even_parity}); odd parity of the vertex associated to the observable $\hat{\Theta}$ with respect to the same symmetry operation $\mathscr{g}$, as in Eq.\@ (\ref{eq:bare_vertex_odd}). For vertices that transform like polar vectors, $\mathscr{g}=\mathscr{I}$ is identified with momentum inversion, $\vec{k} \mapsto -\vec{k}$. According to these criteria, various electrodynamic quantities are universally devoid of vertex corrections at $q=0$ in the local DMFT-like limit: these include particle current, momentum current, and energy current, having corresponding vertices proportional to momentum $\vec{k}$.

Other vertices which transform like rank-2 tensors, with spatial indices $\left\{\alpha,\beta\right\}$, may enjoy vertex-correction cancellations under additional constraints on the symmetry transformation $\mathscr{g}$. These operations include: a mirror symmetry plane $\mathscr{M}_\alpha$ orthogonal to $\alpha$, a two-fold rotation $\mathscr{C}_\alpha$ about the axis $\alpha$, or an improper rotation $\mathscr{S}_n$ that involves a rotation plus a mirror plane perpendicular to $\alpha$. All these options suppress vertex corrections for the components $\alpha \beta$ of kinetic shear-stress vertices $\Gamma_{\hat{\textit{T}},\alpha\beta}(\vec{k},0,i\omega_m, i\Omega_n)$, involving the spatial direction $\alpha$. Furthermore, higher-order rotations $\mathscr{C}_3$, $\mathscr{C}_4$, $\mathscr{C}_6 \ldots$ suppress all off-diagonal components of shear-stress vertex corrections. 

However, other notable quantities are unavoidably affected by ladder vertex corrections: among those are the charge density and the bulk components of the kinetic stress tensor. Thus, even in the DMFT-like limit, neglecting vertex corrections at $q=0$ in two-particle correlation functions is not always rigorously justified; this work provides specific symmetry criteria for those corrections to vanish. 

For density-density interactions, where the two-particle vertex $\Lambda(i\Omega_n)$ depends just on the exchanged frequency $\Omega_n$, the ladder resummation in the Bethe-Salpeter equation (\ref{eq:Bethe_Salpeter_gen}) can be explicitly performed, and leads to the RPA-like structures (\ref{eq:Gamma_Theta_renorm}) of renormalized vertices. 

For the observables $\hat{\Theta}$ enjoying aforementioned cancellation of vertex corrections, the renormalized-bubble form (\ref{eq:chi_Theta_gen_bubble}) of the correlation function is exact at $q=0$. This form specializes to ``Allen-like'' formulae (\ref{eq:Kubo_Theta_Allen_q0}) when the transport function (\ref{eq:gen_transport_function_q0}), encoding all the effects of the dispersion on $\hat{\Theta}$-correlations, is much broader in energy than single-particle spectral functions, i.e.\@, in the coherent/Fermi-liquid scenario. Conversely, the self energy-dependent analytical form (\ref{eq:Kubo_bubble_expl_q0_NFL}) applies to the opposite limit of momentum-independent spectral functions, which models incoherent/non-Fermi liquid electronic states.

For isotropic dispersions (\ref{eq:epsilonk_isotropic}) at arbitrary finite $\vec{q}$, $i\Omega_n$, vertex corrections to the transverse components of vector-like vertices are shown to identically vanish in a charge-conserving but not momentum-conserving configuration. Hence, the nonlocal electrodynamics (transverse response to a vector potential) of isotropically dispersing systems without Galilean invariance can be investigated with the renormalized-bubble form (\ref{eq:Kubo_Theta_Allen_q0}) of two-particle correlation functions. Likewise, for quadratic but anisotropic dispersions, vertex corrections vanish at $q\neq 0$ but $i\Omega_n=0$ (static limit) for vectorial vertices proportional to momentum $\vec{k}$, such as the current vertex. 

Furthermore, if the dispersion is isotropic and quadratic as in Eq.\@ (\ref{eq:xi_quadr}), then it is possible to use the bare (unrenormalized) current vertex while still preserving the charge-conservation Ward identity (\ref{eq:Ward_identity_charge}), provided that the density vertex is renormalized in accordance with Eqs.\@ (\ref{eq:density_current_charge_conserving}). 

Enforcing momentum conservation at each $\vec{q}$ and $i\Omega_n$ is impossible with a purely local two-particle vertex $\Lambda(\omega_m,\omega_m',\Omega_n)$, since the latter lacks the required momentum structure to allow for the needed momentum redistribution among one-particle states. However, implicitly relaxing the hypothesis of a local two-particle vertex, one can directly employ the finite-$\vec{q}$ momentum-conservation Ward identity (\ref{eq:Ward_identity_momentum}) to impose momentum conservation at any $\vec{q}$, either in the continuum or in crystalline lattices when Umklapp and other momentum-conserving scattering channels are frozen or negligible. Thus, from Eq.\@ (\ref{eq:Ward_identity_momentum}) one derives a relation between the renormalized momentum and stress vertices, which in turn leads to the relation (\ref{eq:Chi_JJ_Chi_TT}) between the momentum-momentum and stress-stress correlation functions. For the isotropic quadratic dispersion (\ref{eq:xi_quadr}), the latter relation leads to the known Galilean-invariant connection between the conductivity and viscosity tensors at order $q^2$ \cite{Bradlyn-2012}, as quoted in Eqs.\@ (\ref{eq:sigma_T_quadr_Galilean}) and (\ref{eq:sigma_2_visc_cond_Galilean}). Fig.\@ \ref{fig:scheme_conservation_laws} provides a visual summary of the conceptual steps leading from charge-conserving theories and vertex-corrections vanishing, to exact or approximate momentum conservation, to the viscosity-conductivity relation.   

The $\vec{k}$-parity criteria outlined in this work need to be generalized if the single-band, non-superconducting, or time-reversal symmetry assumptions are relaxed, in accordance with Sec.\@ \ref{Discussion_broken_symmetries}. The case of Landau diamagnetism is considered as an exemplary application, and in Sec.\@ \ref{Hall_viscosity} it is shown that vertex corrections vanish from the Hall viscosity in the local DMFT-like limit. However, the $T=0$ ``topological'' value of the Hall viscosity for noninteracting Landau levels is shown to be altered by a local self-energy and by temperature through spectral-functions broadening and Fermi-Dirac occupation functions, respectively. 

The present results elucidate general and exact symmetry-based criteria for vanishing vertex corrections in local quantum electronic fluids, encompassing Fermi liquids and non-Fermi liquids, and provide theoretical guidance for nonlocal electrodynamics experiments in strongly interacting systems of relevance for condensed matter, quantum simulators, and high-energy physics. This work offers a suitable platform for the analysis of local dynamical correlations, and allows for future generalizations to vertex corrections generated by topological/multiband systems, broken-symmetry states like Eliashberg superconductors, and time-reversal asymmetric configurations. 

\section{Acknowledgements}

I am grateful to R.\@ Aversa, N.\@ Ben-Shachar, C.\@ Berthod, E.\@ Di Salvo, I.\@ Gornyi, A.\@ A.\@ Haghighirad, T.\@ Hazra, T.\@ Holder, J.\@ Hoffmann, G.\@ A.\@ Inkof,  I.\@ Jang, M.\@ Le Tacon, A.\@ Mirlin, P.\@ Moll, D.\@ Schultz, J.\@ Schmalian, D.\@ van der Marel, and J.\@ Zaanen for illuminating discussions. This work was partially supported by the Swiss National Science Foundation (SNSF) through project 200021-162628 and through the SNSF Early Postdoc.Mobility Grant P2GEP2\_181450. This work was also supported by the European Commission’s Horizon 2020 RISE program Hydrotronics (Grant No. 873028), and by the German Research Foundation (DFG) through CRC TRR 288 ``Elasto-Q-Mat'' project A07. Finally, the support of the Wilhelm and Else Heraeus Foundation through inspirational workshops, where part of this work has been conceived and realized, is also acknowledged. 

\appendix

\section{Many-body theory with local self-energy: basic definitions and properties}\label{Many_body_def}

\subsection{One-particle Green's function and self-energy}

Let us consider dispersive electrons endowed with a local (momentum-independent) self-energy $\Sigma(i\omega_n)$, where $\omega_n=(2n+1)\pi k_B T$ are fermionic Matsubara frequencies. The interacting propagator on the imaginary axis is then
\begin{equation}\label{eq:Green_Sigma_imag}
G(\vec{k},i\omega_n)=\frac{1}{i\omega_n-\xi_{\vec{k}}-\Sigma(i\omega_n)},
\end{equation}
where $\xi_{\vec{k}}=\epsilon_{\vec{k}}-\mu$ is the difference between the single-particle energy eigenvalues $\epsilon_{\vec{k}}$ and the chemical potential $\mu$ \cite{Mahan-2000,Bruus-2004mb,Berthod-2018}. The imaginary-axis Green's function (\ref{eq:Green_Sigma_imag}) corresponds to the Fourier transform of the object
\begin{equation}\label{eq:G_k_tau}
G(\vec{k},\tau)=-\left\langle \hat{\mathscr{T}}_{\tau} \left\{\hat{c}_{\vec{k},\sigma}(\tau) \hat{c}^\dagger_{\vec{k},\sigma}(0)\right\} \right\rangle,
\end{equation}
written in terms of imaginary time $\tau \in \left[0,\beta\right]$ with $\beta=1/(k_B T)$, where $\left\langle \cdots \right\rangle=\mathrm{Tr}\left\{e^{-\beta \hat{H}} \cdots\right\}/\mathscr{Z}$ is the thermal average, $\mathscr{Z}=\mathrm{Tr}\left\{e^{-\beta \hat{H}}\right\}$ is the partition function for the Hamiltonian $\hat{H}$, and $\hat{\mathscr{T}}_{\tau}$ is the time-ordering operator. Therefore, $G(\vec{k},i\omega_n)=\int_0^\beta d \tau e^{i \omega_n \tau} G(\vec{k},\tau)$.

Through the analytic continuation $i\omega_n \rightarrow \hbar\omega+i0^+$ of Eq.\@ (\ref{eq:Green_Sigma_imag}), we formally find the real-axis retarded (R) propagator
\begin{equation}\label{eq:Green_Sigma}
G^R(\vec{k},\omega)=\frac{1}{\hbar\omega-\xi_{\vec{k}}-\Sigma^R(\omega)},
\end{equation}
with the retarded self-energy $\Sigma^R(\omega)$. The propagator (\ref{eq:Green_Sigma}) stems from the Dyson equation 
\begin{equation}\label{eq:G_Sigma_Dyson_math}
\left[G^R(\vec{k},\omega)\right]^{-1}=\left[G^R_0(\vec{k},\omega)\right]^{-1}-\Sigma^R(\omega),
\end{equation}
where $G^R_0(\vec{k},\omega)$ is the noninteracting Green's function (with $\Sigma^R(\omega)=0$).

For the local self-energy $\Sigma^R(\omega)$, the electronic spectral function stems from the Lehmann representation \cite{Mahan-2000,Bruus-2004mb,Berthod-2018}
\begin{equation}\label{eq:spectr_rep}
G(\vec{k},i \omega_m)=\int_{-\infty}^{+\infty} d \epsilon \frac{A_G(\vec{k},\epsilon)}{i \omega_m -\epsilon},
\end{equation}
corresponding to Eq.\@ (\ref{eq:Green_Sigma_imag}), which leads to the Lorentzian-like form \cite{Berthod-2013,Berthod-2018}
\begin{align}\label{eq:A_G_gen}
A_G(\vec{k},\epsilon)&=-\frac{1}{\pi} \mathrm{Im}\left\{G^R(\vec{k},\epsilon)\right\} \nonumber \\&=-\frac{1}{\pi} \frac{\Sigma_2(\epsilon)}{\left[\epsilon-\xi_{\vec{k}}-\Sigma_1(\epsilon)\right]^2+\left[\Sigma_2(\epsilon)\right]^2},
\end{align}
where $\Sigma_1(\omega)=\mathrm{Re}\left\{\Sigma^R(\omega)\right\}$ and $\Sigma_2(\omega)=\mathrm{Im}\left\{\Sigma^R(\omega)\right\}$. 

Aside from the single-particle momentum $\vec{k}$ and fermionic Matsubara frequency $i\omega_n$, for two-body correlation functions we consider the bosonic Matsubara frequency $i\Omega_n=2 \pi n k_B T$ and the exchanged momentum $\vec{q}$, which can be defined for two-body correlation functions in particle-hole notation, as detailed in the next section.  

\subsection{Two-particle Green's functions with space- and time-translation invariance, particle-hole representation, three-point vertices, and Bethe-Salpeter equations}

For this reminder of two-particle Green's functions and vertex corrections, I closely follow Refs.\@ \onlinecite{Rohringer-2018}, \onlinecite{Bruus-2004mb}, and \onlinecite{Berthod-2018}. 
As a starting point, let us consider the following definition of the $n$-particle imaginary-time Green's function \cite{Rohringer-2018}:
\begin{align}\label{eq:-particle_G_def}
G_{i_1,\ldots i_{2n}}^{(n)}(\tau_1, \ldots \tau_{2n})&=(-1)^n\left\langle \hat{\mathscr{T}}_{\tau} \left\{ \hat{c}_{i_1}(\tau_1) \hat{c}_{i_2}^\dagger(\tau_2)  \right.\right. \nonumber \\ & \left.\left. \ldots \hat{c}_{i_{2n-1}}(\tau_{2n-1}) \hat{c}_{i_{2n}}^\dagger(\tau_{2n}) \right\}\right\rangle,
\end{align}
where the indices $i_j=\left\{ \vec{r}_j/\vec{k}_j, l_j, \sigma_j \ldots \right\}$ collect all internal degrees of freedom, such as space coordinate $\vec{r}_j$ or momentum $\vec{k}_j$ depending on real/reciprocal space representation, angular momentum $l_j$, spin $\sigma_j$, and so on. The average $\left\langle \cdot \right\rangle=\mathrm{Tr}\left\{ e^{-\beta \hat{H}}\right\}/\mathscr{Z}$ is taken with respect to the Hamiltonian $\hat{H}$, with $\mathscr{Z}=\mathrm{Tr}\left\{e^{-\beta \hat{H}}\right\}$ being the corresponding partition function and $\beta=1/(k_B T)$. The $n$-particle definition (\ref{eq:-particle_G_def}) reduces to Eq.\@ (\ref{eq:G_k_tau}) for $n=1$, i.e.\@, $G(\vec{k},\tau)\equiv G^{(1)}_{\vec{k} \sigma, \vec{k} \sigma}(\tau,0)=-\left\langle \hat{\mathscr{T}}_{\tau} \left\{\hat{c}_{\vec{k},\sigma}(\tau) \hat{c}^\dagger_{\vec{k},\sigma}(0)\right\} \right\rangle$.
Moreover, for $n=2$ Eq.\@ (\ref{eq:-particle_G_def}) formally depends on four momenta and spins, $i_j=\left\{k_j,\sigma_j\right\}$, $j=\left\{1,2,3,4\right\}$. 
\begin{figure*}[ht] \centering
\includegraphics[width=0.85\linewidth]{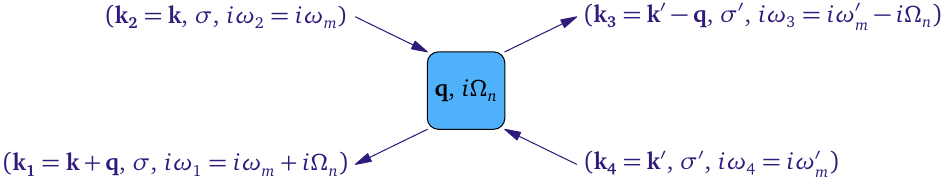}
\caption{\label{fig:transfer} Illustration of a two-particle Green’s function from the particle-hole perspective, described by the corresponding momentum and frequency notations given by Eqs.\@ (\ref{eq:momentum_cons_4}) and (\ref{eq:particle-hole_transfer}). The total momentum and frequency of the particle-hole pair are $\vec{q}$ and $i\Omega_n$, respectively. 
 }
\end{figure*}

Specializing to time-, lattice, and SU(2)-symmetric (paramagnetic) systems, total momentum conservation implies 
\begin{equation}\label{eq:momentum_cons_4}
\vec{k}_1+\vec{k}_2=\vec{k}_3+\vec{k}_4,
\end{equation}
while total spin conservation reduces the $2^4=16$ spin combinations to the two components $\sigma \sigma \sigma' \sigma' \equiv \sigma \sigma'$, which could still be related by SU(2), but is convenient to keep separate in order to not mix momenta and frequencies \cite{Rohringer-2018}. 

Referring to the so-called particle-hole notation depicted in Fig.\@ \ref{fig:transfer}, with momenta labeled as in Eq.\@ (\ref{eq:momentum_cons_4}), one can introduce the transferred momentum and the transferred frequency, 
\begin{subequations}\label{eq:particle-hole_transfer}
\begin{equation}\label{eq:particle-hole_transfer_q}
\vec{q}=\vec{k}_1-\vec{k}_2 =\vec{k}_4-\vec{k}_3,
\end{equation}
\begin{equation}\label{eq:particle-hole_transfer_omega}
i\Omega_n=i\omega_1-i\omega_2=i\omega_4-i\omega_3,
\end{equation}
\end{subequations}
and the two-particle Green's function results
\begin{align}\label{eq:G_2_particle-hole}
& G^{(2)}_{\vec{k}+\vec{q} \sigma, \vec{k}\sigma, \vec{k}'-\vec{q} \sigma', \vec{k}'\sigma'}(\tau_1,\tau_2,\tau_3,0) \nonumber \\ &\equiv G_{\sigma \sigma'}^{(2)}(\vec{k},\vec{k}',\vec{q},\tau_1,\tau_2,\tau_3),
\end{align}
with corresponding Fourier transform in Matsubara-frequency space
\begin{align}\label{eq:G_2_particle-hole_frequency}
&G_{\sigma \sigma'}^{(2)}(\vec{k},\vec{k}',\vec{q},i\omega_m,i\omega_m',i\Omega_n)=\int_0^\beta d\tau_1 \int_0^\beta d\tau_2 \int_0^\beta d \tau_3 \nonumber \\ & \times e^{i (\omega_m+\Omega_n) \tau_1} e^{-i \omega_m \tau_2} e^{i (\omega_m'-\Omega_n) \tau_3} G_{\sigma \sigma'}^{(2)}(\vec{k},\vec{k}',\vec{q},\tau_1,\tau_2,\tau_3).
\end{align}
In this particle-hole notation, the two-particle Green's function (\ref{eq:G_2_particle-hole_frequency}) can be viewed as the scattering amplitude of an incoming particle and hole with total energy $i\Omega_n$ and total momentum $\vec{q}$. Such representation is especially convenient in systems characterized by strong particle-hole fluctuations, e.g.\@, charge or spin fluctuations \cite{Rohringer-2012,Rohringer-2018}. Of course, different notations just correspond to the most convenient ``coordinate system'' for the problem: another choice is to fix the total momentum of two particles, in the so-called particle-particle representation \cite{Rohringer-2012,Rohringer-2018}. 

One then defines the ``momentum-resolved'' connected generalized susceptibilities in the particle-hole notation \cite{Rohringer-2018}, 
\begin{align}\label{eq:chi_sigmasigmaprime}
\chi_{\sigma \sigma'}(\vec{k},\vec{k}',&\vec{q},i\omega_m,i\omega_m',i\Omega_n)\nonumber \\ &=G_{\sigma \sigma'}^{(2)}(\vec{k},\vec{k}',\vec{q},i\omega_m,i\omega_m',i\Omega_n) \nonumber \\ &-\beta G(\vec{k},i\omega_m)G(\vec{k}',i\omega_m') \delta_{\Omega_n,0} \delta_{\vec{q},0}, 
\end{align}
where the second term at the right-hand side represents the subtracted disconnected part. Eq.\@ (\ref{eq:chi_sigmasigmaprime}) is useful to construct the susceptibility given by Eq.\@ (\ref{eq:chi_Theta_gen}), as we will shortly appreciate. From Eq.\@ (\ref{eq:chi_sigmasigmaprime}), one can first define the ``charge'' ($C$) and ``spin'' ($S$) channels, as
\begin{align}\label{eq:chi_gen_charge}
&\chi_{C}(\vec{k},\vec{k}',\vec{q},i\omega_m,i\omega_m',i\Omega_n)=\chi_{\uparrow \uparrow}(\vec{k},\vec{k}',\vec{q},i\omega_m,i\omega_m',i\Omega_n) \nonumber \\ &+ \chi_{\uparrow \downarrow}(\vec{k},\vec{k}',\vec{q},i\omega_m,i\omega_m',i\Omega_n) 
\end{align}
and
\begin{align}\label{eq:chi_gen_spin}
&\chi_{S}(\vec{k},\vec{k}',\vec{q},i\omega_m,i\omega_m',i\Omega_n)=\chi_{\uparrow \uparrow}(\vec{k},\vec{k}',\vec{q},i\omega_m,i\omega_m',i\Omega_n) \nonumber \\ &- \chi_{\uparrow \downarrow}(\vec{k},\vec{k}',\vec{q},i\omega_m,i\omega_m',i\Omega_n). 
\end{align}
In particular, the present work deals with the charge channel, i.e.\@, Eq.\@ (\ref{eq:chi_gen_charge}), although the manipulations that follow can be analogously performed in the spin channel \cite{Rohringer-2018}. Let us now proceed and define the physical susceptibilities $\chi_{\hat{\Theta} \hat{\Theta}}^{\vec{\alpha} \vec{\beta}}(\vec{q},i\Omega_n)$ for the observable $\hat{\Theta}$: to assemble the physical susceptibility, we need to attach the external legs of the charge susceptibility given by Eq.\@ (\ref{eq:chi_gen_charge}) to two three-point vertices for the variable $\hat{\Theta}$, and to sum over the internal variables $\left\{\vec{k},\vec{k}',i\omega_m,i\omega_m'\right\}$. In the following, all these steps and the definitions of the three-point vertices will be specified. 

First, to define the bare three-point vertex, suppose that the observable $\hat{\Theta}$ couples linearly to the set of spatial components $\vec{\alpha}$ of an external field $\hat{\vec{A}}$, according to
\begin{equation}\label{eq:coupling_A_Theta}
\hat{\Theta}_{\vec{\alpha}}(\vec{q})=\sum_{\vec{k}} \Gamma_{\hat{\Theta},\vec{\alpha}}^{(0)}(\vec{k},\vec{q})\hat{c}_{\vec{k}+\vec{q}}^\dagger \hat{c}_{\vec{k}}.
\end{equation}
From Eq.\@ (\ref{eq:coupling_A_Theta}), one reads off the bare vertex $\Gamma_{\hat{\Theta},\vec{\alpha}}^{(0)}(\vec{k},\vec{q})$, with multiple examples reported in Sec.\@ \ref{Vertex_corrections_q0}. Equipped with these bare vertices, we can construct the physical two-particle correlation function, or physical susceptibility, $\chi_{\hat{\Theta} \hat{\Theta}}^{\vec{\alpha} \vec{\beta}}(\vec{q},i\Omega_n)$ for the observable $\hat{\Theta}$ \cite{Berthod-2018}, by summation over all the internal fermionic variables: 
\begin{align}\label{eq:chi_theta_theta_gen}
&\chi_{\hat{\Theta} \hat{\Theta}}^{\vec{\alpha} \vec{\beta}}(\vec{q},i\Omega_n)=-\frac{1}{\mathscr{V}^2}\sum_{\vec{k},\vec{k}'} (k_B T)^2\sum_{i\omega_m, i\omega_m'} \Gamma_{\hat{\Theta},\vec{\alpha}}^{(0)}(\vec{k},\vec{q}) \nonumber \\ &\times \chi_{C}(\vec{k},\vec{k}',\vec{q},i\omega_m,i\omega_m',i\Omega_n) \Gamma_{\hat{\Theta},\vec{\beta}}^{(0)}(\vec{k}',-\vec{q}).
\end{align}
 There are now different but equivalent ways of writing the susceptibility $\chi_{\hat{\Theta} \hat{\Theta}}^{\vec{\alpha} \vec{\beta}}(\vec{q},i\Omega_n)$, to classify all contributing diagrams in conceptually distinct classes \cite{Rohringer-2012,Rohringer-2018}. One route involves separating the susceptibility in a ``particle-hole disconnected part'' (\ref{eq:chi_charge_phys_Theta_disconnect}), which represents independent propagation of the (renormalized) particle and hole, and a ``particle-hole connected part'' part (\ref{eq:chi_charge_phys_Theta_connect}), representing all possible scattering events between the particle and the hole. Mathematically, this amounts to
\begin{subequations}
\begin{align}\label{eq:chi_charge_phys_Theta_connect_disconnect}
\chi_{\hat{\Theta} \hat{\Theta}}^{\vec{\alpha} \vec{\beta}}(\vec{q},i\Omega_n)=\chi_{\hat{\Theta} \hat{\Theta}}^{(0),\vec{\alpha} \vec{\beta}}(\vec{q},i\Omega_n)+\delta \chi_{\hat{\Theta} \hat{\Theta}}^{\vec{\alpha} \vec{\beta}}(\vec{q},i\Omega_n),
\end{align}
\begin{align}\label{eq:chi_charge_phys_Theta_disconnect}
\chi_{\hat{\Theta} \hat{\Theta}}^{(0),\vec{\alpha} \vec{\beta}}(\vec{q},i\Omega_n)&=\frac{1}{\mathscr{V}} \sum_{\vec{k}} k_B T \sum_{i\omega_m} \Gamma_{\hat{\Theta},\vec{\alpha}}^{(0)}(\vec{k},\vec{q}) \nonumber \\ & \times G(\vec{k},i\omega_m) G(\vec{k}+\vec{q},i\omega_m+i\Omega_n)  \nonumber \\ &\times \Gamma_{\hat{\Theta},\vec{\beta}}^{(0)}(\vec{k}+\vec{q},-\vec{q}),
\end{align}
\begin{align}\label{eq:chi_charge_phys_Theta_connect}
\delta\chi_{\hat{\Theta} \hat{\Theta}}^{\vec{\alpha} \vec{\beta}}(\vec{q},i\Omega_n)&=\frac{1}{\mathscr{V}} \sum_{\vec{k},\vec{k}'} (k_B T)^2 \sum_{i\omega_m,i\omega_m'} \Gamma_{\hat{\Theta},\vec{\alpha}}^{(0)}(\vec{k},\vec{q})  \nonumber \\ &\times G(\vec{k},i\omega_m)G(\vec{k}+\vec{q},i\omega_m+i\Omega_n) \nonumber \\ &\times V(\vec{k},\vec{k}',\vec{q},i\omega_m,i\omega_m',i\Omega_n) \nonumber \\ &\times G(\vec{k}',i\omega_m')G(\vec{k}'-\vec{q},i\omega_m'-i\Omega_n) \nonumber \\ &\times  \Gamma_{\hat{\Theta},\vec{\beta}}^{(0)}(\vec{k}',-\vec{q}). 
\end{align}
\end{subequations}
The object $V(\vec{k},\vec{k}',\vec{q},i\omega_m,i\omega_m',i\Omega_n)$ appearing in Eq.\@ (\ref{eq:chi_charge_phys_Theta_connect}) is the (reducible) two-particle vertex function, which contains all Feynman diagrams connecting all four Green’s functions \cite{Bruus-2004mb,Rohringer-2018}; for instance, for Fermi liquids $V(\vec{k},\vec{k}',\vec{q},i\omega_m,i\omega_m',i\Omega_n)$  is proportional to the scattering amplitude between quasiparticles \cite{Abrikosov-2012meth}.

A more detailed classification is obtained by categorizing the Feynman diagrams contributing to $V(\vec{k},\vec{k}',\vec{q},i\omega_m,i\omega_m',i\Omega_n)$ according to their two-particle reducibility. These diagrams fall into four topologically distinct classes: they are either fully two-particle irreducible ($\mathrm{fi}$) or reducible in one of the three channels, namely particle-hole ($\mathrm{ph}$), vertical particle-hole ($\overline{\mathrm{ph}}$), and particle-particle ($\mathrm{pp}$): 
\begin{align}\label{eq:parquet_decomposition}
V(\vec{k},\vec{k}',\vec{q},i\omega_m,i\omega_m',i\Omega_n)&=\Lambda_{\mathrm{fi}}(\vec{k},\vec{k}',\vec{q},i\omega_m,i\omega_m',i\Omega_n) \nonumber \\ & + \Phi_{\mathrm{ph}}(\vec{k},\vec{k}',\vec{q},i\omega_m,i\omega_m',i\Omega_n) \nonumber \\ & + \Phi_{\overline{\mathrm{ph}}}(\vec{k},\vec{k}',\vec{q},i\omega_m,i\omega_m',i\Omega_n) \nonumber \\ & + \Phi_{\overline{\mathrm{pp}}}(\vec{k},\vec{k}',\vec{q},i\omega_m,i\omega_m',i\Omega_n).
\end{align}
The unique decomposition (\ref{eq:parquet_decomposition}) provides the basis of the so-called parquet equations \cite{Diatlov-1956,DeDominicis-1962,DeDominicis-1964a,DeDominicis-1964b,Bickers-1991,Janis-2001,Janis-2017}.

An alternative decomposition of the reducible two-particle vertex function just considers reducibility in a specific channel: 
\begin{align}\label{eq:BSE_decomposition}
V(\vec{k},\vec{k}',\vec{q},i\omega_m,&i\omega_m',i\Omega_n)=\Lambda_{l}(\vec{k},\vec{k}',\vec{q},i\omega_m,i\omega_m',i\Omega_n)\nonumber \\ & + \Phi_{l}(\vec{k},\vec{k}',\vec{q},i\omega_m,i\omega_m',i\Omega_n),
\end{align}
where $l=\left\{\mathrm{ph}, \overline{\mathrm{ph}}, \mathrm{pp} \right\}$. Here $\Lambda_{l}(\vec{k},\vec{k}',\vec{q},i\omega_m,i\omega_m',i\Omega_n)$ refers to the part of the two-particle vertex function which is irreducible in channel $l$. In the present work, we focus on the particle-hole irreducible part ($l=\mathrm{ph}$), which is connected to the reducible vertex function in Eq.\@ (\ref{eq:BSE_decomposition}) through a Bethe-Salpeter equation \cite{Bickers-1991,Bruus-2004mb,Rohringer-2018}: 
\begin{align}\label{eq:BSE_two_particle_vertex}
V(\vec{k},\vec{k}',\vec{q},&i\omega_m,i\omega_m',i\Omega_n)=\Lambda_{\mathrm{ph}}(\vec{k},\vec{k}',\vec{q},i\omega_m,i\omega_m',i\Omega_n) \nonumber \\ + & \frac{1}{\mathscr{V}}\sum_{\vec{k}_l} k_B T \sum_{i\omega_l} \Lambda_{\mathrm{ph}}(\vec{k},\vec{k}_l,\vec{q},i\omega_m,i\omega_l,i\Omega_n) \nonumber \\ & \times G(\vec{k}_l,i\omega_l)G(\vec{k}_l+\vec{q},i\omega_l+i\Omega_n) \nonumber \\ & \times V(\vec{k}_l,\vec{k}',\vec{q},i\omega_l,i\omega_m',i\Omega_n).
\end{align}
Notice that Eq.\@ (\ref{eq:BSE_two_particle_vertex}) corresponds to an infinite summation of ladder diagrams for the given channel $l=\mathrm{ph}$, while the mutual renormalizations between different channels have to be separately included through the parquet decomposition (\ref{eq:parquet_decomposition}). The present work just considers the decomposition (\ref{eq:BSE_decomposition}) in the particle-hole channel. Hence, $\Lambda_{\mathrm{ph}}(\vec{k},\vec{k}',\vec{q},i\omega_m,i\omega_m',i\Omega_n) \equiv \Lambda(\vec{k},\vec{k}',\vec{q},i\omega_m,i\omega_m',i\Omega_n)$ as it appears in Eq.\@ (\ref{eq:Bethe_Salpeter_gen_nonlocal}). 

The physical susceptibility $\chi_{\hat{\Theta} \hat{\Theta}}^{\vec{\alpha} \vec{\beta}}(\vec{q},i\Omega_n)$ can also be written in an equivalent form that encapsulates all diagrams other than the renormalized particle-hole bubble into a renormalized three-point vertex $\Gamma_{\hat{\Theta},\vec{\alpha}}(\vec{k},\vec{q},i\omega_m,i\Omega_n)$, different from the bare vertex $\Gamma_{\hat{\Theta},\vec{\alpha}}^{(0)}(\vec{k},\vec{q})$ stemming from Eq.\@ (\ref{eq:coupling_A_Theta}): 
\begin{align}\label{eq:chi_charge_phys_Theta}
\chi_{\hat{\Theta} \hat{\Theta}}^{\vec{\alpha} \vec{\beta}}(\vec{q},i\Omega_n)&=\frac{1}{\mathscr{V}} \sum_{\vec{k}} k_B T \sum_{i\omega_m}  \Gamma_{\hat{\Theta},\vec{\beta}}^{(0)}(\vec{k}+\vec{q},-\vec{q}) \nonumber \\ &\times G(\vec{k},i\omega_m)G(\vec{k}+\vec{q},i\omega_m+i\Omega_n) \nonumber \\ &\times \Gamma_{\hat{\Theta},\vec{\alpha}}(\vec{k},\vec{q},i\omega_m,i\Omega_n).
\end{align}
Eq.\@ (\ref{eq:chi_charge_phys_Theta}) coincides with Eq.\@ (\ref{eq:chi_Theta_gen}), and is visually represented at the rightmost diagram on the second line of Fig.\@ \ref{fig:Diag_Xthetatheta_VC}. By comparing the equivalent formulations given by Eqs.\@ (\ref{eq:chi_charge_phys_Theta_connect_disconnect}) and (\ref{eq:chi_charge_phys_Theta}), we deduce the relation between the renormalized three-point vertex and the reducible two-particle vertex function \cite{Berthod-2018}: 
\begin{align}\label{eq:Gamma_F_Gamma0}
&\Gamma_{\hat{\Theta},\vec{\alpha}}(\vec{k},\vec{q},i\omega_m,i\Omega_n)=\Gamma_{\hat{\Theta},\vec{\alpha}}^{(0)}(\vec{k},\vec{q}) \nonumber \\ &+\frac{1}{\mathscr{V}} \sum_{\vec{k}'} k_B T\sum_{i\omega_m'} \Gamma_{\hat{\Theta},\vec{\alpha}}^{(0)}(\vec{k}',\vec{q}) G(\vec{k}',i\omega_m') \nonumber \\ & \times G(\vec{k}'+\vec{q},i\omega_m'+i\Omega_n) \nonumber \\ & \times V(\vec{k}',\vec{k}+\vec{q},\vec{q},i\omega_m',i\omega_m+i\Omega_n,i\Omega_n). 
\end{align}
Now, combining Eqs.\@ (\ref{eq:BSE_two_particle_vertex}) and Eq.\@ (\ref{eq:Gamma_F_Gamma0}), one finally recognizes the ladder series that constitutes the Bethe-Salpeter equation for the renormalized three-point vertex $\Gamma_{\hat{\Theta},\vec{\alpha}}(\vec{k},\vec{q},i\omega_m,i\Omega_n)$, in terms of the particle-hole irreducible two-particle vertex function $\Lambda_{\mathrm{ph}}(\vec{k}',\vec{k}+\vec{q},\vec{q},i\omega_m',i\omega_m+i\Omega_n,i\Omega_n)$: the final result is Eq.\@ (\ref{eq:Bethe_Salpeter_gen_nonlocal}).

\subsection{Kubo formulas: conductivity and viscosity tensors}\label{App:conductivity_viscosity_explicit}

The focus of this work is on Kubo formulae of the kind of Eq.\@ (\ref{eq:Kubo_general_Theta}), for correlation functions as in Eq.\@ (\ref{eq:chi_Theta_gen}). Here we explicitly write two notable special cases, for the conductivity and the viscosity tensors. 
The conductivity tensor in linear response for a translationally invariant system satisfies the Kubo formula
\begin{equation}\label{eq:Kubo_conductivity_gen}
\sigma_{\alpha \beta}(\vec{q},\omega)=\frac{i e^2}{\omega}\left[\chi_{\hat{J} \hat{J}}^{R,\alpha\beta}(\vec{q},\omega)+\delta_{\alpha \beta}\frac{\left\langle n\right\rangle}{m}\right],
\end{equation}
where $\chi_{\hat{J} \hat{J}}^{R,\alpha\beta}(\vec{q},\omega)=\lim_{i \Omega_n\rightarrow \hbar\omega+i0^+} \chi_{\hat{J} \hat{J}}^{\alpha\beta}(\vec{q},i\Omega_n)$ is the retarded current-current correlation function, analytically continued to the real axis. Explicitly, the current-current correlation function is a special case of Eq.\@ (\ref{eq:chi_Theta_gen}) for an observable $\hat{\Theta}=\hat{J}$ equal to the current density. Eq.\@ (\ref{eq:Kubo_conductivity_gen}) also includes the diamagnetic static term with average electron density $\left\langle n\right\rangle$ and electron mass $m$. 

The generalized viscosity tensor satisfies another linear-response Kubo formula: 
\begin{align}\label{eq:Kubo_viscosity}
\eta_{\alpha \beta \gamma \delta}(\omega)&=\frac{i}{\omega+i0^+}\left[\chi_{\hat{\textit{T}} \hat{\textit{T}}}^{R,\alpha \beta \gamma \delta}(0,\omega)-\chi_{\hat{\textit{T}} \hat{\textit{T}}}^{R,\alpha \beta \gamma \delta}(0,0)\right],
\end{align}
where $\chi_{\hat{\textit{T}} \hat{\textit{T}}}^{R,\alpha \beta \gamma \delta}(0,0)=k_T^{-1} \delta_{\alpha \beta} \delta_{\gamma \delta}$ with $k_T^{-1}=-\left.\mathscr{V}\partial P /\partial \mathscr{V}\right|_{\mathscr{N},T}$ isothermal compressibility \cite{Bradlyn-2012,Link-2018a}, and $\chi_{\hat{\textit{T}} \hat{\textit{T}}}^{R,\alpha \beta \gamma \delta}(\vec{q},\omega)=\lim_{i \Omega_n \rightarrow \hbar\omega+i 0^+} \chi_{\hat{\textit{T}} \hat{\textit{T}}}^{\alpha \beta \gamma \delta}(\vec{q},i\Omega_n)$ is the (retarded) analytic continuation of the stress-stress correlation function. Formally, the stress-stress correlation function results from Eq.\@ (\ref{eq:chi_Theta_gen}) for an observable $\hat{\Theta}=\hat{\textit{T}}$ equal to the stress tensor. 

Notice that, in general, the stress tensor contains a kinetic part, due to stresses imposed on the fluid by physical flow of electrons, and an interaction (``virial'') portion, which physically stems from metric-dependent interaction contributions contained in the system Hamiltonian $\hat{H}$, which generate stresses without physical electron flow. The interaction contributions can be taken into account by endowing the system action $\mathscr{S}$, constructed from the Hamiltonian $\hat{H}$, with a background metric $g$: the total stress tensor contributions $\hat{\textit{T}}_{t,\alpha \beta}$ are then obtained by varying $\mathscr{S}$ with respect to strain components $\epsilon_{\alpha \beta}$, and will include a kinetic part $\hat{\textit{T}}_{\alpha \beta}$  as well as an interaction $\hat{\textit{T}}_{i,\alpha \beta}$ part \cite{Jang-2026_preprint}. 
The present paper focuses just on the kinetic stress tensor, for which the bare kinetic stress vertex given by Eqs.\@ (\ref{eq:stress_vertex_gen_shear}), (\ref{eq:stress_vertex_gen_Hall}), and (\ref{eq:stress_vertex_gen}) for shear, Hall (odd), and bulk contributions respectively. 

Whether the interaction part of the stress tensor is nonnull depends on the specific interaction acting on the electron liquid. If mechanical strain $\epsilon_{\alpha \beta}$ does not couple to interactions at all, like in a lattice Hubbard model, then $\hat{\textit{T}}_{i,\alpha \beta}=0$ and stresses are carried exclusively by the kinetic-stress sector; in a continuum theory with background flat spatial metric $g$, the interaction contribution is nonnull just for bulk stresses, $\hat{\textit{T}}_{i,\alpha \alpha}\neq0$, while for shear stresses $\hat{\textit{T}}_{i,\alpha \beta}=0$, $\alpha \neq \beta$; for finite-range interactions, in general there will be interaction contributions to the stress vertex both in the shear and bulk stress channels. 

\section{Conditions for dominance of frequency dependence over momentum dependence in the self-energy of many-body systems}\label{App:frequency_over_momentum}

In general, the many-body self-energy $\Sigma(\vec{k},i\omega_m)$ and its analytic-continued retarded counterpart $\Sigma^R(\vec{k},\omega)$ depend on both momentum $\vec{k}$ and frequency $\omega$. However, the relative importance of momentum and frequency arguments depends on the specific physical problem and regime under consideration, and it varies with temperature, internal degrees of freedom, and nature of microscopic interactions. 
In particular, situations where dynamical interactions can be considered (at least approximately) local, i.e.\@, momentum-independent, considerably simplify many-body calculations, such as computations of the correlation functions from Eq.\@ (\ref{eq:chi_Theta_gen}) and interaction vertices as in passing from Eq.\@ (\ref{eq:Bethe_Salpeter_gen_nonlocal}) to Eq.\@ (\ref{eq:Bethe_Salpeter_gen}). In these situations, the self-energy $\Sigma^R(\omega)$ depends only on frequency and the irreducible two-particle vertex $\Lambda(i\omega_m,i\omega_m',i\Omega_n)$ is also momentum-independent. In the following, some exemplary cases are enumerated when dynamical many-body effects can be considered at least approximately local. 

\subsection{Strongly interacting correlated-electron systems and non-Fermi liquids} 

In strongly correlated solid-state systems, complex experimental phase diagrams are accompanied by low-energy intertwined phases, high interaction energy scales compared to characteristic transition temperatures, and collective electron behaviour \cite{Lee-2006,Sachdev-2011quantum,Rozenberg-2019,Chowdhury-2022}. There, experiments and computations find extended parameter regimes where the self-energy is dominated by its frequency dependence, essentially because electron-electron interactions lead to dynamical ($\omega$-dependent) effects that are enhanced at specific energies by predominantly local interactions. For instance, in Mott insulators as described by the Hubbard model with onsite Hubbard repulsion $U$ and hopping parameter $t$ \cite{Capone-2004,Lee-2006,Schaefer-2021,Qin-2022,Menke-2024}, at large $U/t$ charge fluctuations are hindered, and electron motion becomes governed by local temporal fluctuations (doublon-holon processes): in this regime, the self-energy can be strongly frequency-dependent, with features like the formation of a Mott gap, while the momentum dependence can be relatively weaker. This is particularly evident in optical conductivity measurements, where the impact of the frequency dependence of the self-energy can be directly observed \cite{Rozenberg-1995,Shinjo-2021}.

Moreover, in the normal state of high-temperature superconductors frequency-dependent self-energy effects are essential both in the pseudogap regime at underdoping, where a gap appears in the normal-state single-particle spectrum, and in the strange-metal \cite{Zaanen-2019,Chowdhury-2022,Phillips-2022}/bad-metal \cite{Emery-1995,Lee-2006,Vucicevic-2015} regimes near optimal doping characterized by power-law dependence of observables on temperature, as detected through ARPES (Angle-Resolved Photoemission Spectroscopy) \cite{Fedorov-1999,Damascelli-2003,Cooper-2009}, EELS (Electron Energy-Loss Spectroscopy) \cite{Mitrano-2018,Husain-2019,Chen-2024,Guo-2024_preprint}, resistivity \cite{Giraldo-Gallo-2018,Legros-2019,Grissonnanche-2021,Shi-2025}, and optical conductivity \cite{vanderMarel-2003,vanHeumen-2022,Michon-2023} measurements. The momentum dependence of the extracted self-energy is found to be comparatively weaker than its frequency evolution. 

Finally, let us mention quantum $U(1)$ and disordered spin liquids \cite{Nave-2007,Wang-2010,Sachdev-2011quantum,Broholm-2020}, where the self-energy’s frequency dependence becomes crucial because the system is dominated by fluctuating spin correlations rather than by momentum-dependent quasiparticle excitations. Here, the frequency dependence may reflect the scaling of spin excitations at different energy scales.

\subsection{Systems close to a classical or quantum phase transition}\label{App:QCP_patch}

In many materials, especially those with strong correlations, the system might be in a regime where temporal fluctuations are dominant, and the dynamics of quasiparticles are primarily controlled by frequency $\omega$ (or energy $\hbar \omega$) rather than by momentum $\vec{k}$.
For instance, near a Mott transition or quantum critical point (e.g., in systems exhibiting magnetic or charge order), the system can undergo significant changes due to the frequency dependence of interactions, such as in the case of critical slowing down \cite{Goldenfeld-2018RG}. Near such a transition, the self-energy can become nearly independent of momentum but very sensitive to frequency, particularly at long wavelengths or low momenta. In the proximity of the transition, quantum fluctuations of soft collective modes (e.g.\@, spin, charge, loop-current, lattice fluctuations) are crucial for inelastic electron-boson scattering, so that the electron self-energy can be dominated by frequency-dependent effects due to the dynamics of these fluctuations, while momentum dependence may be secondary or weaker.
A microscopic mechanism to explain this sensitivity is provided by patch theories of critical Fermi surfaces \cite{Polchinski-1994,Altshuler-1994,Lee-2009,Mross-2010,Metlitski-2010,Sachdev-2011quantum,Patel-2017b}, which I qualitatively summarize below in one of its most recent incarnations \cite{Esterlis-2021,Guo-2022,Guo-2023}. 

Consider a point on a 2D Fermi surface where the dispersion is linearized in $k_x$ and quadratic in $k_y$ to account for a finite local curvature $\kappa_R$: 
\begin{equation}\label{eq:patch_linearized}
\epsilon_{\vec{k}}\approx v_F k_x + \kappa_R (k_y)^2,
\end{equation}
where $v_F$ is the Fermi velocity. Let us assume that the electrons interact with a soft critical scalar bosonic mode (e.g.\@, a Pomeranchuk instability, a $U(1)$ gauge field, a phonon) with propagator
\begin{equation}\label{eq:boson_soft_QC}
D(\vec{q},i\Omega_n)=\frac{1}{\gamma\frac{\left|\Omega_n\right|}{q}+q^2 +(m_b)^2},
\end{equation}
where $\left|\Omega_n\right|/q$ is the Landau-damping term, $\gamma \in \mathbb{R}^+$, and the renormalized mass $m_b \rightarrow 0$ at the quantum critical point (QCP). The Landau-damping term can be derived from the boson self-energy $\Pi(\vec{q},i\Omega)$ within the patch, assuming (and then self-consistently verifying) that the fermion self-energy $\Sigma(i \omega)$ is momentum-independent \cite{Esterlis-2021}. Within the patch (\ref{eq:patch_linearized}), the momentum exchange in the boson propagator (\ref{eq:boson_soft_QC}) is dominated by small-momentum ($q \rightarrow 0^+$) processes and is mainly transverse, $q \approx q_y$. 

In this limit, the fermion self-energy resulting from boson exchange at $T=0$ (where Matsubara energies $i\omega$ become continuous, and in this section $i\omega$ has units of $i$ times energy) is
\begin{equation}\label{eq:Sigma_exchange_QC}
\Sigma(\vec{k},i\omega)\propto \int d\Omega \int d q_x \int dq_y \frac{D (\vec{q},i\Omega)}{i (\omega+\Omega)-\epsilon_{\vec{k}+\vec{q}}}.
\end{equation}
Since $q \approx q_y$, $D(\vec{q},i\Omega)$ is approximately independent from $q_x$, which allows one to perform the integral over $q_x$ in Eq.\@ (\ref{eq:Sigma_exchange_QC}) with the residue theorem, after inserting the dispersion (\ref{eq:patch_linearized}):
\begin{align}\label{eq:int_dqx}
&\int d q_x \frac{1}{i(\omega+\Omega)-v_F(k_x+q_x)-\kappa_R (k_y+q_y)^2} \nonumber \\ & \sim -i \pi \mathrm{sign}(\omega+\Omega).
\end{align}
Thus, the integral (\ref{eq:int_dqx}) removes the dependence on $k_x$ and $q_x$ entirely. The remaining integral over $q_y$ in Eq.\@ (\ref{eq:Sigma_exchange_QC}) reads
\begin{align}\label{eq:Sigma_exchange_QC_2}
\Sigma(\vec{k},i\omega)& \propto \int d\Omega \int dq_y \frac{\mathrm{sign}(\omega+\Omega)}{\gamma\frac{\left|\Omega\right|}{q_y}+(q_y)^2} \nonumber \\ & \sim \int d\Omega \frac{\mathrm{sign}(\omega+\Omega)}{\left|\Omega\right|^{\frac{1}{3}}}\sim \omega^{\frac{2}{3}}. 
\end{align}
Eq.\@ (\ref{eq:Sigma_exchange_QC_2}) shows that the fermionic self-energy due to soft-boson exchange is dynamical (depends on $\omega^{2/3}$) and momentum-independent within the patch theory. The same conclusions, here presented in a simple one-loop argument, hold in the self-consistent calculation of bosonic and fermionic properties, realized, e.g.\@, in the translationally invariant 2D Yukawa-Sachdev-Ye--Kitaev model \cite{Esterlis-2021}. 
In the renormalization-group sense, the scaling of transverse momentum is $\left[k_x\right]=\mathscr{E}$, the one of tangential momentum is $\left[k_y\right]=\sqrt{\mathscr{E}}$, and the one of the frequency-dependent part (\ref{eq:Sigma_exchange_QC_2}) of the self-energy is $\left[\Sigma\right]\propto \omega^{\frac{2}{3}} \sim \mathscr{E}$. 

Furthermore, the locality of the self-energy (\ref{eq:Sigma_exchange_QC_2}) from the exchange of bosons, in accordance with Eq.\@ (\ref{eq:Sigma_exchange_QC}), implies an approximate locality of the two-particle vertex $\Lambda(i\omega_m,i\omega_m',i\Omega_n)$ from the Kadanoff-Baym conserving condition given by Eq.\@ (\ref{eq:Lambda_Sigma_conserving}), whenever the boson propagator (\ref{eq:boson_soft_QC}) is dominated by low-momentum ($q\rightarrow 0$) and finite-frequency dynamics. This occurs for local inelastic fermion-boson scattering as modeled by spatially disordered theories \cite{Guo-2023,Patel-2023,Li-2024,Valentinis-2026}.

\subsection{High-frequency limit and optical excitations}

At high frequencies $\omega$ (or exchanged energy scales $\hbar\omega$) with respect to all other system energy scales, the frequency dependence of $\Sigma^R(\vec{k},\omega)$ often becomes more important than its momentum dependence. This dominance occurs because at temperatures $k_B T \ll \hbar \omega$ the electron ensemble effectively behaves as a $T\approx 0$ system, and conduction states are essentially sampled at the Fermi wave vector $\vec{k}\approx \vec{k}_F$, while the sum over $\vec{k}$ in the conductivity -- cfr.\@ Eq.\@ (\ref{eq:chi_Theta_gen}) for $\hat{\Theta}=\hat{J}$ and $q=0$ or equivalently Eq.\@ (\ref{eq:Kubo_conductivity_gen}) at $q=0$ -- averages over Fermi-surface anisotropy, especially in the standard Drude/diffusive conductive regime \cite{Dressel-2001,Berthod-2013,Berthod-2018}. The high-frequency regime is also the one where the energy dependence of transport/single-particle scattering rates and effective masses unavoidably matters, while the causal properties of the self-energy dictate its decay at least as $1/\omega$ at high frequencies \cite{Berthod-2013, Berthod-2018}. All the above discussion affects high-frequency, low-momentum probes such as optical conductivity and photoemission experiments, where the self-energy’s frequency dependence directly reflects the spectral properties of the system \cite{Fedorov-1999,Damascelli-2003,Cooper-2009,vanderMarel-2003,vanHeumen-2022,Michon-2023}. 

\section{Derivation of the Ward identity for charge conservation from the ladder Bethe-Salpeter equation}\label{app:Charge_conservation}

In this appendix we derive the Ward identity (\ref{eq:Ward_identity_charge}) relating the density and current vertices, from the general Bethe-Salpeter equation (\ref{eq:Bethe_Salpeter_gen}). Specializing the latter to the density vertex ($\hat{\Theta}\equiv \hat{n}$) and multiplying by $i\Omega_n$, we have 
\begin{widetext}
\begin{align}\label{eq:Bethe_Salpeter_n_iOmega_n}
i \Omega_n\Gamma_{\hat{n}}(\vec{k},\vec{q},i\omega_m,i\Omega_n)&=i\Omega_n \Gamma_{\hat{n}}^{(0)}(\vec{k},\vec{q})+ i\Omega_n k_B T\sum_{i\omega_m'} \Lambda(i\omega_m, i\omega_m',i \Omega_n) \frac{1}{\mathscr{V}}\sum_{\vec{k}'} G(\vec{k}',i\omega_m') G(\vec{k}'+\vec{q},i\omega_m'+i\Omega_n) \nonumber \\ &\times \Gamma_{\hat{n}}(\vec{k}',\vec{q},i\omega_m',i\Omega_n).
\end{align}
In the same way, from the ladder equation for the current vertex ($\hat{\Theta}\equiv \hat{\vec{J}}$) multiplied by $\hbar q_\alpha$, we have
\begin{align}\label{eq:Bethe_Salpeter_J_qalpha}
\hbar q_\alpha \Gamma_{\hat{J},\alpha}(\vec{k},\vec{q},i\omega_m,i\Omega_n)&=\hbar q_\alpha \Gamma_{\hat{J},\alpha}^{(0)}(\vec{k},\vec{q})+ \hbar q_\alpha k_B T\sum_{i\omega_m'} \Lambda(i\omega_m, i\omega_m',i \Omega_n) \frac{1}{\mathscr{V}}\sum_{\vec{k}'} G(\vec{k}',i\omega_m') G(\vec{k}'+\vec{q},i\omega_m'+i\Omega_n) \nonumber \\ &\times \Gamma_{\hat{J},\alpha}(\vec{k}',\vec{q},i\omega_m',i\Omega_n).
\end{align}
Subtracting Eq.\@ (\ref{eq:Bethe_Salpeter_J_qalpha}) from Eq.\@ (\ref{eq:Bethe_Salpeter_n_iOmega_n}), we obtain 
\begin{align}\label{eq:Bethe_Salpeter_subtract}
W(\vec{k},\vec{q},i\omega_m,i\Omega_n)&=\underbrace{i\Omega_n \Gamma_{\hat{n}}^{(0)}(\vec{k},\vec{q})-\hbar \vec{q} \cdot \vec{\Gamma}_{\hat{J}}^{(0)}(\vec{k},\vec{q})}_{\boxed{B}}+k_B T \sum_{i\omega_m'} \Lambda(i\omega_m,i\omega_m',i\Omega_n) \frac{1}{\mathscr{V}} \sum_{\vec{k}'} G(\vec{k}',i\omega_m') G(\vec{k}'+\vec{q},i\omega_m'+i\Omega_n) \nonumber \\ &\times W(\vec{k}',\vec{q},i\omega_m',i\Omega_n),
\end{align}
where we defined the quantity
\begin{equation}\label{eq:W_def}
W(\vec{k},\vec{q},i\omega_m,i\Omega_n)=i \Omega_n\Gamma_{\hat{n}}(\vec{k},\vec{q},i\omega_m,i\Omega_n)- \hbar \vec{q} \cdot \vec{\Gamma}_{\hat{J}}(\vec{k},\vec{q},i\omega_m,i\Omega_n).
\end{equation}
The term $\boxed{B}$ in Eq.\@ (\ref{eq:Bethe_Salpeter_subtract}) is fixed by the noninteracting Ward identity
\begin{align}\label{eq:Ward_charge_nonint}
i\Omega_n \Gamma_{\hat{n}}^{(0)}(\vec{k},\vec{q})-\hbar \vec{q} \cdot \vec{\Gamma}_{\hat{J}}^{(0)}(\vec{k},\vec{q})&=i\Omega_n-\xi_{\vec{k}+\vec{q}}+\xi_{\vec{k}}=G_0^{-1}(\vec{k}+\vec{q},i\omega_m+i\Omega_n)-G_0^{-1}(\vec{k},i\omega_m) \nonumber \\ & =X(\vec{k},\vec{q},i\omega_m,i\Omega_n)+\Sigma(i\omega_m+i\Omega_n)-\Sigma(i\omega_m), 
\end{align}
where
\begin{equation}\label{eq:X_def}
X(\vec{k},\vec{q},i\omega_m,i\Omega_n)=G^{-1}(\vec{k}+\vec{q},i\omega_m+i\Omega_n)-G^{-1}(\vec{k},i\omega_m).
\end{equation}
In the noninteracting Ward identity (\ref{eq:Ward_charge_nonint}), the bare current vertex at finite momentum $\vec{q}$ for arbitrary dispersion $\epsilon_{\vec{k}}$ can be suitably defined through a line integral in momentum space:
\begin{equation}\label{eq:Gamma_J_0_arbitrary}
\Gamma_{\hat{J},\alpha}^{(0)}(\vec{k},\vec{q})=\frac{1}{\hbar}\int_0^1 d\lambda \frac{\partial \epsilon_{\vec{k}+\lambda \vec{q}}}{\partial k_\alpha}. 
\end{equation}
Eq.\@ (\ref{eq:Gamma_J_0_arbitrary}) reproduces the bare current vertex (\ref{eq:current_vertex_quadr_0}) for the quadratic isotropic dispersion of Eq.\@ (\ref{eq:xi_quadr}), and it also yields
\begin{align}
\hbar \vec{q} \cdot \vec{\Gamma}_{\hat{J}}^{(0)}(\vec{k},\vec{q})=\epsilon_{\vec{k}+\vec{q}}-\epsilon_{\vec{k}} = \xi_{\vec{k}+\vec{q}}-\xi_{\vec{k}}.
\end{align}
Eq.\@ (\ref{eq:Gamma_J_0_arbitrary}) is thus compatible with the Ward identity (\ref{eq:Ward_charge_nonint}) and a bare density vertex $\Gamma_{\hat{n}}^{(0)}(\vec{k},\vec{q})=1$. Nevertheless, these choices are not unique, as one can always add a transverse part to the bare current vertex, and still preserve the bare charge Ward identity which just specifies the longitudinal part of the current vertex (parallel to $\vec{q}$).

Now we utilize the DMFT identity corresponding to the Kadanoff-Baym prescription (\ref{eq:Lambda_Sigma_conserving}) for conserving approximations, which here reads
\begin{align}\label{eq:Kadanoff_Baym_Sigma}
\Sigma(i\omega_m+i\Omega_n)-\Sigma(i\omega_m)&=k_B T\sum_{i\omega_m'} \Lambda(i\omega_m,i\omega_m',i\Omega_n) \frac{1}{\mathscr{V}}\sum_{\vec{k}'} G(\vec{k}'+\vec{q},i\omega_m'+i\Omega_n)G(\vec{k}',i\omega_m') \nonumber \\ & \times \left[G^{-1}(\vec{k}',i\omega_m')-G^{-1}(\vec{k}'+\vec{q},i\omega_m'+i\Omega_n)\right] \nonumber \\ & = k_B T\sum_{i\omega_m'} \Lambda(i\omega_m,i\omega_m',i\Omega_n) \frac{1}{\mathscr{V}}\sum_{\vec{k}'}\left[G(\vec{k}'+\vec{q},i\omega_m'+i\Omega_n)-G(\vec{k}',i\omega_m')\right].
\end{align}
Combining Eqs.\@ (\ref{eq:Kadanoff_Baym_Sigma}) and (\ref{eq:X_def}), we realize that $X(\vec{k},\vec{q},i\omega_m,i\Omega_n)$ formally satisfies the same equation as $W(\vec{k},\vec{q},i\omega_m,i\Omega_n)$, i.e.\@, Eq.\@ (\ref{eq:Bethe_Salpeter_subtract}). Explicitly, 
\begin{align}\label{eq:Kadanoff_Baym_W_X}
W(\vec{k},\vec{q},i\omega_m,i\Omega_n)-X(\vec{k},\vec{q},i\omega_m,i\Omega_n)&=k_B T\sum_{i\omega_m'} \Lambda(i\omega_m,i\omega_m',i\Omega_n) \frac{1}{\mathscr{V}}\sum_{\vec{k}'} G(\vec{k}'+\vec{q},i\omega_m'+i\Omega_n)G(\vec{k}',i\omega_m') \nonumber \\ & \times \left[W(\vec{k}',\vec{q},i\omega_m',i\Omega_n)-X(\vec{k}',\vec{q},i\omega_m',i\Omega_n)\right].
\end{align}
Provided that the system is not resonating with a collective mode pole for $\left(\vec{q},i\Omega_n\right)$, we thus have that $W(\vec{k},\vec{q},i\omega_m,i\Omega_n)=X(\vec{k},\vec{q},i\omega_m,i\Omega_n)=G^{-1}(\vec{k}+\vec{q},i\omega_m+i\Omega_n)-G^{-1}(\vec{k},i\omega_m)$. Inserting the latter expression into Eq.\@ (\ref{eq:Bethe_Salpeter_subtract}) with the help of Eq.\@ (\ref{eq:W_def}) on the left-hand side and Eq.\@ (\ref{eq:Kadanoff_Baym_Sigma}) on the right-hand side, we finally achieve the charge-conservation Ward identity
\begin{align}
& i \Omega_n \Gamma_{\hat{n}}(\vec{k},\vec{q},i\omega_n, i\Omega_n)-\hbar \vec{q} \cdot \vec{\Gamma}_{\hat{\vec{J}}}(\vec{k},\vec{q},i\omega_n, i\Omega_n) \nonumber =G^{-1}(\vec{k}+\vec{q},i\omega_n+i\Omega_n) -G^{-1}(\vec{k},i\omega_n), 
\end{align}
which is Eq.\@ (\ref{eq:Ward_identity_charge}).
\end{widetext}

\section{Derivation of the Ward identity for momentum conservation from the ladder Bethe-Salpeter equation}\label{app:Momentum_conservation}

This appendix contains the derivation of the momentum-conservation Ward identity (\ref{eq:Ward_identity_momentum}) from the Bethe-Salpeter equation (\ref{eq:Bethe_Salpeter_gen}).
Specializing the latter to momentum density operator ($\hat{\Theta}\equiv \hat{\vec{\pi}}$), the renormalized momentum vertex is 
\begin{widetext}
\begin{align}\label{eq:Bethe_Salpeter_pi_alpha}
\Gamma_{\hat{\pi},\alpha}(\vec{k},\vec{q},i\omega_m,i\Omega_n)&= \Gamma_{\hat{\pi},\alpha}^{(0)}(\vec{k},\vec{q})+  k_B T\sum_{i\omega_m'} \Lambda(i\omega_m, i\omega_m',i \Omega_n) \frac{1}{\mathscr{V}}\sum_{\vec{k}'} G(\vec{k}',i\omega_m') G(\vec{k}'+\vec{q},i\omega_m'+i\Omega_n) \nonumber \\ &\times \Gamma_{\hat{\pi},\alpha}(\vec{k}',\vec{q},i\omega_m',i\Omega_n).
\end{align}
In the same way, from the ladder equation for the stress operator ($\hat{\Theta}\equiv \hat{\underline{\underline{\textit{T}}}}$) we have the renormalized stress vertex
\begin{align}\label{eq:Bethe_Salpeter_T_alphabeta}
\Gamma_{\hat{\textit{T}},\alpha \beta}(\vec{k},\vec{q},i\omega_m,i\Omega_n)&=\Gamma_{\hat{\textit{T}},\alpha \beta}^{(0)}(\vec{k},\vec{q})+ k_B T\sum_{i\omega_m'} \Lambda(i\omega_m, i\omega_m',i \Omega_n) \frac{1}{\mathscr{V}}\sum_{\vec{k}'} G(\vec{k}',i\omega_m') G(\vec{k}'+\vec{q},i\omega_m'+i\Omega_n) \nonumber \\ &\times \Gamma_{\hat{\textit{T}},\alpha \beta}(\vec{k}',\vec{q},i\omega_m',i\Omega_n).
\end{align}
Let us multiply Eq.\@ (\ref{eq:Bethe_Salpeter_pi_alpha}) by $i\Omega_n$ and Eq.\@ (\ref{eq:Bethe_Salpeter_T_alphabeta}) by $\sum_{\beta}q_\beta$, then subtract the first resulting equation by the second. The outcome can be written as 
\begin{align}\label{eq:Bethe_Salpeter_momentum_subtract}
Y_\alpha(\vec{k},\vec{q},i\omega_m,i\Omega_n)&=C_\alpha(\vec{k},\vec{q},i\omega_m,i\Omega_n)+k_B T \sum_{i\omega_m'} \Lambda(i\omega_m,i\omega_m',i\Omega_n) \frac{1}{\mathscr{V}} \sum_{\vec{k}'} G(\vec{k}',i\omega_m') G(\vec{k}'+\vec{q},i\omega_m'+i\Omega_n) \nonumber \\ &\times Y_\alpha(\vec{k}',\vec{q},i\omega_m',i\Omega_n),
\end{align}
where 
\begin{align}\label{eq:Y_alpha}
Y_\alpha(\vec{k},\vec{q},i\omega_m,i\Omega_n)=i\Omega_n \Gamma_{\hat{\pi},\alpha}(\vec{k},\vec{q},i\omega_m,i\Omega_n)-\sum_\beta q_\beta \Gamma_{\hat{\textit{T}},\alpha\beta}(\vec{k},\vec{q},i\omega_m,i\Omega_n),
\end{align}
and
\begin{align}\label{eq:C_alpha}
C_\alpha(\vec{k},\vec{q},i\omega_m,i\Omega_n)=i\Omega_n \Gamma_{\hat{\pi},\alpha}^{(0)}(\vec{k},\vec{q})-\sum_\beta q_\beta \Gamma_{\hat{\textit{T}},\alpha\beta}^{(0)}(\vec{k},\vec{q}).
\end{align}
\end{widetext}
The bare momentum vertex which enters into Eq.\@ (\ref{eq:C_alpha}) is
\begin{equation}\label{eq:momentum_bare_vertex_finiteq}
\Gamma_{\hat{\pi},\alpha}^{(0)}(\vec{k},\vec{q})=k_\alpha+\frac{q_\alpha}{2},
\end{equation}
while for the bare stress vertex at finite $\vec{q}$ it is convenient to define a line integral in momentum space
\begin{equation}\label{eq:stres_bare_vertex_finiteq}
\Gamma_{\hat{\textit{T}},\alpha \beta}^{(0)}(\vec{k},\vec{q})=\left(k_\beta+\frac{q_\beta}{2}\right) \int_0^1 d \lambda \frac{\partial \epsilon_{\vec{k}+\lambda \vec{q}}}{\partial k_\alpha},
\end{equation}
so that the group velocity $\hbar^{-1}\partial \epsilon_{\vec{k}}/\partial k_\alpha$ is calculated along a path from $\vec{k}$ to $\vec{k}+\vec{q}$. 
For a quadratic isotropic dispersion (\ref{eq:xi_quadr}), i.e.\@, $\epsilon_{\vec{k}}=\hbar^2 k^2/(2m)$ with $m$ band electron mass, Eq.\@ (\ref{eq:stres_bare_vertex_finiteq}) specializes to 
\begin{equation}\label{eq:stress_bare_vertex_finiteq_quadratic}
\Gamma_{\hat{\textit{T}},\alpha \beta}^{(0)}(\vec{k},\vec{q})=\frac{\hbar^2}{m}\left(k_\beta+\frac{q_\beta}{2}\right)\left(k_\alpha+\frac{q_\alpha}{2}\right),
\end{equation}
which correctly yields 
\begin{align}\label{eq:diverg_stress_quadratic}
\sum_{\beta} q_\beta \Gamma_{\hat{\textit{T}},\alpha \beta}^{(0)}(\vec{k},\vec{q})&=\left(k_\alpha+\frac{q_\alpha}{2}\right)\left(\frac{\hbar^2}{m} \vec{k}\cdot \vec{q}+\frac{q^2}{2}\right) \nonumber \\ & \equiv\left(k_\alpha+\frac{q_\alpha}{2}\right) \left(\xi_{\vec{k}+\vec{q}}-\xi_{\vec{k}}\right). 
\end{align}
However, the kinetic stress tensor, which is the Noether current for translations, is not uniquely defined \cite{He-2014}: for instance, one could equivalently arrive at the divergence (\ref{eq:diverg_stress_quadratic}) by using the modified bare stress tensor \cite{He-2014}
\begin{align}\label{eq:stress_bare_vertex_finiteq_quadratic_He-2014}
\Gamma_{\hat{\textit{T}},\alpha \beta}^{(0)}(\vec{k},\vec{q})&=\hbar^2  \frac{(k_\alpha+q_\alpha) k_\beta+(k_\beta+q_\beta) k_\alpha}{2m} \nonumber \\ & + \delta_{\alpha \beta} \frac{\hbar^2 q^2}{4m}. 
\end{align}
Returning to the most general definition (\ref{eq:stres_bare_vertex_finiteq}), from the fundamental theorem of calculus we can calculate the divergence of the bare stress tensor by 
\begin{align}\label{eq:stres_bare_vertex_finiteq_divergence}
\sum_\beta q_\beta &\Gamma_{\hat{\textit{T}},\alpha \beta}^{(0)}(\vec{k},\vec{q})=\sum_\beta q_\beta \left(k_\alpha+\frac{q_\alpha}{2}\right) \int_0^1 d \lambda \frac{\partial \epsilon_{\vec{k}+\lambda \vec{q}}}{\partial k_\beta} \nonumber \\ &= \left(k_\alpha+\frac{q_\alpha}{2}\right)\int_0^1 d\lambda \left(\sum_{\beta} q_\beta \frac{\partial \epsilon_{\vec{k}+\lambda \vec{q}}}{\partial k_\beta}\right) \nonumber \\ &= \left(k_\alpha+\frac{q_\alpha}{2}\right)\int_0^1 d\lambda \frac{d}{d \lambda} \epsilon_{\vec{k}+\lambda \vec{q}} \nonumber \\ &= \left(k_\alpha+\frac{q_\alpha}{2}\right)\left( \epsilon_{\vec{k}+\vec{q}}-\epsilon_{\vec{k}}\right)\nonumber \\ &= \left(k_\alpha+\frac{q_\alpha}{2}\right)\left( \xi_{\vec{k}+\vec{q}}-\xi_{\vec{k}}\right).
\end{align}
Therefore, through Eq.\@ (\ref{eq:stres_bare_vertex_finiteq_divergence}) we realize that the momentum vertex (\ref{eq:momentum_bare_vertex_finiteq}) and the stress vertex (\ref{eq:stres_bare_vertex_finiteq}) satisfy the bare momentum-conservation Ward identity
\begin{widetext}
\begin{align}\label{eq:momentum_Ward_bare}
C_\alpha(\vec{k},\vec{q},i\omega_m,i\Omega_n)&= \left(k_\alpha+\frac{q_\alpha}{2}\right)\left[G_0^{-1}(\vec{k}+\vec{q},i\omega_m+i\Omega_n)-G_0^{-1}(\vec{k},i\omega_m)\right] \nonumber \\ & =\left(k_\alpha+\frac{q_\alpha}{2}\right) \left[i\Omega_n-\xi_{\vec{k}+\vec{q}}+\xi_{\vec{k}}\right] \nonumber \\ &=\left(k_\alpha+\frac{q_\alpha}{2}\right)\left[X(\vec{k},\vec{q},i\omega_m,i\Omega_n)+\Sigma(i\omega_m+i\Omega_n)-\Sigma(i\omega_m)\right], 
\end{align}
where the shorthand notation given by Eq.\@ (\ref{eq:C_alpha}) has been employed, and the quantity $X(\vec{k},\vec{q},i\omega_m,i\Omega_n)$ satisfies Eq.\@ (\ref{eq:X_def}), identically to the case of the charge-conservation Ward identity. Exactly as in the case for charge conservation we adopt the Kadanoff-Baym conserving prescription (\ref{eq:Kadanoff_Baym_Sigma}). Then, from the derivation of the charge-conservation Ward identity, we have
\begin{equation}\label{eq:X_G}
X(\vec{k},\vec{q},i\omega_m,i\Omega_n)=G^{-1}(\vec{k}+\vec{q},i\omega_m+i\Omega_n)-G^{-1}(\vec{k},i\omega_m),
\end{equation}
away from density collective modes. Eq.\@ (\ref{eq:X_G}) also implies that $(k_\alpha+q_\alpha/2)X(\vec{k},\vec{q},i\omega_m,i\Omega_n)$ satisfies the same Bethe-Salpeter equation as $Y_\alpha(\vec{k},\vec{q},i\omega_m,i\Omega_n)$, that is, Eq.\@ (\ref{eq:Bethe_Salpeter_momentum_subtract}), provided that we are not at a momentum-density collective-mode resonance. Away from such collective modes, we then have $(k_\alpha+q_\alpha/2)X(\vec{k},\vec{q},i\omega_m,i\Omega_n)=Y_\alpha(\vec{k},\vec{q},i\omega_m,i\Omega_n)$, to be combined with Eq.\@ (\ref{eq:Bethe_Salpeter_momentum_subtract}) with the result
\begin{align}
& i \Omega_n \Gamma_{\hat{\pi},\alpha}(\vec{k},\vec{q},i\omega_n, i\Omega_n)-\sum_\beta q_\beta \cdot \vec{\Gamma}_{\hat{\textit{T}},\alpha\beta}(\vec{k},\vec{q},i\omega_n, i\Omega_n) \nonumber =\left(k_\alpha+\frac{q_\alpha}{2}\right)\left[G^{-1}(\vec{k}+\vec{q},i\omega_n+i\Omega_n) -G^{-1}(\vec{k},i\omega_n)\right], 
\end{align}
which coincides with Eq.\@ (\ref{eq:Ward_identity_momentum}).
\end{widetext}

Further remarks concerning the symmetry of the proposed bare stress tensor vertex (\ref{eq:stres_bare_vertex_finiteq}) are in order. Physically, the Ward identity (\ref{eq:Ward_identity_momentum}), specialized to the \emph{bare} vertices as in Eq.\@ (\ref{eq:momentum_Ward_bare}), constrains just the longitudinal (i.e.\@, parallel to $\vec{q}$) part of the bare stress vertex, according to
\begin{align}\label{eq:diverg_stress_gen}
\sum_{\alpha} q_\alpha \Gamma_{\hat{\textit{T}},\alpha \beta}^{(0)}(\vec{k},\vec{q})&=\left(k_\beta+\frac{q_\beta}{2}\right) \left(\xi_{\vec{k}+\vec{q}}-\xi_{\vec{k}}\right). 
\end{align}
Therefore, any choice of the bare stress vertex $\Gamma_{\hat{\textit{T}},\alpha \beta}^{(0)}(\vec{k},\vec{q})$ with divergence in momentum space compatible with Eq.\@ (\ref{eq:diverg_stress_gen}) obeys the momentum-conservation Ward identity (\ref{eq:momentum_Ward_bare}) by construction. 
Within this context, let us observe that the stress-vertex choice given by Eq.\@ (\ref{eq:stres_bare_vertex_finiteq}) is not explicitly symmetric under the exchange of indices $\alpha \leftrightarrow \beta$ for generic dispersion $\epsilon_{\vec{k}}$, unless we specialize to quadratic isotropic dispersion (\ref{eq:xi_quadr}), which produces Eq.\@ (\ref{eq:stress_bare_vertex_finiteq_quadratic}). 
In general, the symmetry under index exchange $\alpha \leftrightarrow \beta$ of the stress tensor ensures the conservation of orbital angular momentum for a continuous body \cite{Landau-1987fm,Bradlyn-2012}. In classical systems, if the Cauchy stress tensor were not symmetric a net internal moment (torque) would exist on an infinitesimal element. In the quantum case, the asymmetry of the stress tensor signifies that the system possesses intrinsic angular momentum (spin) or other internal degrees of freedom: thus, the field content transforms non-trivially under the Lorentz group. Hence, the stress-tensor asymmetry in the quantum realm means that orbital angular momentum is not conserved on its own, while the \emph{total} angular momentum (including the spin component) is.
Indeed, in quantum field theory, the ``canonical'' stress tensor derived from Noether's theorem (via translation symmetry), in a similar way as for Eq.\@ (\ref{eq:stres_bare_vertex_finiteq}), is often asymmetric. To make $\Gamma_{\hat{\textit{T}},\alpha \beta}^{(0)}(\vec{k},\vec{q})$ symmetric, one can add a term involving the spin tensor \cite{Principi-2016,Link-2018a}, resulting in the Belinfante–Rosenfeld tensor \cite{Rosenfeld-1938,Belinfante-1940}, which is symmetric and represents the total energy-momentum. The antisymmetric part of the orbital part of the stress tensor directly relates to the divergence of the spin tensor, ensuring that the total angular momentum is conserved.
Explicitly, the Belinfante-Rosenfeld construction modifies the ``canonical'' non-symmetric stress tensor vertex $\Gamma_{\hat{\textit{T}},\alpha \beta}^{(0)}(\vec{k},\vec{q})$ by adding a divergence-free portion which is antisymmetric in its first two indices:
\begin{equation}\label{eq:Belinfante_Rosenfeld}
\Gamma_{\hat{\textit{T}},\alpha \beta}^{(B)}(\vec{k},\vec{q})=\Gamma_{\hat{\textit{T}},\alpha \beta}^{(0)}(\vec{k},\vec{q})+i \sum_{\gamma} q_\gamma \mathscr{B}^{\gamma \alpha \beta}(\vec{k},\vec{q}),
\end{equation}
where the Belinfante modification reads $\mathscr{B}^{\gamma \alpha \beta}(\vec{k},\vec{q})=-\mathscr{B}^{\alpha \gamma \beta}(\vec{k},\vec{q}) \, \forall \left\{\alpha, \beta,\gamma \right\}$. This term is divergence-free (purely transverse) in momentum space, and thus it does not affect the relation (\ref{eq:diverg_stress_gen}) needed for the momentum-conservation Ward identity, while symmetrizing the stress vertex $\Gamma_{\hat{\textit{T}},\alpha \beta}^{(B)}(\vec{k},\vec{q})$ in Eq.\@ (\ref{eq:Belinfante_Rosenfeld}). 

\vspace{0.5cm}

\section{Derivation of the correlation functions at zero momentum without vertex corrections}\label{App:corr_Theta_bubble_q0_deriv}

\subsection{Coherent regime at zero momentum: Allen formula for the linear-response function}\label{App_Kubo_coherent}

In coherent regime, the transport function at zero momentum can be approximated by its zero-energy value, as in Eq.\@ (\ref{eq:constant_transport_function_q0}). Here, using the identity for Lorentzian functions
\begin{align}\label{eq:Lorentzian_identity}
&\frac{1}{\pi} \int_{-\infty}^{+\infty} d\xi \frac{\alpha_1 \alpha_2}{\left[(\epsilon_1-\xi)^2+\alpha_1^2\right]\left[(\epsilon_2-\xi)^2+\alpha_2^2\right]} \nonumber \\ &=\frac{\alpha_1+\alpha_2}{(\epsilon_1-\epsilon_2)^2+(\alpha_1+\alpha_2)^2}
\end{align}
applied to the spectral functions $A_G(\xi,\epsilon_1)A_G(\xi,\epsilon_2)$, we achieve
\begin{widetext}
\begin{equation}\label{eq:Chi:theta_bubble_expl_q0_FL}
\chi_{\hat{\Theta} \hat{\Theta}}^{\vec{\alpha}\vec{\beta},(0)}(0,i\Omega_n)=\frac{\Phi_{\hat{\Theta}}^{\vec{\alpha}\vec{\beta}}(0)}{\pi} \int_{-\infty}^{+\infty} d \epsilon_1 \int_{-\infty}^{+\infty} d\epsilon_2 \frac{f_{FD}(\epsilon_1)-f_{FD}(\epsilon_2)}{i\Omega_n+\epsilon_1-\epsilon_2}\frac{-\Sigma_2(\epsilon_1)-\Sigma_2(\epsilon_2)}{\left[\epsilon_1 -\Sigma_1(\epsilon_1)-\epsilon_2+\Sigma_1(\epsilon_2)\right]^2 \left[\Sigma_2(\epsilon_1)+\Sigma_2(\epsilon_2)\right]^2}. 
\end{equation}
\end{widetext}
In the current set of assumptions, Eq.\@ (\ref{eq:Chi:theta_bubble_expl_q0_FL}) can be recast into an ``Allen formula'' for the corresponding observable $\hat{\Theta}$ \cite{Berthod-2013,Allen-2015} if the latter follows a linear-response Kubo formula given by Eq.\@ (\ref{eq:Kubo_general_Theta}).

Explicitly, from Eq.\@ (\ref{eq:Chi:theta_bubble_expl_q0_FL}), one can formulate an argument formally analogous to App.\@ A of Ref.\@ \onlinecite{Berthod-2013} (where the $q=0$ optical conductivity, $\Theta_{\vec{\alpha}\vec{\beta}}(0,\omega)\equiv \sigma_{\alpha \beta}(0,\omega)$, was considered): using $1/(x \pm i0^+)= \mathscr{P}\mathscr{P} 1/x \mp i \pi \delta(x)$, one first analytically continues with $i\Omega_n \rightarrow \hbar\omega+i 0^+$, and calculates the imaginary part of the correlation function (\ref{eq:Chi:theta_bubble_expl_q0_FL}), which corresponds to the real (dissipative) part $\mathrm{Re}\left\{\Theta_{\vec{\alpha}\vec{\beta}}(0,\omega)\right\}$ of the Kubo formula (\ref{eq:Kubo_general_Theta}); then, one proves
that $\mathrm{Im}\left\{\Theta_{\vec{\alpha}\vec{\beta}}(0,\omega)\right\}$ is formally analogous to $\mathrm{Re}\left\{\Theta_{\vec{\alpha}\vec{\beta}}(0,\omega)\right\}$, because the resulting complex-valued $\Theta_{\vec{\alpha}\vec{\beta}}(0,\omega)$ in Eq.\@ (\ref{eq:Kubo_general_Theta}) is an analytical function of the variable $\omega$ in the upper half of the complex plane and vanishes faster than $1/\omega$ for $|\omega| \rightarrow +\infty$ for a causal self-energy. The result is a generalized ``Allen formula'' \cite{Berthod-2013,Allen-2015}, i.e.\@, Eq.\@ (\ref{eq:Kubo_Theta_Allen_q0}). 

\subsection{Incoherent regime at zero momentum}\label{App_Kubo_incoherent}

Inserting Eq.\@ (\ref{eq:broad_AG_q0}) for the broad, momentum-independent spectral functions in Eq.\@ (\ref{eq:Chi:theta_bubble_expl_q0}), we can extract the integral over $\xi$ of the transport function from the other nested energy integrals, thus obtaining
\begin{widetext}
\begin{equation}\label{eq:Chi:theta_bubble_expl_q0_NFL}
\chi_{\hat{\Theta} \hat{\Theta}}^{\vec{\alpha}\vec{\beta},(0)}(0,i\Omega_n)=\left[\int_{-\infty}^{+\infty} d \xi \Phi_{\hat{\Theta}}^{\vec{\alpha}\vec{\beta}}(\xi)\right] \int_{-\infty}^{+\infty} d \epsilon_1 \int_{-\infty}^{+\infty} d\epsilon_2 \frac{f_{FD}(\epsilon_1)-f_{FD}(\epsilon_2)}{i\Omega_n+\epsilon_1-\epsilon_2}A_G(\epsilon_1) A_G(\epsilon_2). 
\end{equation}
Let us first find an expression for the imaginary part of Eq.\@ (\ref{eq:Chi:theta_bubble_expl_q0_NFL}). To this purpose, we analytically continue with $i\Omega_n \rightarrow \hbar\omega+i 0^+$, and we take the imaginary part of the resulting expression with $1/(x \pm i0^+)= \mathscr{P}\mathscr{P} 1/x \mp i \pi \delta(x)$, thus obtaining 
\begin{align}\label{eq:Chi:theta_bubble_expl_q0_NFL_2}
\mathrm{Im}\left\{\chi_{\hat{\Theta} \hat{\Theta}}^{R,\vec{\alpha}\vec{\beta},(0)}(0,\omega)\right\}&=-\frac{\left[\int_{-\infty}^{+\infty} d \xi \Phi_{\hat{\Theta}}^{\vec{\alpha}\vec{\beta}}(\xi)\right]}{\pi} \int_{-\infty}^{+\infty} d \epsilon_1 \left[f_{FD}(\epsilon_1)-f_{FD}(\epsilon_1+\hbar\omega)\right] \frac{\Sigma_2(\epsilon_1)}{\left[\epsilon_1 -\Sigma_1(\epsilon_1)\right]^2 +\left[\Sigma_2(\epsilon_1)\right]^2} \nonumber \\ & \times \frac{\Sigma_2(\epsilon_1+\hbar\omega)}{\left[\epsilon_1+\hbar\omega -\Sigma_1(\epsilon_1+\hbar\omega)\right]^2+ \left[\Sigma_2(\epsilon_1+\hbar\omega)\right]^2} \nonumber \\ &= -\frac{\left[\int_{-\infty}^{+\infty} d \xi \Phi_{\hat{\Theta}}^{\vec{\alpha}\vec{\beta}}(\xi)\right]}{\pi} \int_{-\infty}^{+\infty} d \epsilon_1 \left[f_{FD}(\epsilon_1)-f_{FD}(\epsilon_1+\hbar\omega)\right] \mathrm{Im}\left\{\overline{G}^R(\epsilon_1)\right\} \mathrm{Im}\left\{\overline{G}^R(\epsilon_1+\hbar\omega)\right\},
\end{align}
where $\overline{G}^R(\epsilon)=1/\left[\epsilon-\Sigma^R(\epsilon)\right]=\int_{-\infty}^{+\infty} d\epsilon_2 A_G(\epsilon_2)/\left[\epsilon+i0^+ -\epsilon_2\right]$ is the local retarded Green's function. Expressing the product of imaginary parts at the last line of Eq.\@ (\ref{eq:Chi:theta_bubble_expl_q0_NFL_2}) as a single real part, using $\mathrm{Im}\left\{a\right\}\mathrm{Im}\left\{b\right\}=\mathrm{Re}\left\{a b^\ast-ab\right\}/2$, we arrive at
\begin{align}\label{eq:Chi:theta_bubble_expl_q0_NFL_3}
\mathrm{Im}\left\{\chi_{\hat{\Theta} \hat{\Theta}}^{R,\vec{\alpha}\vec{\beta},(0)}(0,\omega)\right\}&=-\frac{\left[\int_{-\infty}^{+\infty} d \xi \Phi_{\hat{\Theta}}^{\vec{\alpha}\vec{\beta}}(\xi)\right]}{2\pi} \mathrm{Im}\left\{i\int_{-\infty}^{+\infty} d \epsilon \left[f_{FD}(\epsilon)-f_{FD}(\epsilon+\hbar\omega)\right] \right. \nonumber \\ & \left.
\left\{\frac{1}{\left[\epsilon-\Sigma^R(\epsilon)\right]\left\{\epsilon+\hbar\omega-\left[\Sigma^R(\epsilon+\hbar\omega)\right]^\ast\right\}} -\frac{1}{\left[\epsilon-\Sigma^R(\epsilon)\right]\left[\epsilon+\hbar\omega-\Sigma^R(\epsilon+\hbar\omega)\right]}\right\}\right\}.
\end{align}
\end{widetext} 
Eq.\@ (\ref{eq:Chi:theta_bubble_expl_q0_NFL_3}) determines the real part of the linear-response function (\ref{eq:Kubo_general_Theta}). To obtain the full complex-valued correlation function, one can apply Kramers-Kronig relations to Eq.\@ (\ref{eq:Chi:theta_bubble_expl_q0_NFL_3}) to find the corresponding real part, or one can work directly with the complex-valued Eq.\@ (\ref{eq:Chi:theta_bubble_expl_q0_NFL_2}), where one can separate the integrals over $\epsilon_1$ and $\epsilon_2$ exploiting the momentum independence of $A_G(\epsilon)$. The result is
\begin{align}\label{eq:Chi:theta_bubble_expl_q0_NFL_4}
\chi_{\hat{\Theta} \hat{\Theta}}^{R,\vec{\alpha}\vec{\beta},(0)}(0,\omega)&=\left[\int_{-\infty}^{+\infty} d \xi \Phi_{\hat{\Theta}}^{\vec{\alpha}\vec{\beta}}(\xi)\right] \int_{-\infty}^{+\infty} d \epsilon f_{FD}(\epsilon) A_G(\epsilon) \nonumber \\ & \times \left[ \overline{G}^R(\epsilon+\hbar\omega)+\overline{G}^A(\epsilon-\hbar\omega)\right],
\end{align}
where $\overline{G}^A(\epsilon)=1/\left\{\epsilon-\left[\Sigma^R(\epsilon)\right]^\ast\right\}=\int_{-\infty}^{+\infty} d\epsilon_2 A_G(\epsilon_2)/\left[\epsilon-i0^+ -\epsilon_2\right]$ is the advanced local Green's function.
Now, inserting the explicit expressions for the local Green's functions and the spectral function in Eq.\@ (\ref{eq:Chi:theta_bubble_expl_q0_NFL_4}), we arrive at Eq.\@ (\ref{eq:Kubo_bubble_expl_q0_NFL}).  

\section{Two-band tight-binding toy model for the GaAs bandstructure}\label{app:GaAs_tight_binding}

In order to construct the electronic bands in the example of the GaAs bandstructure of Fig.\@ \ref{fig:GaAs}, I employ the following matrix Hamiltonian:
\begin{align}\label{eq:H_GaAs}
\underline{\underline{\hat{H}}}_{(GaAs)}&=\hat{H}_0(\vec{k}) \otimes \underline{\underline{\mathbb{I}}}_{\sigma} +\underline{\underline{\mathbb{I}}}_{o} \otimes  \hat{H}_D(\vec{k}) \cdot \vec{\sigma} \nonumber \\ & + \underline{\underline{\mathbb{I}}}_{o} \otimes \frac{g_S}{2} \mu_B \vec{B} \cdot \vec{\sigma}.
\end{align}
The unperturbed dispersion in Eq.\@ (\ref{eq:H_GaAs}) results from the $2 \times 2$ Hamiltonian 
\begin{equation}\label{eq:dispersion_fcc_bare}
\hat{H}_0(\vec{k})=\begin{bmatrix}
\epsilon_{\mathrm{Ga}} & t f(\vec{k})\\
t f^\ast(\vec{k}) & \epsilon_{\mathrm{As}} \\
\end{bmatrix}, 
\end{equation}
with the energies $\epsilon_{\mathrm{Ga}}$ and $\epsilon_{\mathrm{As}}$ for Ga and As atoms respectively, as well as the structure factor
\begin{align}\label{eq:struct_factor_GaAs}
f(\vec{k})&=\sum_{i=1}^4 e^{i \vec{k} \cdot \vec{d}_i} \nonumber \\ & =4\left[\cos\left(\frac{k_x a}{4}\right)\cos\left(\frac{k_y a}{4}\right)\cos\left(\frac{k_z a}{4}\right) \right. \nonumber \\ & \left. -i \sin\left(\frac{k_x a}{4}\right)\sin\left(\frac{k_y a}{4}\right)\sin\left(\frac{k_z a}{4}\right)\right],
\end{align}
and the connecting vectors $\vec{d}_i=(a/4)\left\{\left(1,1,1\right), \left(1,-1,-1\right),\left(-1,1,-1\right),\left(-1,-1,1\right)\right\}$. The odd-in-$\vec{k}$ Dresselhaus spin-orbit term in Eq.\@ (\ref{eq:H_GaAs}) satisfies 
\begin{align}\label{eq:Dresselhaus_SOC}
\hat{H}_D(\vec{k})&=\gamma_{D} \left[k_x (k_y^2-k_z^2) \underline{\underline{\sigma}}_{x}+k_y (k_z^2-k_x^2) \underline{\underline{\sigma}}_{y} \right. \nonumber \\ & \left. +k_z (k_x^2-k_y^2) \underline{\underline{\sigma}}_{z}\right],
\end{align}
where $\underline{\underline{\sigma}}_{i}$, and $i=\left\{x,y,z\right\}$, are Pauli matrices Eq.\@ (\ref{eq:H_GaAs}) also includes the Zeeman term $g_S \vec{B} \cdot \vec{\sigma}/2$ , which contains the applied magnetic field $B$, the Land\'{e} g-factor $g_S\approx 2.0023$ for free electrons, the Bohr magneton $\mu_B=e \hbar/(2m)$ with $e$ free electron charge and $m$ free electron mass, and the Pauli-matrices vector $\vec{\sigma}=\left\{\underline{\underline{\sigma}}_{x}, \underline{\underline{\sigma}}_{y}, \underline{\underline{\sigma}}_{z} \right\}$. 
Moreover, in Eq.\@ (\ref{eq:H_GaAs}) we have identity operators in spin $\left\{\left| \uparrow \right\rangle,\left|\downarrow\right\rangle \right\}$ and orbital $\left\{\left| \mathrm{Ga} \right\rangle,\left|\mathrm{As}\right\rangle \right\}$ space respectively, 
\begin{equation}\label{eq:I_spin}
\underline{\underline{\mathbb{I}}}_{\sigma}=\begin{bmatrix} 1 & 0 \\ 0 & 1\end{bmatrix}_{\left\{\left| \uparrow \right\rangle,\left|\downarrow\right\rangle \right\}},
\end{equation}
\begin{equation}\label{eq:I_orb}
\underline{\underline{\mathbb{I}}}_{o}=\begin{bmatrix} 1 & 0 \\ 0 & 1\end{bmatrix}_{\left\{\left| \mathrm{Ga} \right\rangle,\left|\mathrm{As}\right\rangle \right\}},
\end{equation}
and $\otimes$ denotes tensorial product.

For Fig.\@ \ref{fig:GaAs}(b), I have used the following parameters: $\epsilon_{\mathrm{Ga}}=t$, $\epsilon_{\mathrm{As}}=-t$, and $\gamma_D=B=0$. Fig.\@ \ref{fig:GaAs}(c) adds spin-orbit coupling and Zeeman splitting, specifically $\gamma_D=0.01 t a^3$ and $g_S \mu_B B/2=0.4 t$.

\end{document}